# Statistical Dynamics and Subgrid Modelling of Turbulence: From Isotropic to Inhomogeneous

Jorgen S. Frederiksen [1]*, Vassili Kitsios[1] and Terence J. O'Kane[2]

[1] CSIRO Environment, Aspendale, Melbourne, 3195, Australia
[2] CSIRO Environment, Hobart, 7004, Australia
* Correspondence: jorgen.frederiksen@csiro.au

**Abstract:** Turbulence is the most important, ubiquitous, and difficult problem of classical physics. Feynman viewed it as essentially unsolved, without a rigorous mathematical basis to describe the statistical dynamics of this most complex of fluid motion. However, the paradigm shift came in 1959, just 10 years after the establishment of quantum electrodynamics (QED), with the formulation of the Eulerian statistical dynamical theory by Kraichnan, which he called the direct interaction approximation (DIA) closure. It was again based on renormalized perturbation theory, like QED, and is a bare vertex theory that is manifestly realizable. Unprecedented progress followed, over several decades, in understanding, in the skill of closure models, and in direct numerical simulations (DNS) as benchmarks, focused on homogeneous turbulence and subgrid modelling. Here, we review some of these exciting achievements concentrating on theoretical developments and comparisons between closure calculations and DNS ensembles.

We also document in some detail the progress that has been made in extending statistical dynamical turbulence theory to the real world of interactions with mean flows, waves and inhomogeneities such as topography. This is an area that historically has been difficult to progress based on renormalized perturbation theory. The groundwork has now been laid for the implementation of numerically efficient inhomogeneous closures, like the realizable quasi-diagonal direct interaction approximation (QDIA), and even more efficient Markovian inhomogeneous closures (MICs), that have been shown to be skillful in original form or with regularization. Recent developments include the formulation and testing of an eddy damped Markovian anisotropic closure (EDMAC) that is realizable in interactions with transient waves but is as efficient as the eddy damped quasi-normal Markovian (EDQNM). As well a similarly efficient closure, the realizable eddy damped Markovian inhomogeneous closure (EDMIC) has been developed.

Moreover, we present subgrid models that cater for the complex interactions that occur in geophysical flows and are necessary for skillful prognosis in large eddy simulations. Recent progress includes the determination of complete sets of subgrid terms for skillful large eddy simulations of baroclinic inhomogeneous turbulent atmospheric and oceanic flows interacting with Rossby waves and topography. The success of these inhomogeneous closures has also led to further applications in data assimilation and ensemble prediction and generalization to quantum fields.

**Keywords:** Markovian closures; non-Markovian closures; isotropic turbulence; anisotropic turbulence; inhomogeneous turbulence; Rossby waves; realizability; subgrid modelling

# 1. Introduction

## 1.1 Statistical Dynamical Closure Theory

Closure theory is the foundation for the statistical dynamics of general turbulent flows, formulated in terms of low order cumulants. It also provides a basis for determining subgrid scale parameterizations of unresolved eddies which are needed to ensure that large-eddy simulations (LES) perform skillfully at reduced resolutions. The major problem in deriving successful turbulence closure models is truncating the statistical dynamical equations in terms of the lower-order moments or cumulants. Consider for example the common case of quadratic nonlinearity as occurs in the Navier Stokes equations for both two-dimensional (2D) and three-dimensional (3D) flows. Then, the mean field equation is coupled to the second-order moment, and the equation for the second-order moment is in turn coupled to the third-order moment, and so on in an infinite hierarchy of moment equations. Much of statistical closure theory has focused on homogeneous isotropic turbulence (HIT) with zero mean field since this somewhat idealized problem nevertheless captures many of the essential issues of this complex field. Millionshtchikov (1941) [1] proposed such a homogeneous closure based on the quasi-normal approximation. This involves writing the fourth-order moment in terms of the fourth order cumulant (four-point function), taken to be zero, and sums of products of second order cumulants (two-point functions). The third-order moment, that appears in the equation for the second-order term, is then expressed in terms of the second-order terms and this closes the second-order equation. Unfortunately, Ogura (1963) [2] found in numerical integrations of the quasi-normal closure that it was not realizable and led to unphysical negative energy spectra. It should be noted however that historically closures based on cumulant-discard and quasi-normal hypotheses have been used for short term barotropic model predictions and are still in use in studies of eddy-zonal flow interactions (Frederiksen and O'Kane (2019) [3]; Marston et al. (2023) [4] review the literature).

### 1.1.1. Eulerian Non-Markovian Closures for Isotropic Turbulence

The big advance in the statistical dynamical theory of turbulence came with Kraichnan's (1959a, 1961) [5,6] direct interaction approximation (DIA) closure theory for homogeneous turbulence. As noted by Kraichnan (1959) [5]: "This procedure, which we term the direct-interaction approximation has a simple dynamical significance and can be shown to lead to equations which are self consistent in the sense that they yield rigorously realizable second-order moments". This realizability is a consequence of the DIA having an exact stochastic model representation (Kraichnan, 1961 [6])). In the language of modern physics, the DIA equations are structurally the same as the Schwinger-Dyson equations (Schwinger 1948 [7]; Dyson 1949 [8])) in the bare vertex approximation (Frederiksen (2017) [9] reviews the literature). Closely similar non-Markovian closures to the DIA were subsequently developed independently, first in steady state form by Edwards (1964) [10] and then in both steady state and time-dependent forms by Herring (1965, 1966) [11,12] and McComb (1974, 1978) [13,14]. Herring's (1966) [12] self consistent field theory (SCFT) and McComb's (1978) [14] local energy transfer theory (LET) closures in fact only to differ from the DIA in how the two-time cumulant or response function are determined. These three Eulerian non-Markovian closures all have the same single-time two-point closure equation. However, the SCFT and LET effectively invoke a fluctuation dissipation theorem (FDT) (Kraichnan 1959b [15]; Frederiksen at al. 1994 [16]; Kiyani and McComb 2004 [17]). This is the *prior-time FDT* (Carnevale and Frederiksen 1983 [18], Equation (3.5)), defined by

$$C_{\mathbf{k}}(t,t') = R_{\mathbf{k}}(t,t')C_{\mathbf{k}}(t',t') \quad (1)$$

for $t \geq t'$ where $C_{\mathbf{k}}(t,t')$ is the two-time cumulant, $R_{\mathbf{k}}(t,t')$ is the response function and $C_{\mathbf{k}}(t',t')$ is the prior-time single-time cumulant. Note that $C_{\mathbf{k}}(t,t') = C_{-\mathbf{k}}(t',t)$ for $t' > t$. The SCFT and DIA response functions are identical and the SCFT two-time cumulant is obtained from the prior-time FDT. On the other hand, the LET and DIA two-time cumulants are identical and the response function is obtained from Equation (1).

Wyld (1961) [19] and Lee (1965) [20] approached the turbulence closure problem using perturbation theory and renormalization and depicted the expansion terms through Feynman diagrams (Feynman 1949 [21]). Wyld considered diagrams up to fourth order and Lee to sixth order in the nonlinearity. Lee noted that Wyld had mistakenly used the renormalized propagator, or response function, instead of the bare expression in some terms. At second order, where the vertex is bare, the renormalized perturbation theory expressions were shown to reproduce Kraichnan's DIA. Kraichnan (1961) [6] also used a diagrammatic representation of perturbation series in his stochastic model approach to closures and their realizability.

The functional operator approach that led to the Schwinger-Dyson equations for quantum field theories was generalized by Martin et al. (1973) [22] (hereafter MSR) for classical statistical dynamics. They found it necessary to also include the adjoint operator, as well as the dynamical field, to generate the response functions, and the cumulants. Perhaps the most natural and elegant way to formulate statistical field theories is through the equivalent path integral formalism (Feynman and Hibbs 1965 [23], Zee 2010 [24]). This was the approach subsequently taken to classical systems by several authors including Phythian (1977) [25] with the most general, and very clear exposition, by Jensen (1981) [26] who reviews the early literature.

Martin et al. (1973) [21] contended that their functional operator approach gave results that disagreed with Wyld: "We see that beyond fourth order in $\gamma_3$ the renormalization is not as he surmised". However, the important study of Berera et al. (2013) [27] showed that, on correcting some minor errors in both papers, the Wyld diagrammatic approach and the MSR functional formalism in fact give the same results to fourth order, as they should.

As discussed in more detail in Section 4.4, at moderate Reynolds numbers the DIA is in reasonably good agreement with wind tunnel measurements, and with DNS, at the large energy containing scales (Kraichnan 1964a [28]; Herring et al. 1974 [29]; McComb and Shanmugasundaram 1984 [30]; Frederiksen and Davies 2000 [31]). The high Reynolds number spectral behaviour of the DIA is however slightly different from the $k^{-\frac{5}{3}}$ inertial range power laws for 3D turbulence (Kraichnan 1964b) [32], and from the $k^{-3}$ enstrophy cascading inertial range for 2D turbulence (Herring et al. 1974) [29]. The DIA does not account for higher order interactions that renormalize the bare vertex and this is the likely theoretical reason for the slight differences in power laws. The physical basis is that convection (advection) effects are not distinguished from intrinsic distortion effects in the DIA (Kraichnan 1964b [32]; Frederiksen and Davies 2004 [33]). Thus, the higher order terms act to localize the triad interactions. At finite Reynolds numbers, the SCFT and LET closures give very similar results to the DIA for 2D isotropic turbulence (Frederiksen and Davies 2000) [31]. McComb (1990) [34] further analyzed the cause of the differences between the DIA and the Kolmogorov (1941a, 1941b) [35,36] (K41) scaling laws. He notes that the LET closure is consistent with K41 at very high Reynolds numbers and numerical evidence is presented for 3D turbulence by McComb and Shanmugasundaram (1984) [30].

1.1.2. Quasi-Lagrangian Non-Markovian Closures for Isotropic Turbulence

The scaling law deficiency motivated Kraichnan (1965, 1977) [37,38] and Herring and Kraichnan (1979) [39] to formulate quasi-Lagrangian non-Markovian closures for isotropic turbulence. They were followed by alternative formulations by Kaneda (1981) [40] and Gotoh et al. (1988) [41]. The quasi-Lagrangian closures, like the Eulerian DIA, are independent of tuning parameters, however, the abridged versions employ further approximations. Importantly, unlike the Eulerian DIA, the quasi-Lagrangian closures depend on formulations and on norm or basic field variable or representative.

### 1.1.3. Vertex Renormalization

The quasi-Lagrangian closures, like the second order Eulerian non-Markovian closures, do not fundamentally address the vertex renormalization problem. Rather they perform a field variable dependent transformation that attempts to ameliorate some of the deficiencies of second order closure. Martin et al. (1973) [22] emphasized that a fundamental treatment of vertex renormalization would be needed for a parameter free theory of strong turbulence. Unfortunately, this is an extremely difficult problem, as discussed in more detail in Section 12, and has remained elusive for field theories with strong nonlinearity. Before embarking on the impressive, but not wholly fruitful, effort of quasi-Lagrangian closures, Kraichnan (1964b) [32] had recognized that the DIA power law inconsistency with K41 could be overcome by localizing the interactions in the two-time cumulants and response functions. He noted: "Since the modified equation removes the convection effects of large scales on small scales, it represents, in effect, a transformation to the quasi-Lagrangian coordinates called for in the original statement of Kolmogorov's theory (1941)." Moreover, in a prescient observation he noted: "It is difficult to construct a a well-defined transformation to quasi-Lagrangian coordinates in x space which is similarly consistent". Indeed, such a one-parameter regularization, or empirical vertex renormalization, of the Eulerian DIA gives very good agreement with the statistics of DNS for isotropic 2D turbulence (Frederiksen and Davies 2004 [33]). Furthermore, Frederiksen and Davies (2004) [33] found that the regularized DIA performs better than quasi-Lagrangian closures, as discussed further in Section 5. It has also been found for both 2D and 3D isotropic and 2D inhomogeneous turbulence that good agreement of the DIA (Sudan and Pfirsch (1985) [42]; Frederiksen and Davies 2004 [33]) and QDIA (O'Kane and Frederiksen 2004 [43]) closures with DNS occurs with a regularization parameter that is almost universal at $\alpha \approx 4$.

### 1.1.4. Markovian Closures for Homogeneous Turbulence

Orszag (1970) [44] effectively resurrected a variant of the quasi-normal approach to turbulence closure with his Eddy Damped Quasi-Normal Markovian model. The EDQNM specifies an eddy damping term dependent on an empirical parameter and employs a Markovian approximation to stabilize the closure and ensure realizability for isotropic 3D turbulence. Leith (1971) [45] first formulated, implemented and applied the EDQNM closure code for 2D isotropic turbulence to examine atmospheric dynamics and predictability. A generalization of the EDQNM for 2D homogeneous anisotropic turbulence (without waves) was developed by Herring (1975) [46] and used to study the relaxation of anisotropic turbulence to isotropy. As discussed in more detail in Section 6.2, the functional form of the eddy damping in the EDQNM is chosen to be consistent with the K41 inertial range of 3D turbulence and the enstrophy cascading inertial range of 2D isotropic turbulence.

Orszag (1970) [44] arrived at the EDQNM through a modification of the quasi-normal approach to closure but there are several ways of reaching it. One can again take the DIA as the starting point and reduce it from there by using the current-time FDT and a Markov form for the response function with an empirical form for the eddy damping. The *current-time FDT* relates the two-time spectral cumulant $C_\mathbf{k}(t,t')$ to the response function $R_\mathbf{k}(t,t')$ and the current-time single-time cumulant $C_\mathbf{k}(t,t)$ through

$$C_\mathbf{k}(t,t') = R_\mathbf{k}(t,t')C_\mathbf{k}(t,t) \qquad (2)$$

for $t \geq t'$.

In essence, the EDQNM closure consists of just the Markov equation for the single-time cumulant with the triad relaxation function given by an analytical empirical form. It is therefore very computationally efficient compared with the non-Markovian closures with potentially long time-history integrals.

### 1.1.5. Markovian Closures for Homogeneous Turbulence Interacting with Waves

The big advantage of the EDQNM closure is its numerical efficiency since it is a Markovian equation for just the single-time cumulant with a timing that scales like $O(T_{tot})$ where $T_{tot}$ is the total integration interval. In comparison non-Markovian closures scale like $O(T_{tot}^3)$. The analytical form for the triad relaxation function that carries information about triad interactions is a further important reason for the EDQNM computational efficiency. Consequently, it has been widely employed in many analyses and applications [47–52]. There has however been a long-standing limitation of the EDQNM closure and that is the fact that it may not be realizable when the turbulence interacts with transient waves. This is because during the transient evolution phase the oscillations of the waves mean that the real part of the triad relaxation function may become negative resulting in unphysical behaviour. In some studies, the possibility of instability has been avoided by using the steady state form of the triad relaxation function throughout the evolution [52–58]. Another approach has been to work with modified forms of quasi-normal Markovian closures (Sukoriansky and Galperin 2016 [59]; Galperin and Sukoriansky 2021 [60]).

Bowman et al. (1993) [61] in particular, have emphasized the deficiency of the EDQNM in handling transient waves such as drift waves or equivalently Rossby waves. They traced the problem to the used of the current-time FDT in Equation (2). Although lack of guaranteed realizability for the EDQNM was also shown to occur if the prior-time FDT in Equation (1) was used. However, they developed their Realizable Markovian Closure (RMC) by using a FDT that is essentially a half-way house between the current-time and prior-time FDTs. This FDT, that we call the *correlation FDT* is given by:

$$C_{\mathbf{k}}(t,t') = [C_{\mathbf{k}}(t,t)]^{\frac{1}{2}} R_{\mathbf{k}}(t,t') [C_{\mathbf{k}}(t',t')]^{\frac{1}{2}} \tag{3}$$

for $t \geq t'$. The realizability of the RMC is a major achievement but unfortunately the highly efficient analytical form for the triad relaxation function is lost. Instead, its time evolution needs to be calculated by solving an auxiliary system of differential equations which is a nontrivial task. Formally the RMC scales like $O(F_{aux} T_{tot})$, and so depends linearly on $T_{tot}$, like for the EDQNM, but the factor $F_{aux}$ that multiplies $T_{tot}$ to give the computational effort is much larger.

Further analyses and applications of the realizable Markovian closure that cater for anisotropic turbulence with transient waves and use auxiliary differential equations for relaxation functions have been developed by Hu et al. (1997) [62], Bowman and Krommes (1997) [63], and Frederiksen and O'Kane (2023, 2024) [64,65]. Frederiksen and O'Kane (2023, 2024) [64,65] formulated Markovian Anisotropic Closures (MACs) starting from non-Markovian DIA type closures and using each of the three FDTs. They are conveniently combined to the form:

$$C_{\mathbf{k}}(t,t') = [C_{\mathbf{k}}(t,t)]^{1-X} R_{\mathbf{k}}(t,t') [C_{\mathbf{k}}(t',t')]^{X} \tag{4}$$

for $t \geq t'$ and $C_{\mathbf{k}}(t,t') = C_{-\mathbf{k}}(t',t)$ for $t' > t$. Here, the current-time FDT has $X = 0$, the prior-time FDT is for $X = 1$ and the correlation FDT uses $X = \frac{1}{2}$. Their MAC using the current-time FDT is again realizable. Importantly, Frederiksen and O'Kane (2023, 2024) [64,65] also developed the realizable Eddy Damped Markovian Anisotropic Closure (EDMAC) model with analytical triad relaxation function that is as computationally efficient as the EDQNM but caters for transient waves.

1.1.6. Eulerian Non-Markovian Closure for Inhomogeneous Turbulence

Over the past two decades there have been considerable advances in the development and performance of closure theories for inhomogeneous turbulent flows. A particular focus has been on the development of a computationally tractable non-Markovian closure theory. This effort has led to the Quasi-diagonal Direct Interaction Approximation (QDIA) formulated by Frederiksen (1999) [66]. It was applied to tackle the subgrid modeling problem for two-dimensional turbulent flow over topography, and to develop parameterizations of unresolved eddy-eddy, eddy-mean-field, eddy-topographic and mean-field-mean-field interactions. The QDIA has some commonality with the Eulerian DIA of Kraichnan (1959a) [5] in the nonlinear damping and noise (self-energy) terms for the eddy-eddy interactions that appear in the two-point cumulant and response function equations. However, the QDIA equations also contain nonlinear damping and noise (self-energy) terms due to interactions between the eddies, mean-fields and topography. In addition, the mean-field is not zero as in the DIA but it has a prognostic integro-differential equation with self-energy terms due to interactions between eddies, mean-fields and topography. The structure of the QDIA closure theory, with nonlinear damping and forcing represented by self-energy terms, means that it can be used as a transparent, general framework for developing subgrid parameterizations of unresolved scales. It is applicable to a wide range of problems in field theories, including turbulent fluids.

The QDIA statistical closure was first implemented numerically by O'Kane and Frederiksen (2004) [43] for turbulent flow over single realization topography on an $f-$ plane. It was shown to have similar performance to the DIA for homogeneous turbulence, but with the benefit of the mean fields and topography helping to localize interactions, and it was found to be only a few times more computationally demanding.

The QDIA theory was generalized to inhomogeneous turbulent flows interacting with Rossby waves and topography on a $\beta-$ plane by Frederiksen and O'Kane (2005) [67]. In studies of Rossby wave dispersion and predictability in a turbulent environment the QDIA was found to perform remarkably well, with 10-day evolved mean-field pattern correlations as high as $r = 0.9999$ with 1800 member ensembles of DNS. This is a remarkable result for a far from equilibrium process. The QDIA and regularized variants, have been extensively tested in application to problems in dynamics (O'Kane and Frederiksen 2004; Frederiksen and O'Kane 2005) [43,67], predictability (Frederiksen and O'Kane 2005; O'Kane and Frederiksen 2008a) [67,68], data assimilation (O'Kane and Frederiksen 2008b, 2010) [69,70] and subgrid modeling (O'Kane and Frederiksen 2008) [71].

The QDIA theory has been generalized to encompass multi-level and multi-field models for classical and quantum field theories with quadratic nonlinearity (Frederiksen 2017) [9], including quasigeostrophic (QG) baroclinic and three-dimensional (3D) turbulence (Frederiksen 2012a, 2012b) [72,73].

1.1.7. Non-Gaussian Initial Conditions and Restarting of Closure Calculations

The non-Markovian closures reviewed in Sections 1.1.1 and 1.1.6 require the solution of sets of integro-differential equations. The limits in the time-history integrals range from the initial start time to the end time, with the computational task becoming increasingly larger as the integration progresses, and scaling like $O(T_{tot}^3)$. There is a way to ameliorate, to some extent, this considerable computational challenge, which was proposed by Rose (1985) [74]. This involves truncating the integrals, calculating the three-point cumulant, which is then used in the non-Gaussian initial conditions for the restart. Perodically employing this method results in a numerical scheme that scales like $O(N_{tot}T_{tot}^2)$ where $N_{tot}$ is the number of restarts. This three-point cumulant restart method was applied in studies of 2D isotropic turbulence with DIA, SCFT and LET closures by Frederiksen et al. (1994, 2000, 2004) [16,31,33] and resulted in considerable improvements in efficiency. The method was generalized to inhomogeneous turbulent flows by O'Kane and Frederiksen (2004 [43]). This involved calculating updated covariances and using them in periodic restarts in both the mean field and two-point cumulant equations. This generalized restart methodology, described by O'Kane and Frederiksen (2004) [43] and Frederiksen and O'Kane (2005) [67] is based on the properties of the QDIA outlined in Appendix C. It has been applied in studies of predictability, subgrid modelling and data assimilation (Frederiksen and O'Kane [3,67,75]; O'Kane and Frederiksen [68–71]).

1.1.8. Markovian Inhomogeneous Closures

Recently, Markovian Inhomogeneous Closures (MICs) have also been established for turbulent flows interacting with topography and Rossby waves (Frederiksen and O'Kane 2019, 2022) [3,75]. They were developed from the inhomogeneous non-Markovian QDIA closure by assuming any of the three FDTs in Equation (4). It was found that the MICs, and abridged versions of the MICs, in which the mean-field trajectory in the time history integrals is replaced by the current-time mean-field, were able to closely reproduce the statistics of large ensemble DNS simulations. These calculations were performed for the rapid evolution of turbulent flows and Rossby waves interacting with topography on a generalized $\beta-$plane at relatively low Reynolds numbers.

The MIC and abridged MIC models are more computationally efficient than the non-Markovian QDIA closure. However, they are coupled to auxiliary differential equations for two relaxation functions which adds considerably to the computational effort compared with analytical forms as for the EDQNM. Frederiksen and O'Kane (2022) [75] also formulated the Eddy Damped Markovian Inhomogeneous Closure (EDMIC) with analytical forms for the relaxation functions which is very computationally efficient like the EDQNM but caters for inhomogeneous turbulent flows.

*1.2. Subgrid Modelling*

The adoption of statistical dynamical and stochastic methods has greatly contributed to the progress in subgrid-scale parameterisations development in geophysical and canonical flows. Here we cover some of the key historical studies in the deterministic parameterisations of the atmosphere and ocean, statistical dynamical subgrid closures, and stochastic data-driven methods.

1.2.1. Deterministic Parameterizations for Atmospheric Flows

Ever since the very first simulations of the atmosphere, it has long been recognized that the statistical properties of the large-scale flow are very sensitive to the parameterization of the small unresolved subgrid scales. One of the most widely adopted and celebrated subgrid parameterizations is the empirical eddy viscosity model of Smagorinsky (1963) [76]. In its original form, this model is arguably more appropriate for three-dimensional turbulence than for large-scale quasi-two-dimensional geophysical flows. In the latter case, Laplacian operators raised to various powers are generally adopted. Manabe et al. (1979) [77] identified the sensitivity of the kinetic energy spectra in the largest scales to the grid resolution and specification of the subgrid parameterizations. This has persistently been an issue with general circulation models (GCMs) and continues to this day (Laursen and Eliasen 1989 [78]; Koshyk and Boer 1995 [79]; Kaas et al. 1999 [80]; Frederiksen et al. 2003 [81]). A statistical mechanical explanation for this sensitivity was proposed by Frederiksen et al. (1996) [82] based on the ideas of energy and enstrophy conservation.

The influence eddies have on the mean flow have historically been analyzed on the basis of heat and momentum fluxes. In specific flow configurations, Eliassen and Palm (1961) [83] demonstrated that one could instead treat the eddies as inducing an eddy torque on the zonal flow. This was represented by the divergence of their Eliassen-Palm flux. This approach was latter generalized by Andrews and McIntyre (1976) [84] and Boyd (1976) [85] and applied to other equations of fluid motion. Plumb (1979) [86] then showed that for passive scalars in the presence of eddies, the eddy flux consists of symmetric and anti-symmetric (or skew-symmetric) tensors. The symmetric component diffuses the tracers, whilst the anti-symmetric tensor advects them. The zonally averaged Eliassen-Palm flux approach was generalized to three-dimensional time-averaged atmospheric flows by Hoskins et al. (1983) [87]. The utility of this approach in comparison to working in traditional coordinate systems was reviewed by Andrews et al. (1987) [88], Pfeffer (1987) [89] and Grotjahn (1993) [90].

### 1.2.2. Deterministic Parameterizations for Oceanic Flows

For oceanic GCMs the subgrid parameterization problem is arguably even more challenging and important. The formation of mesoscale eddies is intrinsically linked to the process of baroclinic instability. In the ocean, baroclinic instability occurs at scales at least an order of magnitude smaller than scales in the atmosphere. Global climate simulations, run at typical resolutions, do not explicitly resolve these scales nor processes. Early approaches involved the application of empirical tuned diffusion operators in both horizontal and vertical directions (Bryan and Lewis 1979) [91]. Later, Redi (1982) [92] proposed that the diffusion operator should instead mix passive tracers along isopycnals, which are surfaces of constant density. The diffusion operator is diagonal in the isopycnal coordinate system. This is generally not the case in the vertical height system. The Gent and McWilliams (1990) [93] scheme was subsequently developed for the anti-symmetric eddy flux tensor for temperature and salinity. McDougall and McIntosh (1996) [94] generalized the Eliassen-Palm flux approach to oceanic temperature and salinity transport, in three-dimensional time-averaged flows. They obtained a similar parameterization to that of the Gent-McWilliams form (Treguier et al. 1997 [95]; Griffies 1998 [96]). At present, there is no such approach applicable to the momentum equations. In these approaches the resolved scale fields are interpreted as being the mean fields and the subgrid-scale eddies are the transient fluctuations (de Szoeke and Bennett 1993) [97]. However, de Szoeke and Bennett (1993) [97] note: ``This requires the averaging scale to be in a spectral gap between the small-scale fluctuations and the resolved large-scale variations''. This is not a property of the real ocean – see Griffies et al. (2000) [98] for a review of the literature on this topic.

An additional challenge in subgrid parameterization development is to adequately capture the influence of topography and other such inhomogeneities. With application to oceanic flows, Holloway (1992) [99] developed a heuristic parameterization representing the interactions between resolved scale bathymetry and subgrid-scale eddies. This parameterization was based on the notion that when dissipation is weak, quasigeostrophic eddies over bathymetry, or topography, decay through a series of minimum enstrophy states (Bretherton and Haidvogel 1976) [100]. In fact, these stable states are equivalently maximum entropy states or statistical mechanical equilibrium solutions (Frederiksen and Carnevale 1986 [101]; Carnevale and Frederiksen 1987 [102]). The correspondence between nonlinearly stable states and canonical equilibrium solutions extends to baroclinic systems (Frederiksen 1991a, 1991b [103,104]; Majda and Wang 2006 [105]). The idea of Holloway is that, for turbulent eddies over topography, the nonlinear interactions drive the system toward statistical mechanical equilibria. It is this process that determines the subgrid-scale interactions. Holloway (1992) [99] parameterizes this process by an ad hoc eddy tendency term that relaxes the mean flow toward a local canonical equilibrium state. This parameterization has demonstrated improvement in the large-scale flow in inviscid quasigeostrophic models with an idealized topography (Cummins and Holloway, 1993) [106], and in more complex global and regional ocean models (Alvarez et al. 1994) [107]. Subgrid parameterizations for flow over topography have also been developed using entropy production methods in Kazantsev et al. (1998) [108], Polyakov (2001) [109] and Holloway (2004) [110]. See Frederiksen and O'Kane (2008) [111] and Holloway (2009) [112] for reviews of entropy based subgrid modelling methods.

### 1.2.3. Statistical Dynamical Subgrid Modelling

For many subgrid parameterizations it is assumed that the small scales are close to isotropic and that the dissipation and energy production are in balance. Kraichnan (1959a) [5] noted, within the DIA formalism, that for isotropic turbulence the inertial energy transfer can be represented as eddy viscosity and stochastic backscatter terms acting in concert. On average the eddy viscosity drains energy from resolved to the subgrid scales. The backscatter, on the other hand, is a positive semi-definite energy input from the subgrid to the resolved scales. For two-dimensional turbulence, Leith (1971) [45] used the EDQNM closure to derive an eddy dissipation function that generated a $k^{-3}$ kinetic energy spectrum. Kraichnan (1976) [113] then developed an eddy viscosity the theory for both two and three dimensions. In spectral space he identified that the eddy viscosities increased sharply as the wavenumber increased toward the highest resolved cutoff wavenumber. The magnitude of the eddy viscosity is directly proportional to the strength of the interactions between the resolved and subgrid scales. McComb and collaborators have made extensive studies of the subgrid modelling problem and the application to large eddy simulations (McComb and Watt 1990 [114]; Young and McComb 2000 [115]; McComb et al. (2001) [116]; McComb and Johnson 2001 [117]; McComb 2014 [118]). Additional important studies relevant predominantly to three-dimensional homogeneous turbulence include those of Leslie and Quarini (1979) [119], Chollet and Lesieur (1981) [120], Leith (1990) [121], Chasnov (1991) [122], Mason and Thomson (1992) [123], Domaradzki et al. (1993) [124], and Schilling and Zhou (2002) [125].

For barotropic flows on a sphere, Frederiksen and Davies (1997) [126] developed representations of eddy viscosity and stochastic backscatter based on the DIA and EDQNM closures. Atmospheric LES adopting their subgrid parameterizations were in close agreement with higher resolution DNS on the basis of kinetic energy spectra. Implementation of this eddy viscosity parameterization into an atmospheric GCM significantly improved the kinetic energy spectra and general circulation properties (Frederiksen et al. 2003) [81]. General expressions for the eddy-topographic force, eddy viscosity and stochastic backscatter were derived by Frederiksen (1999) [66] based on the QDIA. These terms were calculated by O'Kane and Frederiksen (2008c) [71] for a global time evolving atmospheric flow over topography. The relative magnitude of these terms was quantified and compared with heuristic approaches (Holloway 1992) [99].

Frederiksen (2012a) [72] formulated a multi-field version of the QDIA inhomogeneous closure. This generalization was also formulated for LES with subgrid-scale parameterizations that ensure the large-scale statistical behaviour is independent of resolution. These closures have underpinning non-Markovian stochastic forms, which guarantee realizability of the covariance matrices. These matrices are diagonal in spectral space. The multi-field QDIA was originally formulated for inhomogeneous turbulence and subgrid-scale parameterizations for geophysical flows but is generally applicable to any system of partial differential equations with quadratic nonlinearity.

1.2.4. Stochastic Subgrid Modelling and Data Driven Approaches

Closure models are the natural starting place for developing subgrid scale parameterizations. The complexity of these closures, however, makes it difficult to develop parameterizations for increasingly complex circulation models with many prognostic variables. Data-driven methods offer an alternative. Here the parameterization coefficients governing the subgrid processes are calculated from the subgrid statistics of a high-resolution reference DNS. One such approach is the stochastic modelling method of Frederiksen and Kepert (2006) [127]. Stochastic modelling has had a long and successful history of application to problems in weather and climate (Hasselmann 1976 [128]; Egger 1981 [129]; Palmer 2001 [130]; Majda et al. 2003 [131]; Franzke and Majda 2006 [132]; Shutts 2005 [133]; Seiffert et al. 2006 [134]; Frederiksen et al. 2017 [135]) Stochastic backscatter has been shown to improve ensemble spread and forecast performance in weather prediction (O'Kane and Frederiksen 2008a [68], Berner et al. 2009 [136]).

The wider application of the data-driven stochastic subgrid modelling approach of Frederiksen and Kepert (2006) [128] is discussed in Section 10. Applications include barotropic atmospheres, baroclinic atmospheres, primitive equation atmospheres, baroclinic oceans and fully three-dimensional turbulent boundary layers flows. In Section 11 we illustrate the decomposition of the mean subgrid tendency using closure calculations, and the data-driven approach of Kitsios and Frederiksen (2019) [137].

*1.3. Fluid Dynamical Equations with Quadratic Nonlinearity*

Next, we introduce the types of fluid dynamical equations that will be the focus of this article. We consider quite general equations for inhomogeneous turbulent flows interacting with waves and boundary inhomogeneities like topography. Higher order nonlinearities can also be considered by introducing auxiliary variables as reviewed by Frederiksen (2017) [9]. For these generic types of equations, we then document the closure theories. In detail, the general statistical dynamical QDIA closure equations for multi-field inhomogeneous turbulent flows and their application to the subgrid modeling problem have been derived for generic prognostic equations. They take the form considered by Frederiksen (2012a) [72]:

$$\frac{\partial}{\partial t} q_{\mathbf{k}}^a(t) + D_0^{\alpha\beta}(\mathbf{k}) q_{\mathbf{k}}^\beta(t) \\ = \sum_{\mathbf{p}} \sum_{\mathbf{q}} \left[ K^{abc}(\mathbf{k},\mathbf{p},\mathbf{q}) q_{-\mathbf{p}}^b(t) q_{-\mathbf{q}}^c(t) + A^{abc}(\mathbf{k},\mathbf{p},\mathbf{q}) q_{-\mathbf{p}}^b(t) h_{-\mathbf{q}}^c \right] + f_0^a(\mathbf{k},t). \quad (5)$$

Here the multi-field variable $q_{\mathbf{k}}^a(t)$ represents the flows in spectral space and depends on time $t$, the level or field type $a$ and the wavevector $\mathbf{k}$. The equations contain forcing, $f_0^a(\mathbf{k},t)$, a linear term with coefficients, $D_0^{\alpha\beta}(\mathbf{k})$, and, in addition to the quadratic term, a term bilinear in the dynamical fields $q_{\mathbf{k}}^a(t)$ and in constant fields $h_{\mathbf{k}}^a$ such as topography. We use the convention that repeated superscripts are summed over throughout.

The systems that are of primary interest in this paper have interaction coefficients $A^{abc}(\mathbf{k},\mathbf{p},\mathbf{q})$ and $K^{abc}(\mathbf{k},\mathbf{p},\mathbf{q})$ that satisfy the relationships:

$$K^{abc}(\mathbf{k},\mathbf{p},\mathbf{q}) = \frac{1}{2}[ A^{abc}(\mathbf{k},\mathbf{p},\mathbf{q}) + A^{acb}(\mathbf{k},\mathbf{q},\mathbf{p})] \quad (6)$$

and

$$K^{abc}(\mathbf{k},\mathbf{p},\mathbf{q}) = K^{acb}(\mathbf{k},\mathbf{q},\mathbf{p})]. \quad (7)$$

Typical systems of classical equations to which the theory can be applied are quasi-geostrophic and three-dimensional turbulent flows (Frederiksen 2012a) [72] and the primitive equations (Frederiksen et al. 2015) [138].

1.3.1. Navier Stokes Equations for Two- and Three-Dimensional Turbulent Flow

The incompressible Navies Stokes equations are the iconic classical system for studying turbulence. In the case of three-dimensional (3D) inhomogeneous turbulent flow these equations take the form:

$$\frac{\partial}{\partial t} u^a + u^\beta \frac{\partial u^a}{\partial x^\beta} = -\frac{1}{\rho} \frac{\partial p}{\partial x^a} + \nu_0 \nabla^2 u^a + f_0^a. \quad (8)$$

They describe the evolution of the fluid velocity, $u^a(\mathbf{x},t)$, $a=1,2,3$ at position $\mathbf{x}$ and time $t$. Here, $\rho$ is the density and $p(\mathbf{x},t)$ is the pressure. The viscosity that is used may be the molecular viscosity or an effective bare viscosity which is specified by $\nu_0$, and $f_0^a(\mathbf{x},t)$ are forcing functions that inject energy and can allow the system to arrive at a statistically steady state.

If we relabel $u_{\mathbf{k}}^a(t) \to q_{\mathbf{k}}^a(t)$, then the spectral equations for three-dimensional flow on the periodic domain $0 \leq x = x^1 \leq 2\pi$, $0 \leq y = x^2 \leq 2\pi$, $0 \leq z = x^3 \leq 2\pi$ take the standard form given in Equation (5). This is detailed in Appendix A. The analysis applies equally to the Navier Stokes equations for two-dimensional (2D) flows but of course with just two components $u^a(\mathbf{x},t)$, $a=1,2$.

### 1.3.2. Quasigeostrophic Equations for Turbulent Flow

The baroclinic multi-level equations for quasigeostrophic (QG) flow over topography on an $f$ – plane or $\beta$ – plane may be written in the form:

$$\frac{\partial q^a}{\partial t} = -J(\psi^a, q^a + h^a + \beta y) - D_0^{ab} q^b + f_0^a \tag{9}$$

In planar geometry:

$$J(\psi, q) = \frac{\partial \psi}{\partial x}\frac{\partial q}{\partial y} - \frac{\partial \psi}{\partial y}\frac{\partial q}{\partial x}. \tag{10}$$

Here, $\psi^a$ is the streamfunction, and $q^a$ is the reduced potential vorticity, $\omega^a = \nabla^2 \psi^a$ is the relative vorticity. Also, $D_0^{ab}$ are generalized dissipation operators that may include wave effects to be specified below, and $f_0^a$ are forcing functions.

We shall primarily be interested in the two-level system for which $a = 1$ or $2$, and $q^a = \nabla^2 \psi^a + (-1)^a F_L(\psi^1 - \psi^2)$ with $h^2 = h$, $h^1 = 0$ where $h$ is the scaled topography. The layer coupling parameter $F_L$ (Frederiksen 2012a) [72] is inversely proportional to the static stability.

The spectral equations for $q_\mathbf{k}^a(t)$, describing quasigeostrophic flow on the periodic domain $0 \leq x \leq 2\pi$, $0 \leq y \leq 2\pi$, then again take the form in Equation (5) with $\mathbf{k}$ a two-dimensional wavevector (see Appendix A).

The major aims of this article are to review some of the main theoretical developments, in turbulence closure theory and subgrid scale modelling, that have been achieved since Kraichnan's pioneering effort in developing the DIA closure. Much of our focus will be on 2D and quasigeostrophic turbulence because of our interest in atmospheric and oceanic dynamics; but also, because the formulation of the complex closure models is much more elegant for these equations in vorticity and potential vorticity form. However, we do also consider the major advances in 3D turbulence, but not in as much detail as in the reviews by McComb (1990, 2014) [34,118] or Zhou (2021) [50]. The emphasis will be on approaches based on renormalized perturbation theory and readers interested in renormalization group methods can consult Zhou (2021) [50]. In contrast to most reviews, this article has a strong emphasis on inhomogeneous turbulent flows and the closure and subgrid modelling methods required to tackle this level of complexity.

The plan of this article is as follows. The focus of the turbulence closure analysis will be 2D turbulence represented by the barotropic vorticity equation since this results in a much more elegant formulation than that in terms of the velocities. We present this equation for small-scale circulations interacting with a large-scale wind and topography on a generalized $\beta$ – plane, in Section 2. The focus is on flow on the doubly periodic domain with the large-scale wind interacting with the topography through the form-drag equation. We combine these two equations into a single form in Section 3, by expanding the small-scales in a Fourier series and choosing a suitable representation for the large-scale flow. The theory of Eulerian non-Markovian DIA, SCFT and LET closures is presented in Section 4 for homogeneous turbulence. There we also compare the performance of these closures with ensembles of DNS for HIT. Section 5 follows with a summary of the properties of quasi-Lagrangian closures and regularized DIA closures with localized interactions. In this Section the performance of these closures is also compared with ensembles of DNS and the Eulerian non-Markovian closures for HIT. In Section 6, non-Markovian closures for 2D HAT are presented and three variants of corresponding Markovian closures with auxiliary evolution equations for the triad relaxation functions are derived. In this Section we also document the realizable EDMAC closure for 2D homogeneous anisotropic turbulence (HAT) interacting with transient Rossby waves, and its performance compared with that of the EDQNM for isotropic and anisotropic 2D turbulence of various complexity.

In Section 7, the statistical dynamical equations for the non-Markovian QDIA closure are detailed, for 2D inhomogeneous turbulent flows interacting with Rossby waves and topography, and diagrammatic representations of the closure terms are presented. In Section 8, an abridged simplification of the QDIA is analyzed in which the time-dependent mean-field is approximated by the current-time field in the time-history integrals. There, three variants of abridged MIC models are also determined from the abridged QDIA. This involves using any of the the current-time, prior-time, and correlation FDTs for the two-time cumulants as well as a Markovian form of the response function. Auxiliary Markovian prognostic equations for the relaxation functions needed for the MICs are also derived. The final theoretical development considered in this Section is the eddy damped Markovian inhomogeneous closure (EDMIC), which is a generalization of the EDQNM, that also has analytical forms for the relaxation functions. Thereafter follow results from numerical calculations with non-Markovian and Markovian closures that are compared with very large ensembles of DNS. In Section 9, applications of the QDIA closure to problems of statistical dynamical numerical weather prediction and data assimilation are discussed in the barotropic model context.

In Sections 10 and 11, the problem of deriving subgrid scale parameterizations for unresolved eddies needed for large eddy simulations is considered. In Section 10, the problem is analyzed for homogeneous turbulence with examples given for quasigeostrophic barotropic and baroclinic flows characteristic of the atmosphere and oceans. For both these systems essentially universal scaling laws for the subgrid models are documented. As well, results for turbulent flows in multi-level primitive equation atmospheric models and for high resolution 3D Navier-Stokes boundary layer turbulence are presented. The full complexity of subgrid modelling for inhomogeneous baroclinic atmospheric and oceanic quasigeostrophic flows is considered in Section 11. The basis for the subgrid scale parameterizations is the structure of the multi-level and multi-field QDIA closure and the functional form of the QDIA subgrid terms. This is then used in a data-driven approach involving both stochastic and deterministic parameterizations.

In Section 12, our intention is to put into perspective the achievements in the statistical dynamics of strong interaction turbulence by comparing with progress in some other statistical field theories based on renormalized perturbation theory. Our conclusions and discussion of future prospects including for progress in inhomogeneous turbulence closure theory are summarized in Section 13. Appendix A presents the spectral equations for 2D and 3D Navier Stokes flows, and for quasigeostrophic flows on the plane and on the sphere. The interaction coefficients for barotropic eddies interacting with a large-scale flow and topography are listed in Appendix B. In Appendix C, the integral expressions for the off-diagonal elements of the QDIA two-point cumulants and response functions, and for the three-point cumulants, and expectation of the product of the individual response function and vorticity perturbation, are presented. The multi-field version of the QDIA closure equations is documented in Appendix D and in Appendix E the underpinning Langevin equation that guarantees the realizability of the QDIA multi-field equations is listed. The multiple subgrid scale terms that are implied by the multi-field QDIA closure are described in Appendix F. A general stochastic modelling approach to determining eddy-eddy subgrid scale parameterizations from the results of direct numerical simulations is outlined in Appendix G. In Appendix H, a general data-driven approach for determining the various terms contributing to the mean subgrid tendency in ensembles of direct numerical simulations is presented.

## 2. Barotropic 2D Flow over Topography on a $\beta$–plane

Two dimensional turbulent flows are most elegantly described by the barotropic vorticity equation which combines the two expressions for $u = u^1$ and $v = u^2$ into a single dynamical form. This approach is also most convenient when considering two-dimensional flows over topography on a generalized $\beta$ – plane. Then the flows can be represented by a large-scale flow $U$ together with smaller scales described by the streamfunction $\psi$. The total flow has a streamfunction $\Psi = \psi - Uy$ and it turns out that the spectral equations for these components can also be represented in a single unified form.

*2.1. Large-scale flow equation*

In physical space, the equation for the large-scale flow $U$ is the form-drag equation. The tendency of the large-scale flow is determined by the mountain torque that the scaled topography $h(\mathbf{x})$ exerts through interaction with the smaller scales:

$$\frac{\partial U}{\partial t} = \frac{1}{(2\pi)^2} \int_0^{2\pi} d^2\mathbf{x}\, h(\mathbf{x}) \frac{\partial \psi(\mathbf{x})}{\partial x} + \alpha_U (U_0 - U). \tag{11}$$

Here, we consider the general forced dissipative case where $U$ is relaxed to a mean large-scale flow $U_0$, that could be time-dependent, with strength $\alpha_U$. In Sections 10 and 11 we also consider other geometries and three dimensions. However, for most of the rest of this paper we focus on 2D flows on the doubly periodic plane $0 \leq x \leq 2\pi$, $0 \leq y \leq 2\pi$ with $\mathbf{x} = (x, y)$. Note that without topography, $h(\mathbf{x})$, forcing, $\alpha_U U_0$, and drag, $\alpha_U U$, $U$ is separately conserved.

*2.2. Barotropic vorticity equation for the small scales*

The small scales satisfy a modified version of the barotropic vorticity equation in which the large-scale flow of Equation (5), the topography and the $\beta$-effect also appear:

$$\frac{\partial \zeta}{\partial t} = -J(\psi - Uy, \zeta + h + \beta y + k_0^2 Uy) + \nu_0 \nabla^2 \zeta + f_0. \tag{12}$$

Here, the Jacobian $J$, and the vorticity $\zeta$, are defined by:

$$J(\psi, \zeta) = \frac{\partial \psi}{\partial x}\frac{\partial \zeta}{\partial y} - \frac{\partial \psi}{\partial y}\frac{\partial \zeta}{\partial x} \tag{13}$$

and

$$\zeta = \nabla^2 \psi \equiv \left(\frac{\partial^2}{\partial x^2} + \frac{\partial^2}{\partial y^2}\right)\psi \tag{14}$$

where $\nabla^2$ is the Laplacian.

The structure of the above equations, on the generalized $\beta$-plane, is isomorphic with the barotropic equations on a sphere. The term $k_0^2 \to 2$ on the sphere where it specifies the strength of the vorticity of the solid-body rotation term (Frederiksen and O'Kane 2005) [67]. In Equation (12) $\nu_0$ is the viscosity, $\beta$ specifies the strength of the beta effect, and $f_0$ is the external forcing.

**3. Dynamical Equations in Fourier Space**

The statistical dynamical equations for turbulent flows are most elegantly presented in spectral space from which they can also be back transformed to physical space (Carnevale and Martin 1982 [53]; Carnevale and Frederiksen 1983 [54]). The physical fields are transformed into Fourier space through the representations:

$$\zeta(\mathbf{x}, t) = \sum_{\mathbf{k}} \zeta_{\mathbf{k}}(t) \exp(i\,\mathbf{k}.\mathbf{x}), \tag{15}$$

with the spectral coefficients given by

$$\zeta_{\mathbf{k}}(t) = \frac{1}{(2\pi)^2} \int_0^{2\pi} d^2\mathbf{x}\, \zeta(\mathbf{x}, t) \exp(-i\,\mathbf{k}.\mathbf{x}). \tag{16}$$

Here, the wavevector $\mathbf{k} = (k_x, k_y)$, and the magnitude $k = (k_x^2 + k_y^2)^{1/2}$, where $k_x$ and $k_y$ are the wavenumber components in the $x$ and $y$ directions. The fact that the physical fields are real imposes the complex conjugate symmetry condition $\zeta_{-\mathbf{k}} = \zeta_{\mathbf{k}}^*$ on the spectral fields. The summation in Equation (15) is over the integer wavenumbers in an area (to be specified) that excludes the origin $\mathbf{0}$. Thus $k_x \in \mathbb{Z} \setminus \{0\}$ and $k_y \in \mathbb{Z} \setminus \{0\}$ where $\mathbb{Z}$ is the set of integers  In fact, through a judicious definition of the interaction coefficients we can combine the spectral equations for the 'small-scales' with that for $U$, taken as the zero-wavenumber component. This is achieved (Frederiksen and O'Kane 2005 [67]) by defining the associated vorticity components by

$$\zeta_{-\mathbf{0}} = \mathrm{i} k_0 U, \ \zeta_{\mathbf{0}} = \zeta_{-\mathbf{0}}^* \tag{17}$$

and taking $k_x \in \mathbb{Z}$ and $k_x \in \mathbb{Z}$. The complete set of interaction coefficients $A(\mathbf{k},\mathbf{p},\mathbf{q})$ and $K(\mathbf{k},\mathbf{p},\mathbf{q})$ needed to combine the equations are documented in Appendix B. Finally, the combined equations take the form:

$$\left( \frac{\partial}{\partial t} + \nu_0(\mathbf{k}) k^2 \right) \zeta_\mathbf{k}(t)$$
$$= \sum_{(\mathbf{p},\mathbf{q}) \in \mathcal{T}} \delta(\mathbf{k},\mathbf{p},\mathbf{q}) \left[ K(\mathbf{k},\mathbf{p},\mathbf{q}) \zeta_{-\mathbf{p}}(t) \zeta_{-\mathbf{q}}(t) + A(\mathbf{k},\mathbf{p},\mathbf{q}) \zeta_{-\mathbf{p}}(t) h_{-\mathbf{q}} \right] + f_0(\mathbf{k},t) \tag{18}$$

where

$$\mathcal{T} = \{\mathbf{p},\mathbf{q} | p \leq T, q \leq T\} \tag{19}$$

and we have chosen a circular domain with maximum value $C_T = T$. In Equation (18), $\delta(\mathbf{k},\mathbf{p},\mathbf{q}) = 1$ if $\mathbf{k}+\mathbf{p}+\mathbf{q}=0$ and $\delta(\mathbf{k},\mathbf{p},\mathbf{q}) = 0$ if $\mathbf{k}+\mathbf{p}+\mathbf{q} \neq 0$, and the complex $\nu_0(\mathbf{k}) k^2$ combines the viscosity and the Rossby wave frequency $\omega_\mathbf{k}$

$$D_0(\mathbf{k}) = \nu_0(\mathbf{k}) k^2 = \nu_0(k) k^2 + \mathrm{i} \omega_\mathbf{k} \tag{20}$$

with the Rossby wave frequency given by Equation (23) below. Here we distinguish between the viscosity $\nu_0(k)$ and the generalized dissipation $D_0(\mathbf{k}) = \nu_0(\mathbf{k}) k^2$ by their arguments. The drag and relaxation force in Equation (5) have also been included as the $\mathbf{k}=\mathbf{0}$ components of $f_0(\mathbf{k},t)$ and $\nu_0(\mathbf{k})$ through

$$f_0(\mathbf{0}) = -\mathrm{i} k_0 \alpha_U U_0, \tag{21}$$
$$\nu_0(\mathbf{0}) k_0^2 = \alpha_U. \tag{22}$$

Here, we have generalized the form of the viscosity $\nu_0 \to \nu_0(k)$ which corresponds to more general dissipation operators than the Laplacian in Equation (12).

The Rossby wave frequency is given by the dispersion relationship:

$$\omega_\mathbf{k} \equiv \omega_\mathbf{k}^\beta = -\frac{\beta k_x}{k^2}, \tag{23}$$

and the Doppler shifted Rossby wave frequency is

$$\omega_\mathbf{k}^U(t) = \Omega_\mathbf{k}^U(t) + \omega_\mathbf{k}^\beta = \frac{U(t) k_x (k^2 - k_0^2)}{k^2} - \frac{\beta k_x}{k^2} \tag{24}$$

where

$$\Omega_\mathbf{k}^U(t) = \frac{U(t) k_x (k^2 - k_0^2)}{k^2}. \tag{25}$$

We note that in the case of inviscid flows Rossby waves of the form $\exp \mathrm{i}(\mathbf{k} \cdot \mathbf{x} - \omega_\mathbf{k}^U t)$ are exact solutions to Equation (12).

**4. Euclidean Non-Markovian Closures for Homogeneous Turbulence**

In this Section we formulate the statistical dynamical closure equations for homogeneous turbulent barotropic flows for the case where Rossby waves induce anisotropy. The isotropic case is then a special case in which the $\beta-$effect is set to zero and the forcing is isotropic. The closure theory is formulated in terms of the second order cumulant and the response function and together they describe the evolution of infinite ensembles of flow field. In general, the closure formulation starts by representing a given field $\zeta_\mathbf{k}(t)$ by its mean, $<\zeta_\mathbf{k}(t)>\equiv \overline{\zeta}_\mathbf{k}(t)$, and the deviation from the mean, $\tilde{\zeta}_\mathbf{k}(t)$. As well a specializing to homogeneous turbulence in this Section we also assume that the large-scale flow is zero:

$$U = 0 \; ; \; U_0 = 0. \tag{26}$$

The small scales also have zero mean:

$$<\zeta_\mathbf{k}(t)>\equiv \overline{\zeta}_\mathbf{k}(t) = 0 \; ; \; \zeta_\mathbf{k}(t) = \tilde{\zeta}_\mathbf{k}(t). \tag{27}$$

Thus, the deviation $\tilde{\zeta}_\mathbf{k}$ also satisfies Equation (18) (with $\zeta_\mathbf{k}(t) \to \tilde{\zeta}_\mathbf{k}(t)$) and with the $\zeta_0$ component absent as seen from Equation (17). We also require mean forcing and topography to vanish or they would in general induce inhomogeneity:

$$<f_0(\mathbf{k},t)>\equiv \overline{f}_0(\mathbf{k},t) = 0 \; ; \; f_0(\mathbf{k},t) = \tilde{f}_0(\mathbf{k},t), \tag{28}$$

$$h_\mathbf{k} = 0. \tag{29}$$

*4.1. DIA Closure for Anisotropic Turbulence and Rossby Waves*

The DIA closure for homogeneous anisotropic turbulence consists of non-Markovian statistical dynamical equations for the evolution of the two-time two-point cumulant

$$C_\mathbf{k}(t,t') = <\tilde{\zeta}_\mathbf{k}(t)\tilde{\zeta}_{-\mathbf{k}}(t')> \tag{30}$$

and response function

$$R_\mathbf{k}(t,t') = \left\langle \tilde{R}_\mathbf{k}(t,t') \right\rangle. \tag{31}$$

The response function $R_\mathbf{k}(t,t')$ is the statistical average of individual responses defined by:

$$\tilde{R}_\mathbf{k}(t,t') = \frac{\delta \tilde{\zeta}_\mathbf{k}(t)}{\delta \tilde{f}_0(\mathbf{k},t')}, \tag{32}$$

where $\delta$ denotes the functional derivative. Thus, the individual response function $\tilde{R}_\mathbf{k}(t,t')$ is the change $\delta \tilde{\zeta}_\mathbf{k}(t)$ at a time $t$ in the field caused to an infinitesimal $\delta \tilde{f}_0(\mathbf{k},t')$ forcing perturbation at the earlier time $t'$.

The general multi-field QDIA inhomogeneous closure equations are documented in Appendix D and from them the DIA closure for homogeneous barotropic turbulence can be reduced by using the simplifications in Equations (26) to (29). The DIA closure described in this Section is based on the premise that both the viscosity and the Rossby wave frequency in Equation (20) are of similar magnitude and within perturbation theory both regarded as of zero order. The DIA two-time two-point cumulant equation is determined by the integro-differential equation:

$$\left(\frac{\partial}{\partial t} + \nu_0(\mathbf{k})k^2\right) C_\mathbf{k}(t,t') + \int_{t_0}^{t} ds\, \eta_\mathbf{k}(t,s) C_{-\mathbf{k}}(t',s)$$
$$= \int_{t_0}^{t'} ds\, (S_\mathbf{k}(t,s) + F_0(\mathbf{k},t,s)) R_{-\mathbf{k}}(t',s), \tag{33}$$

for $t > t'$. For $t' > t$ it can be obtained from the property $C_\mathbf{k}(t,t') = C_{-\mathbf{k}}(t',t)$. The last term in Equation (33) comes from the statistical effect of bare random forcing $\tilde{f}_0(\mathbf{k},t)$:

$$<\tilde{f}_0(\mathbf{k},t)\tilde{\zeta}_{-\mathbf{k}}(t')> = \int_{t_0}^{t'} ds\, F_0(t,s) R_{-\mathbf{k}}(t',s). \tag{34}$$

Here, $t_0$ is the initial time and the bare noise covariance

$$F_0(\mathbf{k},t,s) = \ <\tilde{f}_0(\mathbf{k},t)\tilde{f}_0^*(\mathbf{k},s)>. \tag{35}$$

The other terms in Equation (33) are the nonlinear damping and nonlinear noise that are reduced form the multi-field inhomogeneous terms in Appendix D. They are:

$$\eta_{\mathbf{k}}(t,s) = -4 \sum_{(\mathbf{p},\mathbf{q})\in\mathcal{T}^\aleph} \delta(\mathbf{k},\mathbf{p},\mathbf{q}) K(\mathbf{k},\mathbf{p},\mathbf{q}) K(-\mathbf{p},-\mathbf{q},-\mathbf{k}) R_{-\mathbf{p}}(t,s) C_{-\mathbf{q}}(t,s), \tag{36}$$

and

$$S_{\mathbf{k}}(t,s) = 2 \sum_{(\mathbf{p},\mathbf{q})\in\mathcal{T}^\aleph} \delta(\mathbf{k},\mathbf{p},\mathbf{q}) K(\mathbf{k},\mathbf{p},\mathbf{q}) K(-\mathbf{k},-\mathbf{p},-\mathbf{q}) C_{-\mathbf{p}}(t,s) C_{-\mathbf{q}}(t,s) \tag{37}$$

where, since there is no large-scale $U$ term, the $\mathbf{0}$ component is not included and

$$\mathcal{T}^\aleph = \mathcal{T} \setminus \{\mathbf{0}\}. \tag{38}$$

Again, the response function equation for homogeneous barotropic flow can be reduced form the more general closure in Appendix D. It again is given by an integro-differential equation:

$$\left(\frac{\partial}{\partial t} + \nu_0(\mathbf{k}) k^2\right) R_{\mathbf{k}}(t,t') + \int_{t'}^{t} ds\, \eta_{\mathbf{k}}(t,s) R_{\mathbf{k}}(s,t') = \delta(t-t') \tag{39}$$

for $t \geq t'$. Note that the Dirac delta function just ensures that $R_{\mathbf{k}}(t,t) = 1$.

To complete the system of equations for the DIA we also need the tendency equation for the single time two-point cumulant $C_{\mathbf{k}}(t,t)$. This is obtained by taking the limit as $t' \to t$ of $\frac{\partial}{\partial t} C_{\mathbf{k}}(t,t') + \frac{\partial}{\partial t'} C_{\mathbf{k}}(t,t')$. This results in the form:

$$\left(\frac{\partial}{\partial t} + 2\nu_0(k) k^2\right) C_{\mathbf{k}}(t,t) - 2\operatorname{Re} \int_{t_0}^{t} ds\, F_0(\mathbf{k},t,s)) R_{-\mathbf{k}}(t,s)$$
$$= 2N_{\mathbf{k}}(t) = 2\operatorname{Re} \int_{t_0}^{t} ds\, (S_{\mathbf{k}}(t,s) R_{-\mathbf{k}}(t,s) - \eta_{\mathbf{k}}(t,s) C_{-\mathbf{k}}(t,s)). \tag{40}$$

The above system of DIA equations also needs specification of the initial conditions $C_{\mathbf{k}}(t_0,t_0)$ and $R_{\mathbf{k}}(t_0,t_0) = 1$.

*4.2. SCFT and LET Closures for Ansotropic Turbulence and Rossby Waves*

As noted in Section 1.1.1, the SCFT and LET closures can in retrospect be obtained from the DIA closure by invoking the prior-time FDT in Equation (1). For the SCFT this FDT determines the two-time cumulant instead of Equation (33). For the LET closure the FDT determines the response function instead of Equation (39).

*4.3. Performance of Eulerian Non-Markovian Closures for Isotropic Turbulence*

Herring et al. (1974) [29] made a detailed study of the performance of the Eulerian DIA closure compared with the statistics of 10 realizations of DNS for 2D isotropic turbulence. They used the continuous wavenumber formulation of the closure and compared it with the discrete spectra of the DNS for the doubly periodic domain. Frederiksen and Davies (2000) [31] made a similar investigation for the DIA, and for the SCFT and LET, except the closures were formulated for discrete spectra like the DNS. This made the assessment of the closures more direct since the differences are directly attributable to the closure formulations and not to any differences in discretization. However, this comes at the cost of a much larger computational task with the closures.

As examples of the performance of the Eulerian non-Markovian closures we discuss the results of Frederiksen and Davies (2000) [31] initialized with their discrete spectra A and B. These are very similar to the continuous spectra I and II respectively of Herring et al. (1974) [29]. To discuss and compare the performance of discrete and continuous formulations of the closures with that of DNS we first present a few diagnostics.

*4.4. Diagnostics*

In this Subsection we define a set of standard diagnostics for characterizing the behaviour of evolving turbulence. Firstly, we define the total kinetic energy

$$E = \tfrac{1}{2}\sum_{\mathbf{k}} C_{\mathbf{k}}(t,t) k^{-2}, \tag{41}$$

enstrophy

$$F = \tfrac{1}{2}\sum_{\mathbf{k}} C_{\mathbf{k}}(t,t), \tag{42}$$

palinstrophy

$$P = \tfrac{1}{2}\sum_{\mathbf{k}} k^2 C_{\mathbf{k}}(t,t), \tag{43}$$

enstrophy dissipation

$$\eta = \sum_{\mathbf{k}} \nu_0 k^2 C_{\mathbf{k}}(t,t) = 2\nu_0 P, \tag{44}$$

and palintrophy production

$$K = \sum_{\mathbf{k}} k^2 N_{\mathbf{k}}(t). \tag{45}$$

The nonlinear transfer $N_{\mathbf{k}}(t)$ is defined in Equation (40). We also define the large-scale Reynolds number and skewness by:

$$R_L(t) = E \big/ (\nu_0 \eta^{\tfrac{1}{3}}) \tag{46}$$

and

$$S(t) = 2K \big/ (P F^{\tfrac{1}{2}}). \tag{47}$$

The band-averaged transient kinetic energy spectra, and palinstrophy spectra are specified by:

$$E(k_i,t) = \tfrac{1}{2}\sum_{\mathbf{k}\in\mathcal{S}} C_{\mathbf{k}}(t,t) k^{-2}, \tag{48}$$

and

$$P(k_i,t) = \tfrac{1}{2}\sum_{\mathbf{k}\in\mathcal{S}} k^2 C_{\mathbf{k}}(t,t). \tag{49}$$

Here, the set $\mathcal{S}$ is defined by

$$\mathcal{S} = [\mathbf{k} \mid k_i = \mathrm{Int.}(|\mathbf{k}| + \tfrac{1}{2})] \tag{50}$$

so that all elements with $\mathbf{k}$ within a band of unit width, at a radius $k_i$, are averaged.

*4.5. Initial Spectra and Parameter Specifications*

Frederiksen and Davies (2000) [31] essentially replicated the low and moderate Reynolds number DNS and DIA calculations of Herring et al. (1974) [29] but for ensembles of 10 or 40 members rather than 2 and as well for SCFT and LET closures. Their discrete wavenumber closures were much more computationally demanding than the continuous wavenumber closures of Herring et al. (1974) [29] and they therefore used the cumulant update restart scheme, further discussed in Appendix C, for some calculations.

The low Reynolds number calculations used a nondimensional viscosity of $\nu_0 = 5\times 10^{-3}$ with initial, $t_0 = 0$, large-scale Reynolds number of $R_L(t_0) \approx 61$. The runs started from initial spectrum A with

$$C_{\mathbf{k}}(t_0,t_0) = 1.33\times 10^{-4} k^5 \exp\left(-k^2/32\right) \tag{51}$$

which is very similar to spectrum I of Herring et al. (1974) [29]. The initial conditions were Gaussian. The runs were viscous decay integrations with zero forcing and performed at a resolution of circular truncation $C63 \equiv T_{63} = 63$ ( $|\mathbf{k}| = k \leq 63$ ) for 50 timesteps of $\Delta t = 0.016$ for the closures, and 200 timesteps of $\Delta t = 0.004$ for DNS, to a final time $t_{max} = 0.8$. For these low Reynolds number runs, that finished with $R_L(t_f) \approx 52$, the discrete wavenumber closures all performed very well compared with DNS except at the very smallest scales ([31] Figure 1) and better than the continuous wavenumber closures ([29] Figures 22 to 25). The differences in performance are emphasized by the evolution of the skewness, which increases from zero, at the Gaussian initial conditions, to $S(t_{max}) = 0.58$ and $0.47$ for the DNS and the discrete wavenumber DIA closure with cumulant updates ([31] Figure 2c). In contrast, for the continuous wavenumber closures $S(t_{max})$ is only about a half of the value for the DNS ([29] Figure 12).

The moderate Reynolds number runs used a nondimensional viscosity of $v_0 = 2.5 \times 10^{-3}$ with an initial large-scale Reynolds number of $R_L(t_0) \approx 307$. The calculations were initialized with Gaussian conditions from spectrum B, with
$$C_{\mathbf{k}}(t_0, t_0) = 0.18 k^2 \exp\left(-\tfrac{2}{3} k\right), \tag{52}$$
that is closely comparable with spectrum II of Herring et al. (1974) [29]. The DNS and closure calculations were again for decaying unforced turbulence at $C63 \equiv T_{63} = 63$ resolution. Frederiksen and Davies (2000) [31] found that very similar results were obtained at $C48 \equiv T_{48} = 48$ resolution. The runs were again integrated to a final time $t_{max} = 0.8$ using the same closure and DNS timesteps as for spectrum A.

Figure 1 shows the initial and evolved kinetic energy spectra for the $C48 \equiv T_{48} = 48$ resolution DNS ensemble and closure runs. We notice the very similar evolved spectra for the DIA, SCFT and LET closures with underestimation of small-scale kinetic energy starting from around wavenumber 20. The results for the DIA closure at resolution $C63 \equiv T_{63} = 63$, that uses a cumulant update restart scheme (CUDIA), are very comparable ([31] Figure 4) but slightly better than those for the continuous wavenumber DIA of Herring et al. [29] (Figures 22 to 25). The evolved Reynolds numbers were $R_L(t_{max}) = 280$ for DNS and $R_L(t_{max}) = 284$ for the CUDIA closure, with cumulant update restarts, which are much closer than those in Figure 17 of Herring et al. [29]. The evolved DNS skewness $S(t_{max}) \approx 0.7$ and $S(t_{max}) \approx 0.4$ for the spectrum B closures ([31] Figures 3c and 5c), but this is nevertheless twice the value of that for the DIA with continuous wavenumber spectrum ([29] Figure 16).

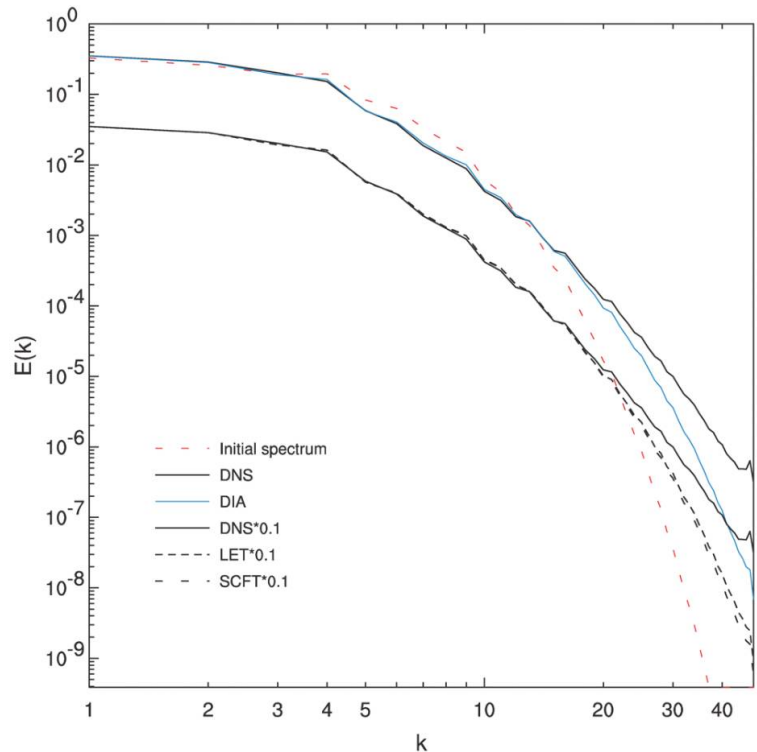

**Figure 1.** Comparison of transient kinetic energy spectra $E(k)$ for the DIA, SCFT and LET closures compared with DNS models for initial spectrum B at the initial ($t=0$) and final times ($t_{\max}=0.8$). Adapted from Frederiksen and Davies (2000) [31].

**5. Quasi-Lagrangian and Regularized Closures for Isotropic Turbulence**

*5.1. Impetus for Development of Quasi-Lagrangian Closures*

The formulation and numerical implementation of the Lagrangian History Direct Interaction (LHDI) closures for HIT by Kraichnan (1965) [37] and (1977) [38] and Herring and Kraichnan (1979) [39] is an extraordinary achievement given their complexity even in abridged form. The subsequent development of the Lagrangian Renormalized Approximation (LRA) closure, that is less complex, by Kaneda and collaborators (Kaneda 1981 [40]; Gotoh et al. 1988 [41]) is no less impressive. These closures overcome some of the deficiencies of the DIA but do not provide unique or complete solutions to problems of strong turbulence. The equations for the quasi-Lagrangian closures depend on whether labelling time derivatives (Kraichnan 1965) [37] or measuring time derivatives (Kaneda 1981) [40] are used in the formulation. They also depend on the choice of variable or representative used unlike the Eulerian DIA which is norm independent. The Eulerian DIA is also guaranteed to be realizable which is not the case for the quasi-Lagrangian closures. The quasi-Lagrangian closures make transformations so that the spurious convection effects of the DIA (Kraichnan 1964b) [32] are to some extent ameliorated. However, they are still second order in perturbation theory and a fundamental theory of strong turbulence needs to address the extremely difficult vertex renormalization problem (Kraichnan 1964b [32]; Martin et al. 1973 [22]).

The power law deficiencies of the Eulerian DIA that the quasi-Lagrangian closures were aimed at solving can in fact be overcome more simply by a one parameter modification of the DIA. Kraichnan (1964b) [32] recognized that the Eulerian DIA closure can be modified to remove the spurious convection (advection) effects of the large scales on the small scales. This is most conveniently done using the approach in his Appendix with the interaction coefficient $K(\mathbf{k},\mathbf{p},\mathbf{q})$ replaced by the modified term

$$\breve{K}(\mathbf{k},\mathbf{p},\mathbf{q}) = \theta(p - k/\alpha)\theta(q - k/\alpha)\, K(\mathbf{k},\mathbf{p},\mathbf{q}) \tag{53}$$

in the equation for the response function and two-point two-time cumulant but not in the single-time cumulant. We call this modification of the DIA closure the Regularized DIA (RDIA) and it corresponds to an empirical vertex renormalization.

*5.2. Comparison of Regularized DIA and Quasi-Lagrangian Closures for 2D Turbulence*

Frederiksen and Davies (2004) [33] compared the performance of the RDIA closure with $\alpha = 6$ for 2D isotropic turbulence with that of the abridged versions of the LHDI of Herring and Kraichnan (1979) [39] and the LRA closure of Gotoh et al. (1988) [41]. The study was carried out for spectra A and B in Equations and (51) and (52) with the respective viscosities of $\nu_0 = 5 \times 10^{-3}$ and $\nu_0 = 2.5 \times 10^{-3}$. The comparison was made with Herring and Kraichnan's result for the continuous spectra I and II and with those of Gotoh et al. (1988) [41] for these same two spectra. The DIA performs reasonably well for the low Reynolds number spectrum A except at the very smallest scales and the bigger impact of regularization is for spectrum B.

As noted in Section 4.5, the initial Reynolds number for the DNS and closures for spectrum A, at resolution $C63 \equiv T_{63} = 63$, is $R_L(t_0) \approx 61$ and the evolved $R_L(t_{\max}) \approx 52$ and this is also the case for the cumulant update version of the RDIA (RCUDIA) for this spectrum. The regularization however improves the evolved skewness, even at these low Reynolds numbers, to $S(t_{\max}) = 0.57$ which is very close to the DNS value of $S(t_{\max}) = 0.58$ while $S(t_{\max}) = 0.47$ for the DIA. In contrast, for the velocity based LRA of Goto et al. (1988) [41] the evolved skewness in their Figure 7 is around 20% larger than their DNS for similar low Reynolds number runs.

The spectrum A initial and evolved palinstrophy spectra for the DNS, CUDIA and RCUDIA closures are shown in Figure 2c of Frederiksen and Davies (2004) [33]. The comparison between the DNS and the regularized closure is very good, and an improvement on the CUDIA, except at the very smallest scales with $k \geq 58$ where the DNS has a tendency to kick up slightly and this is also the case in the results shown in Figure 8a of Herring and Kraichnan (1979) [39]. Even at these low Reynolds numbers the regularized DIA closure results compare more closely with DNS than do the spectrum I velocity based abridged LHDI (ALHDI) and strain based abridged LHDI (SBALHDI) of Herring and Kraichnan (1979) [39] (their Figure 8a). Similarly, for spectrum I the velocity based LRA of Gotoh et al. (1988) [41] (their Figure 6) has palinstrophy spectra that are too large at intermediate wavenumbers.

For spectrum B, at resolution $C63 \equiv T_{63} = 63$, the initial Reynolds number is $R_L(t_0) \approx 307$, as described in Section 4.5, and the evolved values are $R_L(t_{\max}) = 280$ for DNS, and importantly also for RDIA, $R_L(t_{\max}) = 284$ for CUDIA, $R_L(t_{\max}) = 291$ for DIA, and $R_L(t_{\max}) = 276$ for RCUDIA. Thus the regularization has clearly improved the evolved Reynolds number over the DIA particularly for the RDIA without restarts. This improvement is also seen in the skewness shown in Figure 4 of Frederiksen and Davies (2004) [33] which has a time-dependence very similar to the DNS while for the DIA it evolves to a final value that is only about 60% of that for the DNS.

Figure 2 shows the spectrum B initial and evolved kinetic energy spectra for the DNS, DIA and RDIA closures (without restarts) adapted from Figure 1 of Frederiksen and Davies (2004) [33] which also shows the correspondind results for the CUDIA and RCUDIA closures. The agreement between the regularized closures and DNS is noteworthy and much better than for the DIA closures. The same high level of agreement between regularized closures and DNS is seen in the evolved ($t_{\max} = 0.8$) palinstrophy spectra in Figures 2a and 2b of Frederiksen and Davies (2004) [33]. These results contrast with the poorer corespondence between spectrum II evolved ALHDI and SBALHDI palinstrophy spectra in Figure 8b of Herring and Kraichnan (1979) [39]. For spectrum II, the velocity based LRA closure of Gotoh et al. (1988) [41] also has deficiencies with underestimation at $t_{\max} = 0.8$ of the evolved palinstrophy (their Figure 8) at the smallest and largest scales and overestimation at intermediate scales.

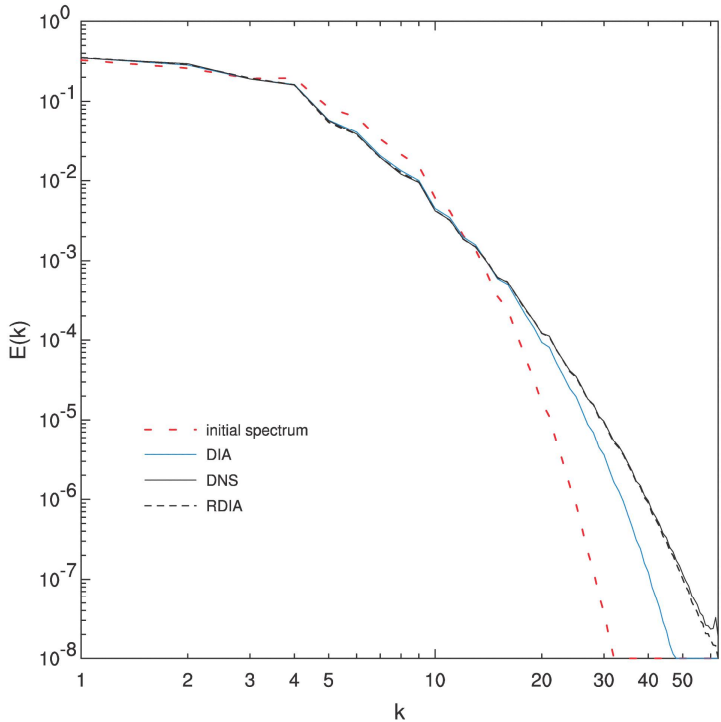

**Figure 2.** Comparison of transient kinetic energy spectra $E(k)$ for the DIA and RDIA closures compared with DNS models for initial spectrum B at the initial ($t = 0$) and final times ($t_{\max} = 0.8$). Adapted from Frederiksen and Davies (2004) [33].

## 6. Statistical Dynamical Equations for Markovian Anisotropic Closures

### 6.1. General Formulation of Markovian Anisotropic Closures

In this Section, we present Markovian Anisotropic Closures (MACs), which are of intrinsic interest, but are also the next stage towards devising the realizable EDMAC model of Frederiksen and O'Kane (2023, 2024) [64,65]. The EDMAC has an analytical form for the relaxation function like the EDQNM. For the MACs the triad relaxation functions are determined by integral equations that can be converted to differential equations. The Markovian version of the DIA single-time two-point cumulant equations, with the auxiliary differential equations for the triad relaxation functions, constitute the MACs. The formulation of the three MACS uses any of the FDTs in Equation (4) and a Markovian reduction of the DIA in Section 4. We denote these closures as $MAC^X$ where $X = 0, \frac{1}{2}, 1$ for the three FDTs.

To develop the $MAC^X$ models we start with the DIA single-time two-point cumulant equation:

$$\left(\frac{\partial}{\partial t} + 2\operatorname{Re}(D_r(\mathbf{k},t))\right) C_\mathbf{k}(t,t) = 2\operatorname{Re}(F_r(\mathbf{k},t)). \tag{54}$$

The renormalized dissipation and noise functions in Equation (54) are defined by

$$D_r(\mathbf{k},t) = D_0(\mathbf{k}) + D_\eta(\mathbf{k},t) \; ; \; F_r(\mathbf{k},t) = F_0(\mathbf{k},t) + F_S(\mathbf{k},t). \tag{55}$$

Here, the generalized dissipation terms are defined by:

$$D_0(\mathbf{k}) = \nu_0(\mathbf{k})k^2 \equiv \nu_0(k)k^2 + i\omega_\mathbf{k}, \tag{56}$$

and the bare noise term is given by:

$$F_0(\mathbf{k},t) = \int_{t_0}^{t} ds\, F_0(\mathbf{k},t,s) R_{-\mathbf{k}}(t,s). \tag{57}$$

For white noise $2\operatorname{Re} F_0(\mathbf{k},t) = F_0(\mathbf{k},t,t)$. The nonlinear damping, with any of the FDTs imposed, is given by

$$D_\eta(\mathbf{k},t) = -4 \sum_{(\mathbf{p},\mathbf{q})\in \mathcal{T}^\mathbf{k}} \delta(\mathbf{k},\mathbf{p},\mathbf{q}) K(\mathbf{k},\mathbf{p},\mathbf{q}) K(-\mathbf{p},-\mathbf{q},-\mathbf{k})$$
$$\times C_{-\mathbf{q}}^{1-X}(t,t) C_{-\mathbf{k}}^{-X}(t,t) \Theta^X(-\mathbf{p},-\mathbf{q},-\mathbf{k})(t). \tag{58}$$

and the nonlinear noise term

$$F_S(\mathbf{k},t) = 2 \sum_{(\mathbf{p},\mathbf{q})\in \mathcal{T}^\mathbf{k}} \sum_\mathbf{p} \sum_\mathbf{q} \delta(\mathbf{k},\mathbf{p},\mathbf{q}) K(\mathbf{k},\mathbf{p},\mathbf{q}) K(-\mathbf{k},-\mathbf{p},-\mathbf{q})$$
$$\times C_{-\mathbf{p}}^{1-X}(t,t) C_{-\mathbf{q}}^{1-X}(t,t) \Theta^X(-\mathbf{k},-\mathbf{p},-\mathbf{q})(t). \tag{59}$$

In these expressions the triad relaxation function

$$\Theta^X(-\mathbf{k},-\mathbf{p},-\mathbf{q})(t) = \int_{t_0}^t ds\, R_{-\mathbf{k}}(t,s) R_{-\mathbf{p}}(t,s) R_{-\mathbf{q}}(t,s) C_{-\mathbf{p}}^X(s,s) C_{-\mathbf{q}}^X(s,s). \tag{60}$$

For consistency with the two-point cumulant equation, the response function equation is taken as:

$$\frac{\partial}{\partial t} R_\mathbf{k}(t,t') + D_r(\mathbf{k},t) R_\mathbf{k}(t,t') = \delta(t-t'). \tag{61}$$

The triad relaxation function can then be expressed by the Markov differential equation:

$$\frac{\partial}{\partial t}\Theta^X(\mathbf{k},\mathbf{p},\mathbf{q})(t) + (D_r(\mathbf{k},t) + D_r(\mathbf{p},t) + D_r(\mathbf{q},t))\Theta^X(\mathbf{k},\mathbf{p},\mathbf{q})(t)$$
$$= C_\mathbf{p}^X(t,t) C_\mathbf{q}^X(t,t) \tag{62}$$

where $\Theta^X(\mathbf{k},\mathbf{p},\mathbf{q})(0) = 0$.

The $MAC^X$ models are defined by Equations (54) and (62) with $D_r$ and $F_r$ given in Equation (55). The auxiliary Markovian differential equations for $\Theta^X$ are more efficient for long time simulations but of course carry the same information as the time history integrals. The three $MAC^X$ are realizable if $D_r$ are all real so that $\Theta^X$ is real and this is the case without transient wave phenomena. We note however, as established by Bowman et al. (1993) [63], that the MAC model with $X = \frac{1}{2}$ is also realizable when transient Rossby waves (or drift waves) are present.

*6.2. Realizable Eddy Damped Markovian Anisotropic Closure*

The EDMAC model is derived from the $MAC^X$ model when $X = 0$ corresponding to the current-time FDT. The single-time two-point cumulant in Equation (54) then reduces to:

$$\left(\frac{\partial}{\partial t} + 2\nu_0(k)k^2\right)C_{\mathbf{k}}(t,t) - 2\operatorname{Re}F_0(\mathbf{k},t)$$
$$= 2N_{\mathbf{k}}(t) = 8\sum_{(\mathbf{p},\mathbf{q})\in\mathcal{T}^{\mathbf{k}}}\delta(\mathbf{k},\mathbf{p},\mathbf{q})K(\mathbf{k},\mathbf{p},\mathbf{q})K(\mathbf{p},\mathbf{q},\mathbf{k})\operatorname{Re}\Theta^{X=0}(\mathbf{k},\mathbf{p},\mathbf{q})(t) \qquad (63)$$
$$\times C_{\mathbf{q}}(t,t)\left[C_{\mathbf{k}}(t,t) - C_{\mathbf{p}}(t,t)\right].$$

In the special case of white noise, $2\operatorname{Re}F_0(\mathbf{k},t) = F_0(\mathbf{k},t,t)$. Here, $\Theta^{X=0}$ is given in Equation (62) which is replaced by $\Theta^{EDMAC}$ for the EDMAC model. In Equation (63) we have used the properties of the interaction coefficients that $K(\mathbf{k},\mathbf{p},\mathbf{q}) = K(\mathbf{k},\mathbf{q},\mathbf{p})$, and $K(\mathbf{k},\mathbf{p},\mathbf{q}) + K(\mathbf{p},\mathbf{q},\mathbf{k}) + K(\mathbf{q},\mathbf{k},\mathbf{p}) = 0$. We have also used the facts that the single-time cumulants are real and $\Theta^{X=0}$ is symmetric in the three wavenumbers when $X=0$.

In the EDQNM closure, Leith (1971) [45] parameterized the eddy damping $D_\eta$ by the form:

$$D_\eta(\mathbf{k},t) = \mu_{\mathbf{k}}^{eddy}(t) = \gamma\left[k^2 C_{\mathbf{k}}(t,t)\right]^{\frac{1}{2}} \qquad (64)$$

that is consistent with the $k^{-3}$ enstrophy cascading inertial range for 2D turbulence. Thus, the total EDQNM damping is:

$$\mu_{\mathbf{k}}(t) = \nu_0(k)k^2 + \mu_{\mathbf{k}}^{eddy}(t) = \nu_0(k)k^2 + \gamma\left[k^2 C_{\mathbf{k}}(t,t)\right]^{\frac{1}{2}}, \qquad (65)$$

and

$$D_r^{EDQNM}(\mathbf{k},t) \to \mu_{\mathbf{k}}(t) + i\omega_{\mathbf{k}}(t) \qquad (66)$$

The parameter $\gamma$ is positive and dimensionless. For 3D turbulence Orszag (1970) [44] used a similar local form for $\mu_{\mathbf{k}}^{eddy}(t)$ that is consistent with the $k^{-\frac{5}{3}}$ forward energy cascading inertial range. Pouquet et al. (1975) [139] use instead an integral over wavenumbers to estimate $\mu_{\mathbf{k}}^{eddy}(t)$.

Frederiksen and O'Kane (2023) [64] proposed a modification of the EDQNM closure that is realizable in the presence of transient waves such as Rossby waves and plasma drift waves. The theoretical basis for this EDMAC model was provided by Frederiksen and O'Kane (2024) [65]. Their analysis is based on a systematic simplification of a slight generalization of the DIA closure for homogeneous anisotropic turbulent flows in the presence of transient Rossby waves. They also allowed for the inclusion of a large-scale mean flow $U(t)$ which generalized the dispersion relation of the Rossby waves from $\omega_{\mathbf{k}} = \omega_{\mathbf{k}}^\beta$ to the Doppler shifted frequency $\omega_{\mathbf{k}}^U(t)$ defined in Equation (24). The analysis followed the approach of Frederiksen (1999) [66] for deriving the QDIA in that the anisotropic (and inhomogeneous) terms were assumed to be first order in formal perturbation theory. The subsequent Markovianization, following closely the approach in this Section arrived at the realizable EDMAC model with the replacement:

$$D_r^{EDMAC}(\mathbf{k},t) \to \rho_{\mathbf{k}}(t) + i\omega_{\mathbf{k}}^U(t) \qquad (67)$$

where

$$\rho_{\mathbf{k}}(t) = \mu_{\mathbf{k}}(t) + c\frac{[\omega_{\mathbf{k}}^U(t)]^2}{\mu_{\mathbf{k}}(t)} > 0 \qquad (68)$$

has a frequency renormalized term added to the EDQNM damping. The EDMAC response function is therefore given by:

$$R_{\mathbf{k}}(t,t') \approx R_{\mathbf{k}}^{EDMAC}(t,t') = \exp\left(-[\rho_{\mathbf{k}}(t) + i\omega_{\mathbf{k}}^U(t)](t-t')\right), \qquad (69)$$

with the EDMAC triad relaxation function taking the form:

$$\Theta^{EDMAC}(\mathbf{k},\mathbf{p},\mathbf{q})(t)$$

$$=\frac{1-\exp\left(-\left[\rho_{\mathbf{k}}(t)+\rho_{\mathbf{p}}(t)+\rho_{\mathbf{q}}(t)+\mathrm{i}(\omega_{\mathbf{k}}^{U}(t)+\omega_{\mathbf{p}}^{U}(t)+\omega_{\mathbf{q}}^{U}(t))\right](t-t_{0})\right)}{\rho_{\mathbf{k}}(t)+\rho_{\mathbf{p}}(t)+\rho_{\mathbf{q}}(t)+\mathrm{i}(\omega_{\mathbf{k}}^{U}(t)+\omega_{\mathbf{p}}^{U}(t)+\omega_{\mathbf{q}}^{U}(t))}. \quad (70)$$

Again, Equation (63), is the prognostic equation for the EDMAC cumulant but with $\Theta^{X=0}$ in Equation (70) substituted by $\Theta^{EDMAC}$. Frederiksen and O'Kane (2024) [65] find that the expected value of the parameter $c$ in Equation (68) is $c=\tfrac{1}{2}$. In fact, however, $c\geq\tfrac{1}{4}$ is a sufficient condition for the EDMAC to be realizable for general time dependent $\omega_{\mathbf{k}}^{U}(t)$. The EDMAC has basically the same computational efficiency as the EDQNM since $\rho_{\mathbf{k}}$ just replaces $\mu_{\mathbf{k}}$ in Equation (70) for the triad relaxation function. It has, however, the big advantage of guaranteed realizability with transient waves. In this it conforms to Kraichnan's (1959a, 1961) [5,6] important insight, discussed in more detail in Section 12.2, that closure approximations should correspond to a physically realizable dynamical system.

*6.3. Generalizations of the Realizable EDMAC Model*

As noted by Frederiksen and O'Kane (2024) [65] there are many possible extensions of the EDMAC model to different geometries, to higher dimension than $d=2$ and to higher order nonlinearity. This includes wave-turbulence interactions for barotropic flows on the differentially rotation sphere (Frederiksen et al. 1996) [82], for 3D quasigeostrophic flows on a $\beta$–plane and on the sphere (Frederiksen 2012a) [72], and for internal gravity waves and turbulence in the diagonal dominance approximation (Carnevale and Frederiksen 1983) [54]. As well, generalization to 3D Navier–Stokes flows with wave-turbulence interactions, such as that of Cambon and Jacquin (1989) [140] is expected to be feasible. Just as the EDQNM closure has been studied in general dimensions, $d\geq 2$, by Rose and Sulem (1978) [141] and Clark et al. (2021) [142], for isotropic turbulence, it may be possible to generalize this to anisotropic turbulence and waves with the realizable EDMAC model.

*6.4. Comparison of EDQNM and EDMAC Model Integrations*

Frederiksen and O'Kane (2024) [65] performed a set of closure calculations for the evolution of turbulence with the EDQNM and EDMAC models. They also allowed for interactions with Rossby waves and possibly also a large-scale mean flow $U$ in some integrations. The aim was to examine the extent to which the extra frequency renormalized contribution to the eddy damping influenced the results. Four runs were made, with a typical $\gamma=0.6$ in Equation (64), which are summarized in Table 1. The EDQNM closure was used for the isotropic run *I*, and the anisotropic run *A1*, with $\beta=0.5$, $c=0$, $U=0$. The EDMAC model was used for the anisotropic run *A2* with $\beta=0.5$, $c=0.5$, $U=0$ and the anisotropic run *A3* with $\beta=0.5$, $c=0.5$, $U=0.065$. Here the nondimensional $\beta=0.5$ corresponds to the earth's differential rotation at $60°$ latitude on using a length scale of one half the earth's radius, $a_e/2$, and time scale of the inverse of the earth's rotation rate, $\Omega^{-1}$. The nondimensional mean zonal flow $U=0.065$ converts to $U=15\text{ ms}^{-1}$ with this same scaling. The runs were unforced with a nondimensional viscosity of $\nu_0=2.5\times 10^{-3}$ and $k_0^2=\tfrac{1}{2}$ in Equation (12) and of course without topography ($h=0$). A resolution of $C64\equiv T_{64}=64$ was used with a nondimensional time step of $\Delta t=0.004$. The runs started with Gaussian initial conditions from spectrum B of Frederiksen and Davies (2000) [31] that is also defined in our Equation (52) and proceeded to a final time of $t_{\max}=0.4$.

Table 2 shows that the evolved large-scale Reynolds numbers and skewness are nearly identical for runs $I$, $A1$ and $A2$. For run $A3$, with the presence of the large-scale zonal flow $U = 0.065$, there is a slight increase in the evolved Reynolds number and a decrease in the evolved skewness indicative of less deviation from Gaussian statistics. These close results for the evolved Reynolds number and skewness are reflected in the evolved kinetic energy spectra shown in Figure 3. Despite the rapid development from the initial conditions the evolved spectra are essentially coincident and have therefore been scaled by successive factors of 10. Importantly, all the results for anisotropic run $A1$ with the EDQNM and anisotropic run $A2$ with the EDMAC model are nearly identical. Here the EDQNM model remained stable even though it is not guaranteed to be realizable in the presence of Rossby waves.

Frederiksen and O'Kane (2024) [65] further examined the broad generality of these results for smaller and larger values of $\gamma$.

**Table 1.** Non-dimensional parameters specifying the isotropic and three anisotropic closure runs.

| Closure Runs | $\beta$ | $C$ | $U$ |
|---|---|---|---|
| Isotropic Run $I$ | 0 | 0 | 0 |
| Anisotropic Run $A1$ | 0.5 | 0 | 0 |
| Anisotropic Run $A2$ | 0.5 | 0.5 | 0 |
| Anisotropic Run $A3$ | 0.5 | 0.5 | 0.065 |

**Table 2.** Initial and evolved Reynolds number and skewness for the isotropic and three anisotropic closure runs.

| Closure Runs | $R_L(0)$ | $R_L(0.4)$ | $S(0)$ | $S(0.4)$ |
|---|---|---|---|---|
| Isotropic Run $I$ | 304.8 | 263.7 | 0 | 0.735 |
| Anisotropic Run $A1$ | 304.8 | 263.8 | 0 | 0.734 |
| Anisotropic Run $A2$ | 304.8 | 263.8 | 0 | 0.735 |
| Anisotropic Run $A3$ | 304.8 | 265.7 | 0 | 0.690 |

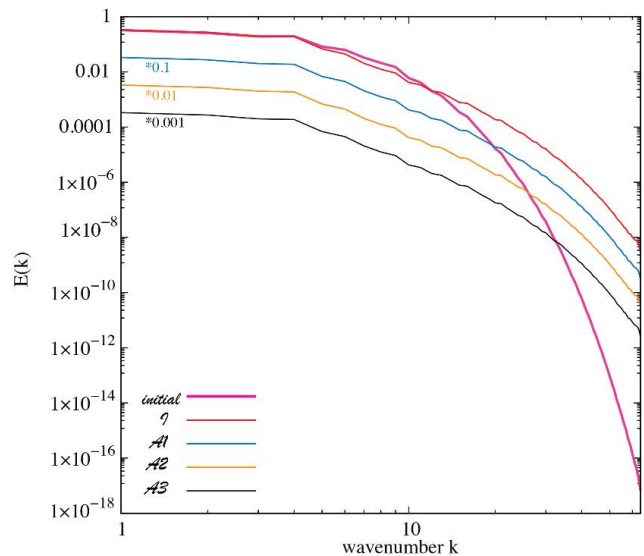

**Figure 3.** Comparison of transient kinetic energy spectra $E(k)$ for the EDQNM and ED-MAC models for the runs in Table 2 at the initial ($t = 0$) and final times ($t_{max} = 0.4$). Adapted from Frederiksen and O'Kane (2024) [65].

## 7. Statistical Dynamical Equations for Inhomogeneous QDIA Closure

The theoretical formulation of closures is based on the statistical dynamics of an infinite ensemble of individual flows, which in our case satisfy the DNS in Equations (5) and (6) in physical space and Equation (10) in Fourier space. Thus, closures have the important advantage of propagating the population statistics rather than relying on the calculation of sufficient sampling statistics obtained via a finite number of DNS ensemble members. The means of ensembles of DNS may result in large errors particularly when the fluctuations are large (O'Kane and Frederiksen 2004 [43], Frederiksen and O'Kane 2005 [67]). In general, considerable care must be taken when setting up even very large ensembles of DNS. These errors are considerably compounded when estimating the higher order moments.

Any given member of the infinite ensemble the vorticity spectral field at wavenumber $\mathbf{k}$ can be written in terms of its mean $<\zeta_\mathbf{k}(t)> \equiv \overline{\zeta}_\mathbf{k}(t)$ and deviation from the ensemble mean $\tilde{\zeta}_\mathbf{k}(t)$:

$$\zeta_\mathbf{k}(t) = <\zeta_\mathbf{k}(t)> + \tilde{\zeta}_\mathbf{k}(t) \equiv \overline{\zeta}_\mathbf{k}(t) + \tilde{\zeta}_\mathbf{k}(t). \tag{71}$$

From Equation (18) the coupled equations for $<\zeta_\mathbf{k}>$ and $\tilde{\zeta}_\mathbf{k}$ then follow as:

$$\left(\frac{\partial}{\partial t} + \nu_0(\mathbf{k})k^2\right)<\zeta_\mathbf{k}(t)>$$
$$= \sum_{(\mathbf{p},\mathbf{q})\in\mathcal{T}} \delta(\mathbf{k},\mathbf{p},\mathbf{q})[K(\mathbf{k},\mathbf{p},\mathbf{q})\{<\zeta_{-\mathbf{p}}(t)><\zeta_{-\mathbf{q}}(t)> + C^{\otimes}_{-\mathbf{p},-\mathbf{q}}(t,t)\} \tag{72}$$
$$+ A(\mathbf{k},\mathbf{p},\mathbf{q})<\zeta_{-\mathbf{p}}(t)>h_{-\mathbf{q}}] + \overline{f}_0(\mathbf{k},t),$$

and

$$\left(\frac{\partial}{\partial t} + \nu_0(\mathbf{k})k^2\right)\tilde{\zeta}_\mathbf{k}(t)$$
$$= \sum_{(\mathbf{p},\mathbf{q})\in\mathcal{T}} \delta(\mathbf{k},\mathbf{p},\mathbf{q})[K(\mathbf{k},\mathbf{p},\mathbf{q})\{<\zeta_{-\mathbf{p}}(t)>\tilde{\zeta}_{-\mathbf{q}}(t) + \tilde{\zeta}_{-\mathbf{p}}(t)<\zeta_{-\mathbf{q}}(t)> \tag{73}$$
$$+ \tilde{\zeta}_{-\mathbf{p}}(t)\tilde{\zeta}_{-\mathbf{q}}(t) - C_{-\mathbf{p},-\mathbf{q}}(t,t)\} + A(\mathbf{k},\mathbf{p},\mathbf{q})\tilde{\zeta}_{-\mathbf{p}}(t)h_{-\mathbf{q}}] + \tilde{f}_0(\mathbf{k},t).$$

In these and subsequent equations, wavenumbers $\mathbf{p},\mathbf{q}$ are in the domain $\mathcal{T} = \{\mathbf{p},\mathbf{q} | p \leq T, q \leq T\}$ defined in Equation (19) as in Section 3. The bare forcing function consists of deterministic and random contributions

$$f_0(\mathbf{k},t) = \overline{f}_0(\mathbf{k},t) + \tilde{f}_0(\mathbf{k},t) \equiv <f_0(\mathbf{k},t)> + \tilde{f}_0(\mathbf{k},t), \tag{74}$$

and the two-time cumulant elements are given by:

$$C^{\oplus}_{-\mathbf{p},-\mathbf{q}}(t,s) \equiv C_{-\mathbf{p},-\mathbf{q}}(t,s) = <\tilde{\zeta}_{-\mathbf{p}}(t)\tilde{\zeta}_{-\mathbf{q}}(s)>. \tag{75}$$

Here, we have placed the superscript $\oplus$ on $C^{\oplus}_{-\mathbf{p},-\mathbf{q}}(t,s)$ for later convenience where it will be replaced by QDIA and the QDIA expression inserted to close the equations.

### 7.1. General Formulation of QDIA Closure

We see from Equation (72) that to close the mean field equation an expression for the off-diagonal elements of the two-point cumulant is required. The QDIA closure equation for $<\zeta_\mathbf{k}(t)>$ is obtained by replacing $C^{\oplus}_{-\mathbf{p},-\mathbf{q}}(t,s) \equiv C_{-\mathbf{p},-\mathbf{q}}(t,s)$ in Equation (72) by $C^{QDIA}_{-\mathbf{p},-\mathbf{q}}(t,s)$ defined in Equation (A25) of Appendix C. We note that the expression for $C^{QDIA}_{-\mathbf{p},-\mathbf{q}}(t,s) \equiv C^{QDIA}_{-\mathbf{p},-\mathbf{q}}(t,s)[C_{\mathbf{k}}, R_{\mathbf{k}}, <\zeta_{\mathbf{k}}>, h_{\mathbf{k}}]$ is a functional of the diagonal two-point cumulant $C_{\mathbf{k}} \equiv C_{\mathbf{k},-\mathbf{k}}$, the diagonal response function $R_{\mathbf{k}} \equiv R_{\mathbf{k},\mathbf{k}}$, mean field $<\zeta_{\mathbf{k}}>$ and topography $h_{\mathbf{k}}$. In turn, we see that the prognostic equations for $C_{\mathbf{k}}$ and $R_{\mathbf{k}}$ are required.

The response function matrix element $\tilde{R}_{\mathbf{k},\mathbf{l}}(t,t')$, defined for $t \geq t'$, is a measure of the change in an ensemble member field $\tilde{\zeta}_{\mathbf{k}}(t)$ at time $t$ due to an infinitesimal change in the forcing $\tilde{f}_0(\mathbf{l},t')$ at the earlier time $t'$. It is defined by through the functional derivative

$$\tilde{R}_{\mathbf{k},\mathbf{l}}(t,t') = \frac{\delta \tilde{\zeta}_{\mathbf{k}}(t)}{\delta \tilde{f}_0(\mathbf{l},t')}, \tag{76}$$

where $\delta$ denotes the functional derivative. The expectation of Equation (76) defines the ensemble average response function

$$R_{\mathbf{k},\mathbf{l}}(t,t') = \left\langle \tilde{R}_{\mathbf{k},\mathbf{l}}(t,t') \right\rangle. \tag{77}$$

We once again use a shortened notation for the diagonal elements of the response function and two-point cumulant:

$$R_{\mathbf{k}}(t,t') \equiv R_{\mathbf{k},\mathbf{k}}(t,t'), \tag{78}$$

$$C_{\mathbf{k}}(t,t') \equiv C_{\mathbf{k},-\mathbf{k}}(t,t') = <\zeta_{\mathbf{k}}(t)\zeta_{-\mathbf{k}}(t')>. \tag{79}$$

To obtain the equation for the diagonal two-time cumulant $C_{\mathbf{k}}$ we multiply each term in Equation (73) by $\tilde{\zeta}_{-\mathbf{k}}(t')$ and average over the infinite ensemble. Similarly, we derive the statistical dynamical equation for the diagonal response function $R_{\mathbf{k}}$ from Equation (73) by taking the functional derivative with respect to $\tilde{f}_0(\mathbf{k},t')$ and ensemble averaging. Employing the relationships described in Appendix C we then arrive at the QDIA closure equations.

The QDIA mean field equation is:

$$\left(\frac{\partial}{\partial t} + \nu_0(\mathbf{k})k^2\right) <\zeta_{\mathbf{k}}(t)> + \int_{t_o}^{t} ds\, \eta_{\mathbf{k}}(t,s) <\zeta_{\mathbf{k}}(s)>$$
$$= \sum_{(\mathbf{p},\mathbf{q})\in\mathcal{T}} \delta(\mathbf{k},\mathbf{p},\mathbf{q}) A(\mathbf{k},\mathbf{p},\mathbf{q}) <\zeta_{-\mathbf{p}}(t)> h_{-\mathbf{q}} \tag{80}$$
$$+ \sum_{(\mathbf{p},\mathbf{q})\in\mathcal{T}} \delta(\mathbf{k},\mathbf{p},\mathbf{q}) K(\mathbf{k},\mathbf{p},\mathbf{q}) <\zeta_{-\mathbf{p}}(t)><\zeta_{-\mathbf{q}}(t)> + f_\chi(\mathbf{k},t) + \overline{f}_0(\mathbf{k},t).$$

The diagonal two-time two-point cumulant equation is:

$$\left(\frac{\partial}{\partial t} + \nu_0(\mathbf{k})k^2\right) C_{\mathbf{k}}(t,t') + \int_{t_0}^{t} ds\, (\eta_{\mathbf{k}}(t,s) + \pi_{\mathbf{k}}(t,s)) C_{-\mathbf{k}}(t',s)$$
$$= \int_{t_0}^{t'} ds\, (S_{\mathbf{k}}(t,s) + P_{\mathbf{k}}(t,s) + F_0(\mathbf{k},t,s)) R_{-\mathbf{k}}(t',s) \tag{81}$$

where $t > t'$ with $C_{\mathbf{k}}(t,t') = C_{-\mathbf{k}}(t',t)$ for $t' > t$. The diagonal response function equation is

$$\left(\frac{\partial}{\partial t} + \nu_0(\mathbf{k})k^2\right) R_{\mathbf{k}}(t,t') + \int_{t'}^{t} ds(\eta_{\mathbf{k}}(t,s) + \pi_{\mathbf{k}}(t,s)) R_{\mathbf{k}}(s,t') = \delta(t-t') \tag{82}$$

for $t \geq t'$ and $R_{\mathbf{k}}(t,t') = 1$.

In Equation (81) note that

$$<\tilde{f}_0(\mathbf{k},t)\tilde{\zeta}_{-\mathbf{k}}(t')> = \int_{t_0}^{t'} ds\, F_0(\mathbf{k},t,s)R_{-\mathbf{k}}(t',s) \quad (83)$$

where $t_0$ is the initial time and

$$F_0(\mathbf{k},t,s) = <\tilde{f}_0(\mathbf{k},t)\tilde{f}_0^*(\mathbf{k},s)>. \quad (84)$$

As well, in Equations (80) to (82), the following terms are needed:

$$f_\chi(\mathbf{k},t) = h_{\mathbf{k}}\int_{t_o}^{t} ds\, \chi_{\mathbf{k}}(t,s), \quad (85)$$

with

$$\chi_{\mathbf{k}}(t,s) = 2\sum_{(\mathbf{p},\mathbf{q})\in \mathcal{T}} \delta(\mathbf{k},\mathbf{p},\mathbf{q})K(\mathbf{k},\mathbf{p},\mathbf{q})A(-\mathbf{p},-\mathbf{q},-\mathbf{k})R_{-\mathbf{p}}(t,s)C_{-\mathbf{q}}(t,s), \quad (86)$$

and

$$\eta_{\mathbf{k}}(t,s) = -4\sum_{(\mathbf{p},\mathbf{q})\in \mathcal{T}} \delta(\mathbf{k},\mathbf{p},\mathbf{q})K(\mathbf{k},\mathbf{p},\mathbf{q})K(-\mathbf{p},-\mathbf{q},-\mathbf{k})R_{-\mathbf{p}}(t,s)C_{-\mathbf{q}}(t,s), \quad (87)$$

$$S_{\mathbf{k}}(t,s) = 2\sum_{(\mathbf{p},\mathbf{q})\in \mathcal{T}} \delta(\mathbf{k},\mathbf{p},\mathbf{q})K(\mathbf{k},\mathbf{p},\mathbf{q})K(-\mathbf{k},-\mathbf{p},-\mathbf{q})C_{-\mathbf{p}}(t,s)C_{-\mathbf{q}}(t,s), \quad (88)$$

$$\pi_{\mathbf{k}}(t,s) = -\sum_{(\mathbf{p},\mathbf{q})\in \mathcal{T}} \delta(\mathbf{k},\mathbf{p},\mathbf{q})R_{-\mathbf{p}}(t,s)[2K(\mathbf{k},\mathbf{p},\mathbf{q})<\zeta_{-\mathbf{q}}(t)> +A(\mathbf{k},\mathbf{p},\mathbf{q})h_{-\mathbf{q}}]$$
$$\times [2K(-\mathbf{p},-\mathbf{k},-\mathbf{q})<\zeta_{\mathbf{q}}(s)> +A(-\mathbf{p},-\mathbf{k},-\mathbf{q})h_{\mathbf{q}}], \quad (89)$$

$$P_{\mathbf{k}}(t,s) = \sum_{(\mathbf{p},\mathbf{q})\in \mathcal{T}} \delta(\mathbf{k},\mathbf{p},\mathbf{q})C_{-\mathbf{p}}(t,s)[2K(\mathbf{k},\mathbf{p},\mathbf{q})<\zeta_{-\mathbf{q}}(t)> +A(\mathbf{k},\mathbf{p},\mathbf{q})h_{-\mathbf{q}}]$$
$$\times [2K(-\mathbf{k},-\mathbf{p},-\mathbf{q})<\zeta_{\mathbf{q}}(s)> +A(-\mathbf{k},-\mathbf{p},-\mathbf{q})h_{\mathbf{q}}]. \quad (90)$$

The terms Equations (85) to (90) renormalize the damping and or forcing in the mean field, two-point cumulant and response function equations. The eddy-topographic force $f_\chi(\mathbf{k},t)$ is a product of the topography $h_{\mathbf{k}}$ and the time history integral of $\chi_{\mathbf{k}}(t,s)$, which measures of the strength of the interaction of the transient eddies with the topography. The nonlinear damping, $\eta_{\mathbf{k}}(t,s)$, due to eddy-eddy interactions modifies dampens the mean field and, as well, two-point cumulant and response function. The term $\pi_{\mathbf{k}}(t,s)$ stems from the interaction of transient eddies with the mean flow and topography and further acts to dampen the two-point cumulant and response functions. The nonlinear noise terms, $S_{\mathbf{k}}(t,s)$ and $P_{\mathbf{k}}(t,s)$ renormalize the specified bare noise spectrum $F_0(\mathbf{k},t,s)$ in the two-point cumulant equation. The term $S_{\mathbf{k}}(t,s)$ occurs due to eddy-eddy interactions whereas $P_{\mathbf{k}}(t,s)$ arises from eddy-mean flow and eddy-topographic interactions. The noise terms are all positive semi-definite and represent energy and enstrophy injection.

Finally, to close the system, we require the integro-differential equation for the single-time two-point cumulant:

$$\left(\frac{\partial}{\partial t}+2\nu_0(k)k^2\right)C_{\mathbf{k}}(t,t)+2\mathrm{Re}\int_{t_0}^{t} ds(\eta_{\mathbf{k}}(t,s)+\pi_{\mathbf{k}}(t,s))C_{-\mathbf{k}}(t,s)$$
$$= 2\mathrm{Re}\int_{t_0}^{t} ds(S_{\mathbf{k}}(t,s)+P_{\mathbf{k}}(t,s)+F_0(\mathbf{k},t,s))R_{-\mathbf{k}}(t,s) \quad (91)$$

where the initial conditions $C_{\mathbf{k}}(t_0,t_0)$ are to be specified.

As shown above, there is some commonality between the QDIA and DIA closures, in their structures, and also in the appearance of the the nonlinear noise $S_{\mathbf{k}}(t,s)$ and nonlinear damping $\eta_{\mathbf{k}}(t,s)$ in the equations for the two-point cumulant $C_{\mathbf{k}}(t,t')$ and response function $R_{\mathbf{k}}(t,t')$. The QDIA closure equations for inhomogeneous flows contain three additional propagators involving the mean-field and topography. These are the noise and damping terms $P_{\mathbf{k}}(t,s)$ and $\pi_{\mathbf{k}}(t,s)$ in the cumulant and response function equations. As well, the in mean-field tendency equation, the term $\eta_{\mathbf{k}}(t,s)<\zeta_{\mathbf{k}}(s)>$ couples the nonlinear damping to the mean-field, and the additional eddy topographic force $f_\chi(\mathbf{k},t)$ is the effect of the eddy-topographic interaction term $\chi_{\mathbf{k}}(t,s)h_{\mathbf{k}}$.

The associated propagators can then be used in a graphical depiction of the integral terms in the QDIA closures for the two-time cumulant, response function and the mean-field. They belong to the class of closed one-loop expressions, with renormalized propagators and bare vertices, whose topological structures are a generalization of the graphs characterizing the homogeneous DIA. In Figures 4, 5 and 6, we show the diagrammatic representation of the integral terms in Equations (81), (82) and (80) respectively with $v_0(\mathbf{k})k^2 = v_0(k)k^2 + i\omega_{\mathbf{k}}$. These graphs are clearly a generalization of the second order renormalized perturbation theory graphs of Wyld (1961) [19] and Lee (1965) [20] that correspond to the DIA.

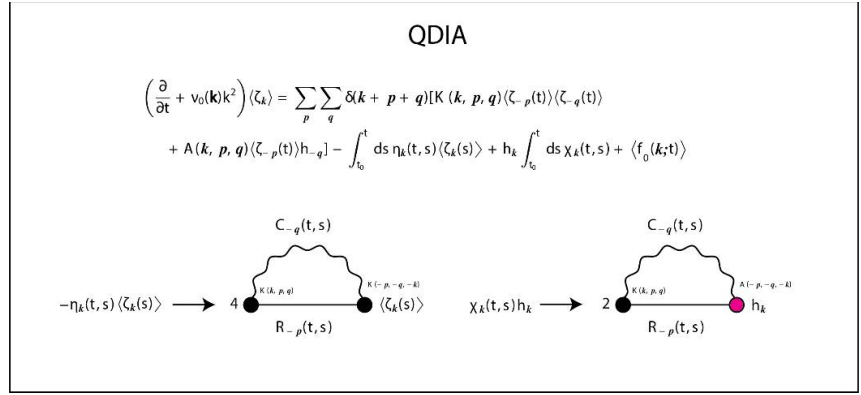

**Figure 4.** The diagrammatic representation of the integral terms in the QDIA mean-field $<\zeta_{\mathbf{k}}(t)>$ tendency equation. Note the Jacobian requires no diagrammatic representation. For the DIA the mean-field is zero.

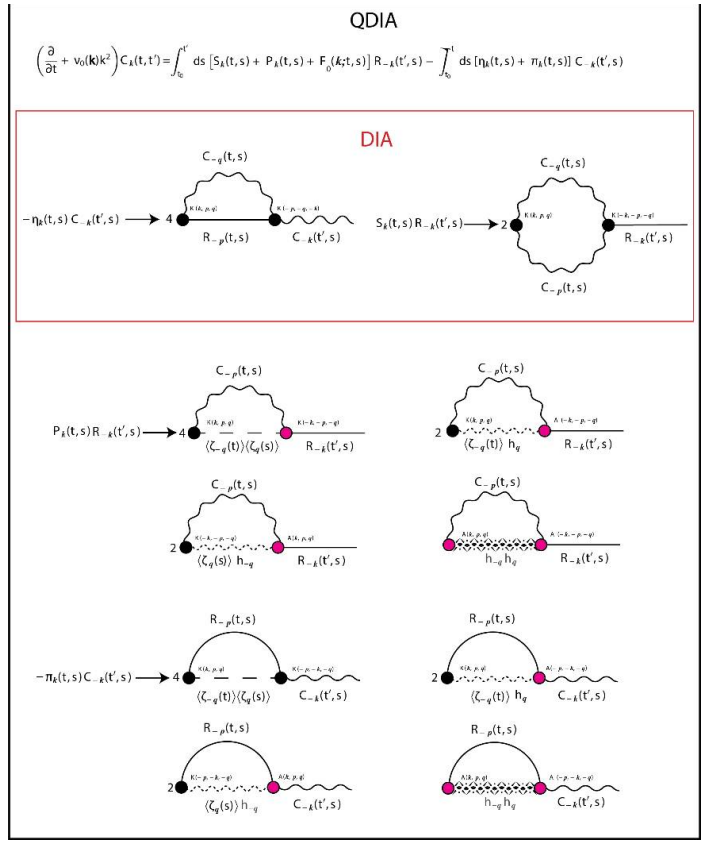

**Figure 5.** The diagrammatic representation of the integral terms in the QDIA diagonal two-point two-time cumulant $C_{\mathbf{k}}(t,t')$ tendency equation. The terms that the QDIA has in common with the DIA are within the red rectangle.

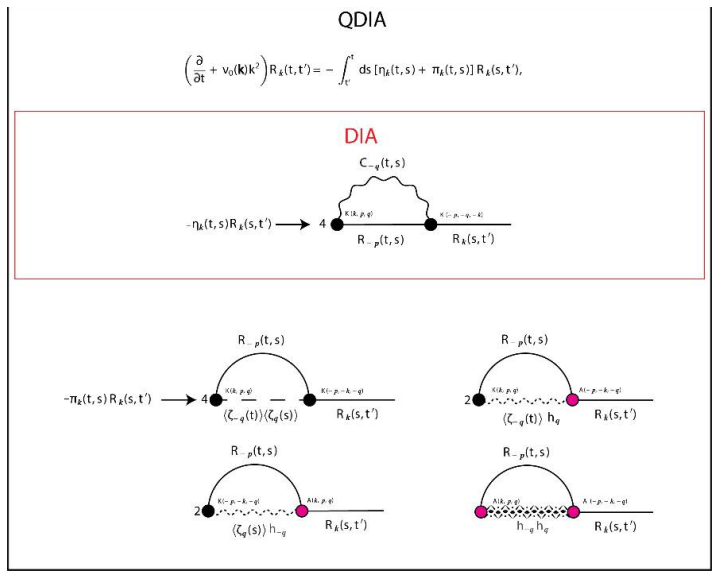

**Figure 6.** The diagrammatic representation of of the integral terms in the QDIA diagonal response function $R_{\mathbf{k}}(t,t')$ tendency equation. The terms that the QDIA has in common with the DIA are within the red rectangle.

## 8. Statistical Dynamical Equations for Markovian Inhomogeneous Closures

*8.1. General Formulation of Markovian Inhomogeneous Closures*

As noted in the Introduction, it possible to derive associated Markovian Inhomogeneous Closures (MICs) from the QDIA closure equations by employing any of the FDTs in Equation (4). Starting with the full QDIA equations of Section 7, Frederiksen and O'Kane (2019) [3] derived these MICs. Subsequently Frederiksen and O'Kane (2022) [75] established abridged MICs form modified the QDIA equations in which the mean-field $<\zeta_{\mathbf{k}}(s)>$ is replaced by the current-time mean field $<\zeta_{\mathbf{k}}(t)>$ in the time history integrals:

$$\overline{\zeta}_{\mathbf{k}}(s) \equiv <\zeta_{\mathbf{k}}(s)> \quad \to \quad <\zeta_{\mathbf{k}}(t)> \equiv \overline{\zeta}_{\mathbf{k}}(t). \tag{92}$$

Since the abridged MICs are slightly simpler, involving two relaxation functions rather than three for the full MICs, we present them here in this review. We note that the numerical performance for the MICs and the abridged MICs was shown to be very similar (Frederiksen and O'Kane 2022) [75]. Further, the abridged MICs using the current-time FDT ($X = 0$) is a steppingstone to deriving the EDMIC model with analytical relaxation functions. Complete details of the formulations of the full MICs are given in Frederiksen and O'Kane (2019) [3].

The single-time two-point cumulant equation for the abridged QDIA, common to each of the abridged MIC models can be written as:

$$\left(\frac{\partial}{\partial t} + 2\operatorname{Re}(D_r(\mathbf{k},t))\right) C_{\mathbf{k}}(t,t) = 2\operatorname{Re}(F_r(\mathbf{k},t)). \tag{93}$$

The renormalized dissipation and noise functions are defined by:

$$D_r(\mathbf{k},t) = D_0(\mathbf{k}) + D_\eta(\mathbf{k},t) + D_\pi(\mathbf{k},t), \tag{94}$$

and

$$F_r(\mathbf{k},t) = F_0(\mathbf{k},t) + F_S(\mathbf{k},t) + F_P(\mathbf{k},t). \tag{95}$$

Here the bare dissipation and noise functions are specified as follows:

$$D_0(\mathbf{k}) = \nu_0(k)k^2 + i\omega_{\mathbf{k}} = \nu_0(\mathbf{k})k^2, \tag{96}$$

and

$$F_0(\mathbf{k},t) = \int_{t_0}^{t} ds F_0(\mathbf{k},t,s) R_{-\mathbf{k}}(t,s). \tag{97}$$

The nonlinear damping given by

$$D_\eta(\mathbf{k},t) = -4 \sum_{(\mathbf{p},\mathbf{q}) \in \mathcal{T}} \delta(\mathbf{k},\mathbf{p},\mathbf{q}) K(\mathbf{k},\mathbf{p},\mathbf{q}) K(-\mathbf{p},-\mathbf{q},-\mathbf{k}) \\ \times C_{-\mathbf{q}}^{1-X}(t,t) C_{-\mathbf{k}}^{-X}(t,t) \Theta^X(-\mathbf{p},-\mathbf{q},-\mathbf{k})(t), \tag{98}$$

and

$$D_\pi(\mathbf{k},t) = -\sum_{(\mathbf{p},\mathbf{q}) \in \mathcal{T}} \delta(\mathbf{k},\mathbf{p},\mathbf{q})[2K(\mathbf{k},\mathbf{p},\mathbf{q}) <\zeta_{-\mathbf{q}}(t)> + A(\mathbf{k},\mathbf{p},\mathbf{q})h_{-\mathbf{q}}] \\ \times [2K(-\mathbf{p},-\mathbf{k},-\mathbf{q}) <\zeta_{\mathbf{q}}(t)> + A(-\mathbf{p},-\mathbf{k},-\mathbf{q})h_{\mathbf{q}}] C_{-\mathbf{k}}^{-X}(t,t) \Psi^X(-\mathbf{p},-\mathbf{k})(t). \tag{99}$$

The nonlinear noise term

$$F_S(\mathbf{k},t) = 2 \sum_{(\mathbf{p},\mathbf{q}) \in \mathcal{T}} \delta(\mathbf{k},\mathbf{p},\mathbf{q}) K(\mathbf{k},\mathbf{p},\mathbf{q}) K(-\mathbf{k},-\mathbf{p},-\mathbf{q}) \\ \times C_{-\mathbf{p}}^{1-X}(t,t) C_{-\mathbf{q}}^{1-X}(t,t) \Theta^X(-\mathbf{k},-\mathbf{p},-\mathbf{q})(t), \tag{100}$$

and

$$F_P(\mathbf{k},t) = \sum_{(\mathbf{p},\mathbf{q}) \in \mathcal{T}} \delta(\mathbf{k},\mathbf{p},\mathbf{q})[2K(\mathbf{k},\mathbf{p},\mathbf{q}) <\zeta_{-\mathbf{q}}(t)> + A(\mathbf{k},\mathbf{p},\mathbf{q})h_{-\mathbf{q}}] \\ \times [2K(-\mathbf{k},-\mathbf{p},-\mathbf{q}) <\zeta_{\mathbf{q}}(t)> + A(-\mathbf{k},-\mathbf{p},-\mathbf{q})h_{\mathbf{q}}] C_{-\mathbf{p}}^{1-X}(t,t) \Psi^X(-\mathbf{k},-\mathbf{p})(t). \tag{101}$$

In these equations, the relaxation functions take the form:

$$\Theta^X(-\mathbf{k},-\mathbf{p},-\mathbf{q})(t) = \int_{t_0}^{t} ds\, R_{-\mathbf{k}}(t,s) R_{-\mathbf{p}}(t,s) R_{-\mathbf{q}}(t,s) C_{-\mathbf{p}}^X(s,s) C_{-\mathbf{q}}^X(s,s), \quad (102)$$

and

$$\Psi^X(-\mathbf{k},-\mathbf{p})(t) = \int_{t_0}^{t} ds\, R_{-\mathbf{k}}(t,s) R_{-\mathbf{p}}(t,s) C_{-\mathbf{p}}^X(s,s). \quad (103)$$

For consistency with the structure of the single-time two-point cumulant equation the response function in Equation (82) is modified to:

$$\frac{\partial}{\partial t} R_{\mathbf{k}}(t,t') + D_r(\mathbf{k},t) R_{\mathbf{k}}(t,t') = \delta(t-t'). \quad (104)$$

This means that the triad relaxation function can be expressed by the Markov differential equation:

$$\frac{\partial}{\partial t}\Theta^X(\mathbf{k},\mathbf{p},\mathbf{q})(t) + (D_r(\mathbf{k},t) + D_r(\mathbf{p},t) + D_r(\mathbf{q},t))\Theta^X(\mathbf{k},\mathbf{p},\mathbf{q})(t) \\ = C_{\mathbf{p}}^X(t,t) C_{\mathbf{q}}^X(t,t) \quad (105)$$

where $\Theta^X(\mathbf{k},\mathbf{p},\mathbf{q})(0) = 0$ and $D_r$ is given in Equation (94). The second relaxation function again satisfies a Markov differential equation:

$$\frac{\partial}{\partial t}\Psi^X(\mathbf{k},\mathbf{p})(t) + (D_r(\mathbf{k},t) + D_r(\mathbf{p},t) + D_r(\mathbf{q},t))\Psi^X(\mathbf{k},\mathbf{p})(t) = C_{\mathbf{p}}^X(t,t) \quad (106)$$

with $\Psi^X(\mathbf{k},-\mathbf{k})(0) = 0$.

To formulate manifestly Markovian equations for the mean-field, and hence establish a closed system of coupled Markovian equations for the mean and two-point cumulant, we again approximate the mean-field $<\zeta_{\mathbf{k}}(s)>$ by the current-time mean-field $<\zeta_{\mathbf{k}}(t)>$ in the time history integrals as in Equation (71). Then, from Equation (80) we have

$$\left(\frac{\partial}{\partial t} + D_0(\mathbf{k}) + D_M(\mathbf{k},t)\right) <\zeta_{\mathbf{k}}(t)> \\ = \sum_{(\mathbf{p},\mathbf{q}) \in \mathcal{T}} \delta(\mathbf{k},\mathbf{p},\mathbf{q})[K(\mathbf{k},\mathbf{p},\mathbf{q}) <\zeta_{-\mathbf{p}}(t)><\zeta_{-\mathbf{q}}(t)> \\ + A(\mathbf{k},\mathbf{p},\mathbf{q}) <\zeta_{-\mathbf{p}}(t)> h_{-\mathbf{q}}] + f_\chi(\mathbf{k},t) + \overline{f}_0(\mathbf{k},t). \quad (107)$$

Here,

$$D_M(\mathbf{k},t) \\ = -4\sum_{(\mathbf{p},\mathbf{q})\in\mathcal{T}} \delta(\mathbf{k},\mathbf{p},\mathbf{q}) K(\mathbf{k},\mathbf{p},\mathbf{q}) K(-\mathbf{p},-\mathbf{q},-\mathbf{k}) C_{-\mathbf{q}}^{1-X}(t,t)\Psi^X(-\mathbf{p},-\mathbf{q})(t), \quad (108)$$

and the expression for the eddy-topographic force reduces to

$$f_\chi(\mathbf{k},t) \\ = 2h_{\mathbf{k}}\sum_{(\mathbf{p},\mathbf{q})\in\mathcal{T}} \delta(\mathbf{k},\mathbf{p},\mathbf{q}) K(\mathbf{k},\mathbf{p},\mathbf{q}) A(-\mathbf{p},-\mathbf{q},-\mathbf{k}) C_{-\mathbf{q}}^{1-X}(t,t)\Psi^X(-\mathbf{p},-\mathbf{q})(t). \quad (109)$$

Again, $\Psi^X$ is determined by Equation (106).

*8.2. Eddy Damped Markovian Inhomogeneous Closure*

From the abridged $MIC^{X=0}$ model, which uses the FDT with $X = 0$, it is possible to derive a very computationally efficient closure, the Eddy Damped Markovian Inhomogeneous Closure (EDMIC). It is a suitable generalization of the EDQNM model for homogeneous isotropic turbulence in that it too uses analytical forms for the relaxation functions. For the current-time FDT when $X = 0$, Equation (93) for the single-time cumulant simplifies as the relaxation functions in Equations (105) and (106) replace $C_{\mathbf{k}}^{x=0} \to 1$. As shown by Frederiksen and O'Kane (2022) [75] it is then possible to approximate the eddy damping by an analytical expression as for the EDQNM closure which is further discussed in Section 6.2.

*8.3. Comparison of Non-Markovian and Markovian Closure Calculations with DNS*

In this Section, we discuss the results of Frederiksen and O'Kane (2022) [75] on the performance of the non-Markovian abridged $QDIA[\overline{\zeta}(t)]$ closure, and the three abridged $MIC^X[\overline{\zeta}(t)]$ Markovian closures compared with an ensemble of 1800 DNS. These results are further contrasted to those of Frederiksen and O'Kane (2019) [3] for the $QDIA[\overline{\zeta}(s)]$ and $MIC^X[\overline{\zeta}(s)]$ models. Here, the arguments specify the abridged version by $\overline{\zeta}(t)$ and the full versions by $\overline{\zeta}(s)$. Taken as a whole, these numerical simulations provide insights into the robustness of the inhomogeneous closure calculations to Markovianization, with respect to the three versions of the FDT. As well they throw light on the accuracy of the abridged versions of QDIA and MIC models and the extent to which they are impacted by the full trajectory of $\overline{\zeta}(s) \equiv <\zeta(s)>$ for $t_0 \leq s \leq t$ or not.

Once again, we employ the setup of Frederiksen and O'Kane (2005) [67] where a large-scale mean eastward flow $U$ interacts with topography in a turbulent environment, which is a severe test of the different closure formulations. We consider turbulent flows on a $\beta$ – plane, where the the large-scale flow $U$ interact with the topography to generated Rossby waves with the eddies providing a form drag on $U$. The associated rapid spin up induces strong transfers of energy from the large-scale flow to the smaller scale mean-field, including wave-turbulence interactions and changes in the wavenumber distribution of transient energy.

The experimental configuration uses length and time scales of one half the earth's radius, $a_e/2$, and the inverse of the earth's rotation rate, $\mathbf{\Omega}^{-1}$, respectively. The initial mean eastward flow $U$ has a speed of $7.5 \text{ ms}^{-1}$ (non-dimensional $U = 0.0325$), the $\beta$ – effect is $1.15 \times 10^{-11} \text{ m}^{-1}\text{s}^{-1}$ (non-dimensional $\beta = \frac{1}{2}$) representative of the earth at $60°$ latitude, and the coefficient of viscosity $v_0$ is $2.5 \times 10^4 \text{ m}^2\text{s}^{-1}$ (non-dimensional $\hat{v} = 3.378 \times 10^{-5}$). The bare forcing $f_0 = 0$, the drag on the large-scale flow $\alpha_U = 0$ and $k_0^2 = \frac{1}{2}$. The topography is a 2.5 km high cone centered at $60°\text{N}$, $180°\text{E}$ with and a diameter of $45°$ latitude Frederiksen and O'Kane (2005) [67] (their Figure 1). This may be viewed as an idealized representation of the Himalayas. The DNS and closures are integrated over 10 days, with a time step of $\Delta t = 1/30$ day (non-dimensional $\Delta t = 0.21$), and performed at a resolution of circular truncation $C15 \equiv T_{15} = 15$ where $|\mathbf{k}| = k \leq 16$.

Table 1 specifies the initial transient isotropic spectrum and initial 'small-scale' mean field, which is localized over the topography and of small amplitude. The closure calculations start from Gaussian initial conditions with $C_{\mathbf{k}}(t_0, t_0) = C_{\mathbf{k}}(0,0)$ and with the initial mean-field, $\overline{\zeta}_{\mathbf{k}}(t_0) = <\zeta_{\mathbf{k}}(t_0)> = <\zeta_{\mathbf{k}}(0)>$, specified as shown. The DNS comprises an ensemble of 1800 simulations started from the initial mean-field to which perturbations are sampled from an isotropic Gaussianly distribution with spectrum $C_{\mathbf{k}}(t_0, t_0) = C_{\mathbf{k}}(0,0)$ also defined in Table 1. Further details on the setup of the DNS perturbations are given in Ref. [67]. A predictor-corrector procedure is used for the time stepping in both DNS and closures and the the trapezoidal rule for calculating the time history integrals in the non-Markovian closures (Frederiksen et al. 1994 [16]; Frederiksen and O'Kane 2005 [67]).

**Table 1.** Initial conditions used in DNS and closure calculations in the comparison of Non-Markovian and Markovian Closure Calculations with DNS.

| $C_\mathbf{k}(0,0)$ | $<\zeta_\mathbf{k}(0)>$ | $a$ | $b$ |
|---|---|---|---|
| $\dfrac{k^2 \times 10^{-2}}{a+bk^2}$ | $-10bh_\mathbf{k}C_\mathbf{k}(0,0)$ | $4.824\times 10^4$ | $2.511\times 10^3$ |

We compare the evolved closure and DNS results by the similarity of the mean flow fields and quantified by their pattern correlations, as well as their mean and transient kinetic energy and palinstrophy spectra. The mean, transient, and total kinetic energy spectra, averaged over circular bands, are defined by:

$$\overline{E}(k_i,t) = \tfrac{1}{2}\sum_{\mathbf{k}\in\mathcal{S}} \langle \zeta_\mathbf{k}(t)\rangle \langle \zeta_{-\mathbf{k}}(t)\rangle k^{-2}, \tag{110}$$

with

$$\tilde{E}(k_i,t) = \tfrac{1}{2}\sum_{\mathbf{k}\in\mathcal{S}} C_\mathbf{k}(t,t) k^{-2}, \tag{111}$$

and the total is

$$E(k_i,t) = \overline{E}(k_i,t) + \tilde{E}(k_i,t). \tag{112}$$

Here, the set $\mathcal{S}$ is defined by Equation (50).

The initial mean non-zonal streamfunction for the simulations is shown in Figure 7a. The other five panels in Figure 7 show the day 10 evolved mean streamfunctions for the non-Markovian QDIA (Figure 7b) and Markovian abridged $MIC^X$ models (Figures 7c, 7d and 7e) and the ensemble of DNS (Figure 7f). All the closures and the direct numerical simulation reveal substantial evolution from the initial mean-field in terms of magnitude and structure, with large scale Rossby wavetrains developed primarily downstream of the conical mountain. There is little evidence to distinguish between the 10-day evolved fields apart from slight variations in the magnitude of the peaks and troughs of the wavetrains.

As presented in Table 2, of the abridged closures, the largest pattern correlations of the DNS mean non-zonal streamfunctions occur for the $MIC^0[\overline{\zeta}(t)]$ at 0.9999, followed closely by the other two, with $MIC^{1/2}[\overline{\zeta}(t)]$ at 0.9994 and $MIC^1[\overline{\zeta}(t)]$ at 0.9995. The lowest pattern correlation is for the $QDIA[\overline{\zeta}(t)]$ at 0.9789, which is still excellent. The corresponding pattern correlations for the DNS evolved field with $MIC^0[\overline{\zeta}(s)]$ is identical to that with $MIC^0[\overline{\zeta}(t)]$ at 0.9999, and each of the three closures $QDIA[\overline{\zeta}(s)]$, $MIC^{1/2}[\overline{\zeta}(s)]$ and $MIC^1[\overline{\zeta}(s)]$ have common value for of 0.9998. It is interesting that the replacement $\overline{\zeta}(s)\to\overline{\zeta}(t)$ has most effect on the non-Markovian QDIA closure and least impact on the $MIC^0$ which implements the current-time FDT. In all cases, given the dramatic evolution that the flow field has undergone, the pattern correlations are amazingly high.

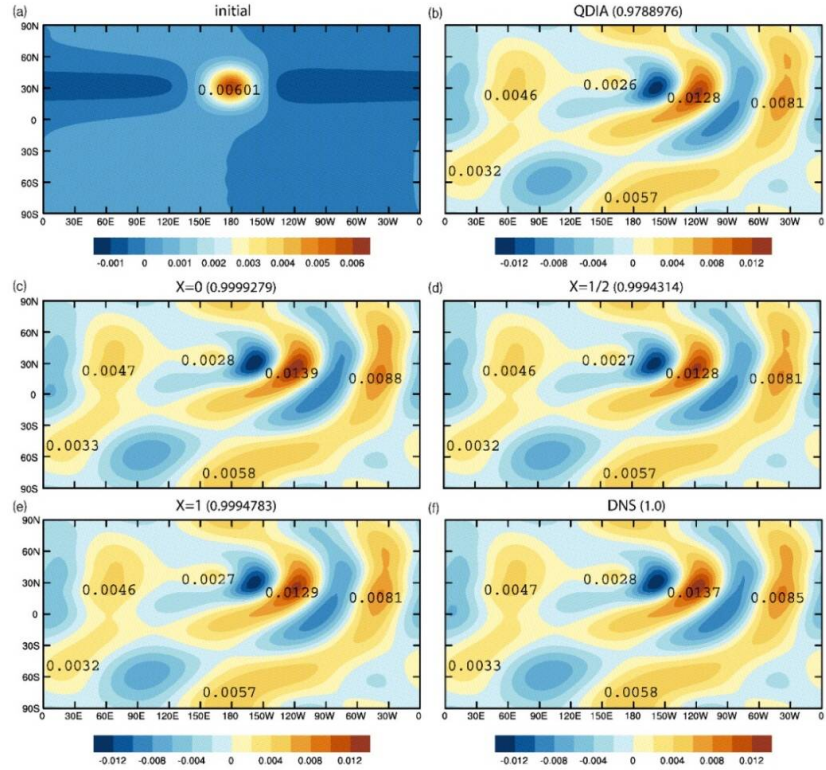

**Figure 7.** The initial (a) and evolved day 10 (b) $QDIA[\overline{\zeta}(t)]$, (c) $MIC^{X=0}[\overline{\zeta}(t)]$, (d) $MIC^{X=\frac{1}{2}}[\overline{\zeta}(t)]$, (e) $MIC^{X=1}[\overline{\zeta}(t)]$, and (f) ensemble of DNS Rossby wave streamfunction in non-dimensional units. Multiply values by $\frac{1}{4}a_e^2\Omega^{-1} = 740 \text{ km}^2\text{s}^{-1}$ to convert to units of $\text{km}^2\text{s}^{-1}$. From Frederiksen and O'Kane (2022) [75].

**Table 2.** Correlations of day 10 non-zonal streamfunction for closures with DNS

| $QDIA[\overline{\zeta}(t)]: 0.9789$ | $MIC^{X=0}[\overline{\zeta}(t)]: 0.9999$ | $MIC^{X=\frac{1}{2}}[\overline{\zeta}(t)]: 0.9994$ | $MIC^{X=1}[\overline{\zeta}(t)]: 0.9995$ |
|---|---|---|---|
| $QDIA[\overline{\zeta}(s)]: 0.9998$ | $MIC^{X=0}[\overline{\zeta}(s)]: 0.9999$ | $MIC^{X=\frac{1}{2}}[\overline{\zeta}(s)]: 0.9998$ | $MIC^{X=1}[\overline{\zeta}(s)]: 0.9998$ |

In Figure 8 we depict the initial, and 10-day evolved, mean and transient kinetic energy spectra of the DNS, and the abridged closures, the $QDIA[\overline{\zeta}(t)]$ and each of the Markovian $MIC^X[\overline{\zeta}(t)]$ variants with $X = 0, \frac{1}{2}, 1$. The $QDIA[\overline{\zeta}(t)]$, $MIC^{\frac{1}{2}}[\overline{\zeta}(t)]$ and $MIC^1[\overline{\zeta}(t)]$ reveal slightly larger evolved transient energy near the peak at the wavenumber $k = 4$ while for $MIC^0[\overline{\zeta}(t)]$ it is marginally underestimated. The wavenumber $k = 4$ corresponds to the peak Rossby wave amplitude in the mean-field. Hence the slight differences are representative of resonant interactions between the large-scale flow and the transient eddy field. In general, all the closures closely compare with the ensemble of DNS. Somewhat surprisingly, the findings reveal that using the current-time mean field in the time history integrals does not significantly degrade the abridged closure results compared with those of the full closures in Frederiksen and O'Kane (2019) [3].

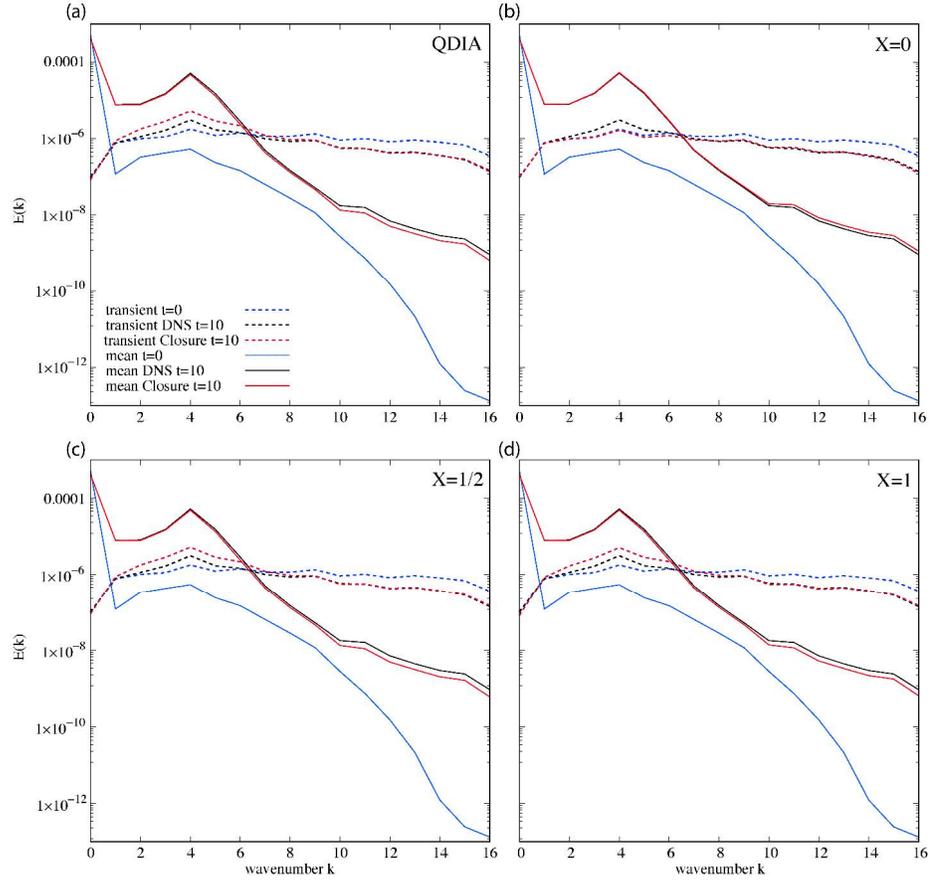

**Figure 8.** The initial and evolved non-dimensional kinetic energy spectra for the four inhomogeneous abridged closures in panels (a) $QDIA[\overline{\zeta}(t)]$, (b) $MIC^{X=0}[\overline{\zeta}(t)]$, (c) $MIC^{1/2}[\overline{\zeta}(t)]$, (d) $MIC^{X=1}[\overline{\zeta}(t)]$, as compared to ensemble mean of 1800 DNSs. In each panel we show: initial mean energy (solid blue), initial transient energy (dashed blue), evolved DNS mean energy (solid black), evolved DNS transient energy (dashed black), evolved closure mean energy (solid red) and evolved closure transient energy (dashed red). Multiplication by $\frac{1}{4}a_e^2\Omega^{-2} = 5.4 \times 10^4 \text{ m}^2\text{s}^{-2}$ is required to convert to units of $\text{m}^2\text{s}^{-2}$. From Frederiksen and O'Kane (2022) [75].

For higher resolution, and higher Reynolds numbers, the expectation is that the Markovian closures, like the Eulerian homogeneous non-Markovian DIA (Frederiksen and Davies 2004) [33] and inhomogeneous QDIA (O'Kane and Frederiksen 2004) [43], require regularization, or empirical vertex renormalization, to produce the correct inertial transfers in the small-scale spectra. The regularization localizes the interaction coefficients by replacing $A(\mathbf{k},\mathbf{p},\mathbf{q})$ with $\theta(p-k/\alpha)\theta(q-k/\alpha)A(\mathbf{k},\mathbf{p},\mathbf{q})$ and, respecting Equation (6), $K(\mathbf{k},\mathbf{p},\mathbf{q})$ by $\theta(p-k/\alpha)\theta(q-k/\alpha)K(\mathbf{k},\mathbf{p},\mathbf{q})$ in the response function and two-time cumulant equations, but not in the single-time cumulant or mean field equations. Here, $\theta$ is the Heaviside step function and $\alpha$ is a wavenumber cut-off parameter. It plays similar role as the $\gamma$ in the eddy damping for the EDQNM closure in Equation (64) of Section 6. As noted in the Introduction, a value of $\alpha \approx 4$ appears to be essentially universal, or only weakly flow dependent, for 2D homogeneous and inhomogeneous turbulence and for 3D HIT.

The high level of correspondence between DNS results and those for all the abridged Markovian inhomogeneous closures is remarkable. This is particularly the case given the rapidly evolving flows and the potential for numerical instability. It encourages us to pursue future numerical studies with the EDMIC model of Section 8.2, that is further simplified, and very efficient, like the EDQNM, with analytical forms for the relaxation functions.

## 9. Applications of the Inhomogeneous QDIA Closure

In this Section we review applications of the QDIA to problems of numerical weather prediction and data assimilation. In both applications the formal problem manifests itself as estimation of covariances and approximation of higher order moments in inhomogeneous systems with quadratic nonlinearities. Our aim is to attempt to reveal the deep insights that closure theory can bring to practical application of wide interest.

### 9.1. Ensemble Prediction

In addition to analysis errors, and model deficiencies, perhaps the greatest limitation to skillful long-range numerical weather prediction (NWP) arises due to the instability properties of the atmospheric flow itself. Key to establishing the theoretical limits to atmospheric predictability is accurate representation of the growth of instabilities or errors. Instabilities are characterized by the divergence of pairs of initially close states. Their evolved configurations arise from the convergence of initially random perturbations onto coherent structures spanned by local or finite time normal modes. In modern ensemble weather forecasting, one seeks to calculate the moments of the relevant meteorological variables. In this way, the problem is recast in terms of statistical quantities, where uncertainties due to systematic errors in the initial conditions, may be estimated and minimized.

In ensemble NWP, initial perturbations are projected onto the subspace of the fast-growing instabilities of the day. Many approaches exist for their generation based on various dynamical vectors such as singular vectors (Leutbecher et al. 2008 [143]; Quinn et al. 2022 [144]), finite time normal modes (Wei and Frederiksen) [145] and bred vectors (Toth and Kalnay 1993 [146]), as reviewed by Frederiksen (2023) [147]. The bred perturbations are simply constructed and are essentially stochastically modified evolved, or backward, leading Lyapunov vectors. In this approach initial forecast perturbations are "bred" from an initially Gaussian isotropic error field through a process of periodic rescaling and reinitializing the forward model. The evolving perturbation is scaled after every time step or cycling over a fixed window in time. In an examination of weather prediction O'Kane and Frederiksen (2008a) [68] applied the QDIA closure, and variants thereof, in order to quantify the contributions of inhomogeneous perturbations and non-Gaussian terms to better estimate forecast error growth. They applied the breeding approach to define a global scaling parameter $g(t)$, in terms of the streamfunction, adjusting the perturbation field with

$$g(t) = \left[ \frac{\sum_{\mathbf{k}} C_{\mathbf{k}}(t_0, t_0) k^{-4}}{\sum_{\mathbf{k}} C_{\mathbf{k}}(t, t) k^{-4}} \right]^{\frac{1}{2}}. \tag{113}$$

The perturbation field is scaled through the expression

$$\zeta_{\mathbf{k}}(t_+) = g(t) \zeta_{\mathbf{k}}(t_-) \tag{114}$$

and the equal time cumulant by

$$C_{\mathbf{k}}(t_+, t_+) = g(t) g(t) C_{\mathbf{k}}(t_-, t_-) \tag{115}$$

while the two-time cumulant is updated by

$$C_{\mathbf{k}}(t_+, t'_+) = g(t) g(t') C_{\mathbf{k}}(t_-, t'_-). \tag{116}$$

Here, the + and - subscripts indicate the prior and posterior fields, respectively. The off-diagonal elements of the covariances are updated through their dependence on the diagonal cumulants as shown in Equation (A25). The breeding cycle evolves isotropic initial perturbations, with uniform structures, allowing their organization, such that they contain information about the fast-growing errors, that capture instabilities present in the underlying dynamics of the flow.

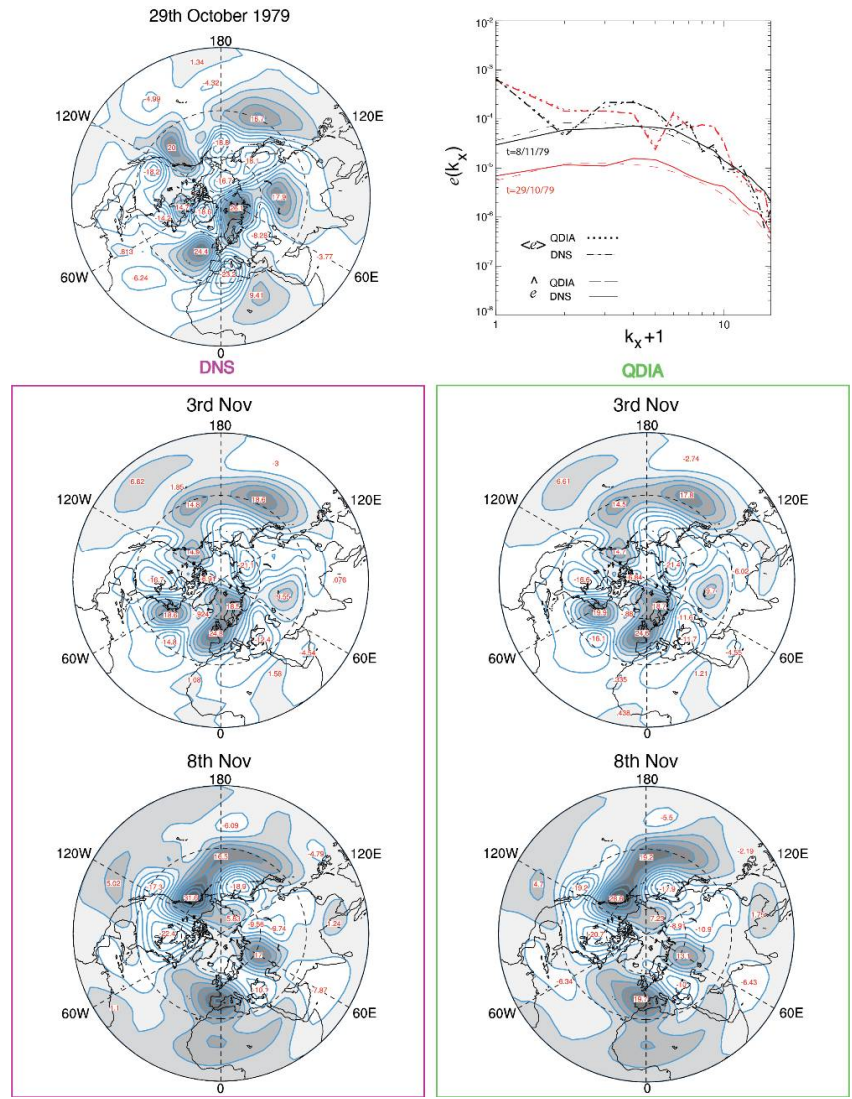

**Figure 9.** The physical space plots of the eddy streamfunction ($m^2 s^{-1}$) for both DNS and QDIA models at the initial day (29 Oct) and day 5 (3 Nov) of the breeding cycle, and after five forecast days (8 Nov). Adapted from O'Kane and Frederiksen (2008a) [68].

In their ensemble prediction studies O'Kane and Frederiksen (2008a) [68] considered 5-day breeding periods starting on 26, 27, 28, and 29 October 1979, respectively followed by 5-day forecast periods. This sequentially stepped through the lifecycle of a large-scale high–low blocking dipole that formed over the Gulf of Alaska developing on 5 November and persisting until 12 November 1979. The formation of large-scale coherent synoptic structures in the mid-troposphere are typically associated with increased flow instability and significantly lower predictability. They found close agreement comparing a regularised variant of the QDIA closure and DNS (3600 realizations), including using a measure of small-scale differences between closure and DNS, namely, the palinstrophy production $K$ of Equation (45). The major finding was the identification of four distinct periods of error organization and growth. Specifically:
1. The organization of error structures and the growth of off-diagonal covariances starting from isotropic initial perturbations;
2. The development of organised flow-dependent error correlations;
3. Rapid initial forecast error growth during this organizational period subsequent to breeding;
4. Reduced error growth as the error field saturates the mean field amplitude and predictability is lost.

They also used stochastic backscatter forcing as a proxy for model noise, or error, perturbing the small-scale transients leading to decreased amplitudes for $K$. This is due to energy injected at the small scales increases isotropy and retards the organization of coherent transient structures at the smallest resolved scales.

*9.2. The Application of the QDIA to Data Assimilation*

Turbulence closures and data assimilation methodologies have a range of similar technical difficulties. The most important is the calculation of the error covariance of fields described by an evolving nonlinear system and the infinite hierarchy of moment equations that need to be closed in terms of low order cumulants. In applications to weather prediction, the aim of data assimilation is to obtain a near optimal estimate of the state of the atmosphere, based on observations and on short term forecasts, which provide the so-called background states, with information in the data void areas. Data assimilation methods are filters used to extract the signal from noisy observations, and biased numerical models, to produce analysed states. These are required to initialize forecast models at the required resolution and time. The crucial goal is the estimation of the current state of the system given observations of the past. Such an estimate is typically obtained recursively in time, utilizing new observations, as they become available either synchronously or collected over a given time window (asynchronously). In this regard, the accuracy of numerical weather prediction models is critically dependent upon the accuracy of the specified initial conditions. For Kalman filter variants, the central problem is how best to model the forecast, or background, error covariance matrices used to generate the analyses which, for large numerical prediction models, with millions of variables as they evolve with time remains an intractable problem.

9.2.1. Kalman Filter Variants

For systems with Gaussian statistics, the Kalman filter is an optimal method of estimating the state of the system given noisy observations. The Kalman filter theory (Kalman 1960) [148] implicitly assumes that both the observations and priori (forecast) distributions, based on the background state, are Gaussianly distributed. The posteriori distribution for the analysis is then derived based on the products of these two Gaussians and yields simply the Kalman filter equations (Lorenc 1986 [149]; Anderson 2001 [150]). In the standard Kalman filter theory it is also assumed that the background or forecast field variable $\zeta^f(t)$ satisfies a linear equation.

The extended Kalman filter (EKF) (Ghil et al. 1981 [151]; Evensen 1992 [152]) is formulated entirely in terms of covariances and completely neglects higher order moments. This approach is effectively a method of successive linearization about the evolving nonlinear trajectory. It is a tangent linear approximation for calculating the covariance $C^f(t,t)$ between $\zeta^f(t)$ at different spatial or spectral space points. In practice, a tangent linear model is used to transform a perturbation $\zeta^f(t)$ at some initial time $t$ to a final perturbation $\zeta^f(t+\Delta t)$ at time $t+\Delta t$ (Kalnay 2003) [153]. Linearizing about the nonlinear model trajectory will however be inaccurate whenever the perturbations are large. In order to tackle strongly nonlinear systems, where bifurcations may occur, Miller et al. (1994) [154] developed the generalized EKF (GEKF). The GEKF uses a moment expansion method in terms of Taylor series to estimate the contributions of the higher order moments to the Kalman gain. This extends the EKF methodology to strongly nonlinear systems by including estimates of the third and fourth order moment contributions to the error. Where regime transitions occur, tracking the transitions requires either reducing the timestep and increasing the frequency of observations or alternately increasing the estimated Kalman gain. The latter requires inclusion of approximations to the third and fourth moments into the EKF to track the reference solution and to capture the bifurcation behavior of chaotic trajectories. In general, the GEKF has been found to be impractical for higher dimensional models.

9.2.2. Spectral Kalman Filter Equations

O'Kane and Frederiksen (2008b, 2010) [69,70] extended the EnKF to specifically account for higher order moments beyond assumptions of Gaussianity. They introduced the statistical dynamical Kalman filter (SDKF) by formulating the Kalman filter equations directly in terms of the spectral coefficients of the forecast model, rather than in terms of physical space grid point values and employing the QDIA to calculate the covariances and higher order cumulants.

The statistical dynamical Kalman filter (SDKF) of O'Kane and Frederiksen (2008b, 2010) [69,70] is analogous to the EnKF, however the SDKF implements the equations:

$$<\zeta_\mathbf{k}^a(t)> = <\zeta_\mathbf{k}^f(t)> + K_\mathbf{k}\left(<d_\mathbf{k}(t)> - H_\mathbf{k}<\zeta_\mathbf{k}^f(t)>\right), \quad (117)$$

and

$$C_\mathbf{k}^a(t,t) = \left(1 - K_\mathbf{k} H_\mathbf{k}\right) C_\mathbf{k}^f(t,t) \quad (118)$$

Here, $<\zeta_\mathbf{k}^a(t)>$ is the analyzed spectral field, $<d_\mathbf{k}(t)>$ the observational field and

$$K_\mathbf{k}(t) = C_\mathbf{k}^f(t,t) H_\mathbf{k}^*(t) \left(K_\mathbf{k}(t) C_\mathbf{k}^f(t,t) H_\mathbf{k}^*(t) + D_\mathbf{k}(t,t)\right)^{-1} \quad (119)$$

is the Kalman filter spectral gain. We factorize the associated observational operator $\mathbf{D}$, and the Kalman gain $\mathbf{K}$, into transformations to and from spectral and model grid point space, and similarly, to and from model grid and observational grid space. The factorizations of the operators are summarized in the table below. We similarly define the operator $\mathbf{H}^0$ and the spectral space transform of the observations $\mathbf{d}$ in terms of the observations $\mathbf{d}^{\text{obs grid}}$ on the observational grid, where $K_\mathbf{k}$ and $H_\mathbf{k}$ are the spectral diagonal elements of $\mathbf{K}^0$ and $\mathbf{H}^0$.

Table 1: Description of forward and backward transformation matrices and vectors.

| $\mathbf{K} = \mathbf{K}^0 \mathbf{K}^1 \mathbf{K}^2$ | $\mathbf{H} = \mathbf{H}^2 \mathbf{H}^1$ |
|---|---|
| $\mathbf{K}^0$ scaling | $\mathbf{H}^0 = \mathbf{K}^1 \mathbf{K}^2 \mathbf{H}^2 \mathbf{H}^1$ |
| $\mathbf{K}^1$ grid to spectral transform | $\mathbf{H}^1$ inverse spectral to grid transform |
| $\mathbf{K}^2$ transform from obs to model grid | $\mathbf{H}^2$ transforms from model to obs grid |
| $\mathbf{d} = \mathbf{K}^1 \mathbf{K}^2 \mathbf{d}^{\text{obs grid}}$ | $\mathbf{D} = \langle \mathbf{d}, \mathbf{d} \rangle$ |

The analysis covariance $C_\mathbf{k}^a(t,t)$ is modelled by Equation (118). Equations (117) and (118) flow through the QDIA equations during integration of the time history integrals automatically updating the off-diagonal covariance $C_{\mathbf{k},-\mathbf{l}}(t,t)$ and three-point terms $\left\langle \tilde{\zeta}_\mathbf{k}(t) \tilde{\zeta}_{-\mathbf{l}}(t) \tilde{\zeta}_{(\mathbf{l}-\mathbf{k})}(t) \right\rangle$. At each cycle, the off-diagonal covariances and three-point cumulant terms, are directly analyzed as follows:

$$C_{\mathbf{k},-\mathbf{l}}^a(t,t) = \left(1 - K_\mathbf{k} H_\mathbf{k}\right) C_{\mathbf{k},-\mathbf{l}}^f(t,t) \left(1 - K_{-\mathbf{l}} H_{-\mathbf{l}}\right) + K_\mathbf{k} D_{\mathbf{k},-\mathbf{l}}(t,t) K_{-\mathbf{l}}, \quad (120)$$

and

$$\left\langle \tilde{\zeta}_{\mathbf{k}}(t)\tilde{\zeta}_{-\mathbf{l}}(t)\tilde{\zeta}_{(\mathbf{l-k})}(t)\right\rangle^{a}$$
$$= (1-K_{\mathbf{k}}H_{\mathbf{k}})(1-K_{-\mathbf{l}}H_{-\mathbf{l}})(1-K_{(\mathbf{l-k})}H_{(\mathbf{l-k})})\left\langle \tilde{\zeta}_{\mathbf{k}}(t)\tilde{\zeta}_{-\mathbf{l}}(t)\tilde{\zeta}_{(\mathbf{l-k})}(t)\right\rangle^{f} \quad (121)$$
$$+ K_{\mathbf{k}}K_{-\mathbf{l}}K_{(\mathbf{l-k})}\left\langle \tilde{d}_{\mathbf{k}}(t)\tilde{d}_{-\mathbf{l}}(t)\tilde{d}_{(\mathbf{l-k})}(t)\right\rangle.$$

Here the prognostic equations for the statistics $<\zeta_{\mathbf{k}}>$, $C_{\mathbf{k}}$, $C_{\mathbf{k},-\mathbf{l}}$, and $\left\langle \tilde{\zeta}_{\mathbf{k}}\tilde{\zeta}_{-\mathbf{l}}\tilde{\zeta}_{(\mathbf{l-k})}\right\rangle$ are determined by the QDIA closure formulated in Section 7. It is worth noting that for ensemble-based methods, such as the EnKF, sampling errors degrade the background covariances. As well, if, as is typical, one assumes that observation errors are Gaussian, with $<\tilde{d}_{\mathbf{k}}>=0$, and diagonal with $D_{\mathbf{k},-\mathbf{l}}=\delta_{\mathbf{kl}}D_{\mathbf{k},-\mathbf{k}}$, sampling error will also introduce spurious observational error covariances and a non-zero $\left\langle \tilde{d}_{\mathbf{k}}(t)\tilde{d}_{-\mathbf{l}}(t)\tilde{d}_{(\mathbf{l-k})}(t)\right\rangle$ which in turn requires an expression for the three-point cumulant and so on. As we have seen in Section 8, the QDIA theory closes the moment hierarchy through prognostic integral equations for the off-diagonal two-point and three-point cumulants. The basis of the QDIA closure is the representation of the off-diagonal elements of the two- and three-point cumulant and response functions in terms of the diagonal elements.

O'Kane and Frederiksen (2008b, 2010) [69,70] extended the approach such that at each cycle non-Gaussianities and inhomogeneities in the initial conditions for the forward model may be incorporated by generalizing the cumulant update approach of Rose (1985) [74]. This is detailed in the study of O'Kane and Frederiksen (2010) [70]. In this way, the SDKF generalizes the EnKF to fully inhomogeneous and non-Gaussian flows. In contrast, as with all methods based on a cumulant discard approach, the GEKF of Miller et al. (1994) [154] cannot guarantee realizability in the presence of waves and or strong nonlinearities. The SDKF on the other hand leverages all the advantages of the QDIA in that it guarantees realizability, is applicable to situations where the observations display non-Gaussian probability distributions, and where strong inhomogeneities are present in the large-scale flow.

## 10. Subgrid Modelling for Homogeneous Turbulence

For geophysical and engineering flows it is not possible to explicitly resolve all scales of motion on a computational grid. One instead resorts to explicitly resolving only the large scales and parameterizing the influence of the small unresolved eddies. The required correction to the instantaneous time derivative of the state variables is referred to as the subgrid tendency. In its most basic form, the subgrid parameterization problem is the determination of the subgrid tendency, and its dependence on the resolved flow.

### 10.1. Data Driven Stochastic Subgrid Parameterization Method

In the data-driven stochastic modelling approaches discussed below, the functional form of the subgrid tendency is motivated by the closures described in the previous sections. One can then establish the subgrid parameterisation coefficients in this functional form, given a sufficiently large database of subgrid tendencies and associated resolved fields. The process by which one can generate such a database is as follows. For a particular initial condition of state vector $\mathbf{q}(t) \equiv \mathbf{q}^{T}(t)$, a high-resolution direct numerical simulation (DNS) is undertaken for one time step, with the associated time derivative denoted by $\mathbf{q}_{t}(t) \equiv \mathbf{q}_{t}^{T}(t)$. This same initial condition is then copied and filtered to remove the small scales. A time step is then taken from this filtered initial state, with its time derivative denoted by $\mathbf{q}_{t}^{R}(t)$. Since the small scales have been filtered out of the initial conditions this time derivative only includes the contributions of the explicitly resolved scales. The subgrid tendency, or the required correction, is then by definition given by $\mathbf{q}_t^S(t) = \mathbf{q}_t^T(t) - \mathbf{q}_t^R(t)$. This process is repeated starting from each evolved state of the DNS, until sufficient samples are generated. By persistently returning to the DNS this ensures one only calculates the subgrid corrections associated with one timestep, and hence does not allow the filtered model to drift onto its own trajectory. The interested reader is referred to Figure 3 of Kitsios et al. (2023) [155] for a schematic illustrating this approach.

In the closure equations the subgrid coefficients are functions of wavenumber, that determines the spatial scale, and are in general time varying. In the data-driven approaches, the coefficients are again scale dependent; in general, statistical steady states have been studied to generate sufficient samples for this method. The subgrid terms are parameterised through the following process:

$$\mathbf{q}_t^S(t) = \tilde{\mathbf{q}}_t^S(t) + \overline{\mathbf{q}}_t^S(t) = \tilde{\mathbf{q}}_t^S(t) + \overline{\mathbf{f}}(t), \tag{122}$$

where

$$\tilde{\mathbf{q}}_t^S(t) = -\mathbf{D}_d \tilde{\mathbf{q}}(t) + \tilde{\mathbf{f}}(t), \tag{123}$$

and

$$\overline{\mathbf{f}} = \overline{\mathbf{q}}_t^S(t) = -\overline{\mathbf{D}}\,\overline{\mathbf{q}} + \chi \mathbf{h} + \mathbf{J}. \tag{124}$$

The tilde notation denotes fluctuating components about a mean. Also $\overline{\mathbf{q}}_t^S(t) = \overline{\mathbf{f}}(t)$ is the mean of the subgrid tendency which for general inhomogeneous turbulence is expressed in terms of the mean field $\overline{\mathbf{q}}$, the topography $\mathbf{h}$, and the residual Jacobian $\mathbf{J}$. The fluctuating component represents the eddy-eddy interactions, capturing the interactions between resolved and subgrid eddies. This component is parameterised by the drain dissipation operator $\mathbf{D}_d$ and the zero-mean stochastic backscatter force $\tilde{\mathbf{f}}(t)$, which is governed by a time independent covariance matrix, as shown in Equation (A72). The fluctuating component can also be parameterised by a statistically equivalent deterministic form given by $-\mathbf{D}_{net}\tilde{\mathbf{q}}$, where the operator $\mathbf{D}_{net} = \mathbf{D}_n$ represents the net effect of the drain dissipation and stochastic forcing, as in Equation (A78). The backscatter dissipation is given by $\mathbf{D}_b = \mathbf{D}_d - \mathbf{D}_{net}$.

Importantly, matrix forms of these dissipations are used in dimensions that are not representable by wavenumbers, such as vertical levels, or across multiple state variables, such as velocity components, to explicitly represent these interactions. The dissipations can also be scaled into eddy viscosities by dividing by the discrete form of the Laplacian operator. These coefficients are calculated from the data with no arbitrary parameters, using the approaches outlined in Appendix F, following Frederiksen and Kepert (2006) [127]. In this Section we concentrate on the parameterisation of the dominant fluctuating subgrid tendency component. The parameterisation of the mean component, $\overline{\mathbf{q}}_t^S(t) = \overline{\mathbf{f}}(t)$, is the focus of Section 11 with application to inhomogeneous flows, where this term becomes relatively more important.

Once the subgrid coefficients have been established, the associated parameterisations are introduced into a low-resolution large eddy simulation (LES) to account for the influence of the missing small scales. Comparisons are then made between the DNS and LES within the large, resolved scales. This approach has been applied to simulations in a wide range of complex flow configurations. In the following subsections we consider analyses and simulations of turbulent flows in situations of increasing complexity. They include: quasi-geostrophic flows in a barotropic atmospheric model; two-level quasi-geostrophic flows in a baroclinic atmospheric model; two-level quasi-geostrophic flows in a baroclinic oceanic model; primitive equation flows in a nine-level atmospheric model; and a three-dimensional Navier-stokes turbulent channel flow in a boundary layer model with more than 300 vertical levels. In each of these examples, we demonstrate excellent agreement between the DNS and LES with the data-driven subgrid parameterizations based on the QDIA formalism.

*10.2. Atmospheric Barotropic Quasi-Geostrophic Turbulent Flows*

The stochastic subgrid modelling approach introduced above was first presented by Frederiksen and Kepert (2006) [127]. In their study a January 1979 barotropic atmospheric flow was simulated with a DNS at resolution $T = 63$. It was truncated back to an LES with resolution of $T_R = 31$, and the subgrid terms determined by the stochastic modelling approach. Figure 10b illustrates the associated drain ($v_d$), backscatter ($v_b$) and net ($v_n = v_{net}$) eddy viscosities. These eddy viscosity terms are isotropised, that is they are averaged over all zonal wavenumbers $m$ for a given total wavenumber $n$. Figure 10a illustrates the same terms, but as defined using the EDQNM closure. The stochastic modelling approach has produced coefficients that are very similar to those of the EDQNM, with differences due in part to sampling issues. The kinetic energy spectra of LES adopting these stochastic modelling coefficients are compared to the high-resolution DNS in Figure 11b. The spectra are averaged over a period of 100 days. The middle set of spectra compare the DNS to LES adopting a stochastic subgrid parameterisation. The bottom set of spectra compare the DNS to LES adopting a purely deterministic representation of the subgrid interactions. Both stochastic and deterministic LES are in good agreement with the DNS. The level agreement is also comparable with the those for the EDQNM illustrated in Figure 11a, and the DIA results of Frederiksen and Davies (1997) [126] (their Figure 11).

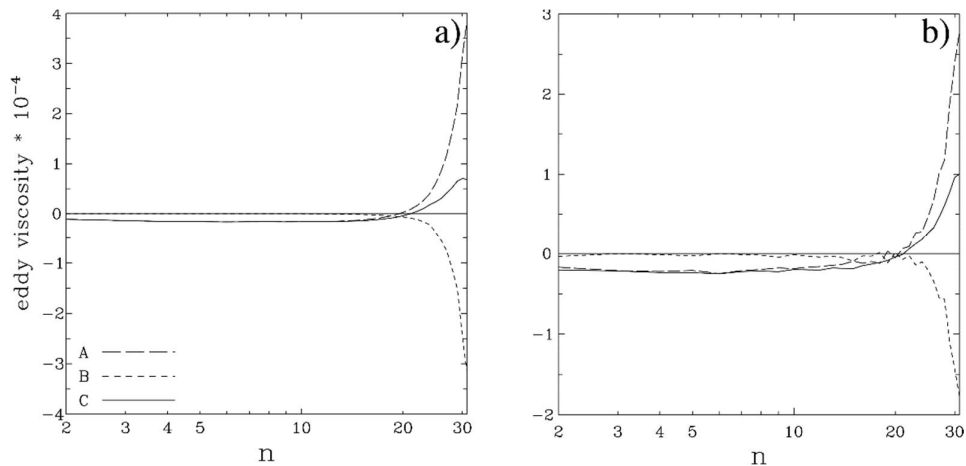

**Figure 10.** (a) EDQNM-based nondimensional subgrid scale parameterisations for T31 from Frederiksen and Davies (1997) [126], for the drain (line A), backscatter (line B) and net (line C) eddy viscosities. (b) as in (a) for DNS-based subgrid terms calculated using the stochastic modelling approach of Frederiksen and Kepert (2006) [127]. Adapted from Frederiksen and Davies (1997) [126] and Frederiksen and Kepert (2006) [127].

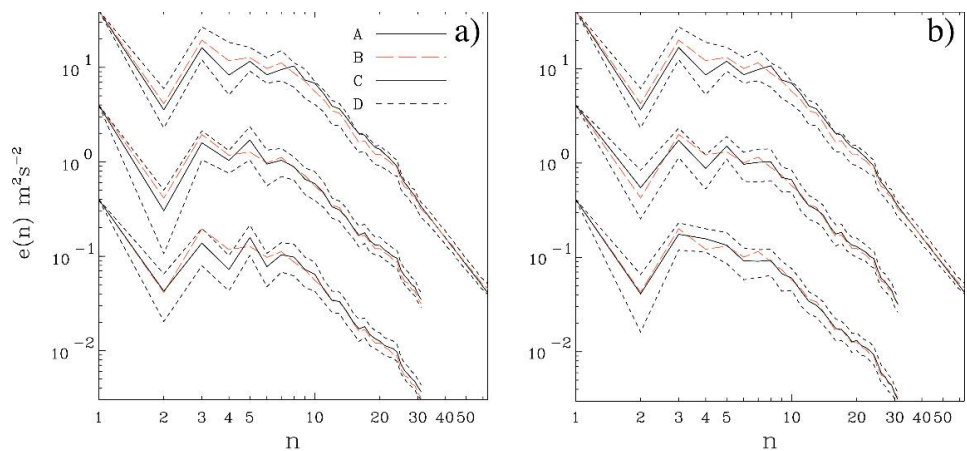

**Figure 11.** (a) Kinetic energy spectra from Frederiksen and Davies (1997) [126], for isotropised observations from January 1979 (line A), DNS at T63 (line B), LES at T31 (line C), and kinetic energy plus and minus one standard deviation (line D). LES with EDQNM-based stochastic backscatter and drain eddy viscosity is scaled by 10$^{-1}$, and LES with EDQNM-based net eddy viscosity is scaled by 10$^{-2}$. (b) as in (a) for DNS-based subgrid terms calculated using the stochastic modelling approach of Frederiksen and Kepert (2006) [127]. Adapted from Frederiksen and Davies (1997) [126] and Frederiksen and Kepert (2006) [127].

*10.3. Atmospheric Baroclinic Quasi-Geostrophic Turbulent Flows*

In the phenomenological view of two-layer QG turbulence, energy is injected near the deformation scale, with central wavenumber $k_R$. This energy is transferred from $k_R$, with a constant energy flux, to the larger scales up to the wavenumber $k_E$. Enstrophy is transferred from $k_R$, with a constant enstrophy flux to smaller scales (Salmon 1978, 1998) []. This divides the scales into three main regimes. The first is the energy containing range for $n < k_E$, comprising of flow configuration specific properties; here for flow on the sphere $n$ is the total wavenumber and $m$ the zonal wavenumber. The second is the inverse energy cascade, for $k_E < n < k_R$, with a kinetic energy spectrum proportional to $n^{-\frac{5}{3}}$, by virtue of the constant energy flux. The third is a forward enstrophy cascade for $n > k_R$, with a kinetic energy spectrum proportional to $n^{-3}$, due to the constant enstrophy flux. In the atmosphere, the values of $k_E$ and $k_R$ are very similar and are of order 10. This means the inverse energy cascade is the atmosphere is not detectable. Note, this is not the case in the ocean, as expanded on the following Section.

Zidikheri and Frederiksen (2009) [156] applied the data-driven approach of Frederiksen and Kepert (2006) [127] to two-level quasi-geostrophic baroclinic atmospheric flows on the sphere with differential rotation and zonal jets. A high-resolution DNS at $T126$ was truncated to various lower resolutions. The wavenumber associated with Rossby radius of deformation is of order $n \approx 10$. The largest wavenumber in the LES was $n = 63$, indicating that that the dominant source of energy injection, which is the baroclinic instability, was explicitly resolved, and did not need to be parameterised. This means that the subgrid parameterisations were required to only capture the physics of the forward enstrophy cascade.

Since the model had two vertical levels, for each wavenumber pair $(m,n)$ the subgrid tendency was a 2-element column vector, and the associated subgrid dissipation and the stochastic backscatter covariances were $2 \times 2$ matrices. When no subgrid parameterisation is employed the tail of the energy spectrum rises with a corresponding drop in energy at larger scales ($n \approx 10$) – see top spectra of Figure 12. This is due to the system attempting to globally conserve energy and enstrophy, referred to the "tail wagging the dog effect", as discussed in Frederiksen et al. (1996) [82]. LES adopting both deterministic and stochastic representations of the fluctuating subgrid tendency are in excellent agreement with the DNS, based on the kinetic energy spectra – see the middle and bottom sets of spectra in Figure 12. The diagonal elements of the dissipation coefficients have very similar cusp-like properties to those calculated in the barotropic study of Frederiksen and Kepert (2006) [127]. The magnitude of the off-diagonal elements is negligible in comparison, due to the insignificant subgrid interactions between the vertical levels. This is the case when baroclinic instability is completely resolved in the LES.

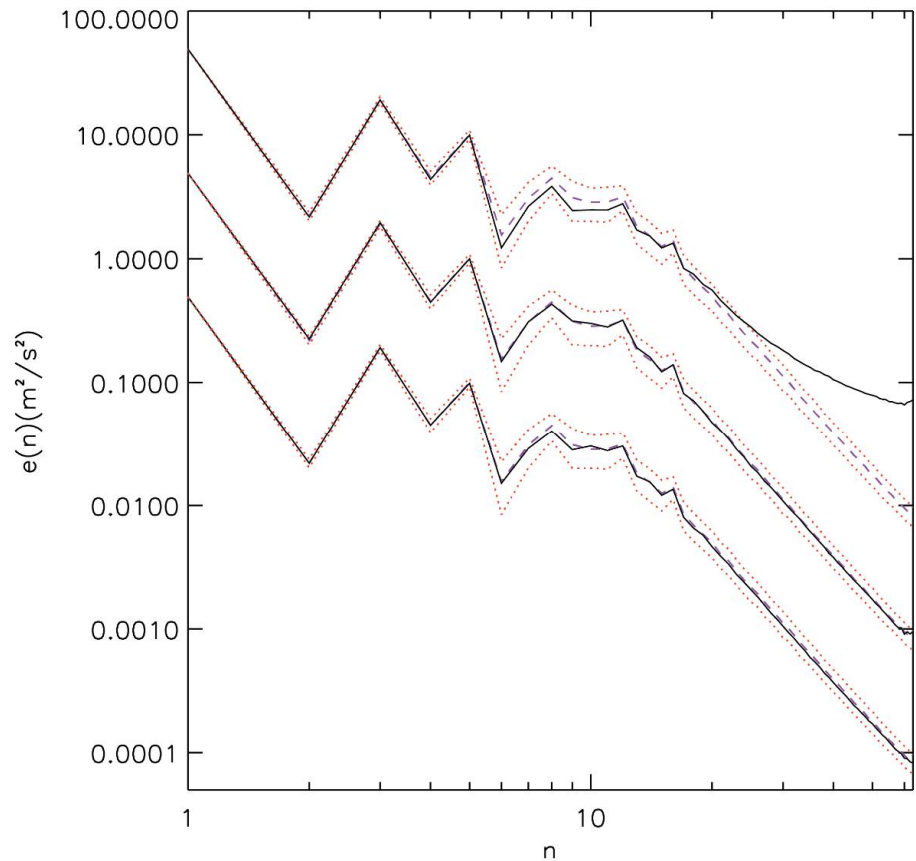

**Figure 12.** Kinetic energy spectra at 250 for LES at T63 (solid black line) for DNS at T126 truncated back to T63 (black dashed line) and DNS plus and minus one standard deviation (red dotted lines). Top set of spectra are for LES with only bare dissipation. Middle set, shifted down one decade from true value, are for LES with deterministic net dissipation. Bottom set, shifted down two decades, are for LES with drain dissipation and stochastic backscatter. Adapted from Zidikheri and Frederiksen (2009) [156].

Building upon these previous results, Kitsios et al. (2012) [157] developed a set of scaling laws for the parameterization coefficients governing the fluctuating subgrid tendency across a range of QG baroclinic atmospheric flows. Two different atmospheres were simulated with starkly different kinetic energy spectra. Many LES were undertaken with varying truncations, the subgrid coefficients calculated, and the dependence of the coefficients upon resolution determined. For truncation between $T_R = 31$ and $T_R = 252$ the eddy viscosity matrices were essentially diagonal in level space, and largely isotropic – independent of wavenumber $m$. The upper-diagonal element of the anisotropic drain and backscatter eddy viscosities are depicted in Figure 13a and 13b, respectively, at a truncation of $T_R = 63$. These coefficients were then isotropised by averaging them across $m$. Isotropised eddy viscosities at multiple resolutions are illustrated in Figure 13c. The diagonal elements of these isotropised coefficients are represented by the form

$$v_\bullet^{jj}(\mathbf{k}) = v_\bullet^{jj}(\mathbf{k}) \left( \frac{n}{T_R} \right)^{\rho_\bullet^j - 2} \tag{125}$$

for each vertical level, $j = 1, 2$. Here, $\bullet = d, b$ or $net$ denotes the drain, backscatter and net viscosities. The value of the eddy viscosities at truncation $v^{jj}(T_R)$ has been determined for each truncation. The above functional form has been fitted to the isotropised eddy viscosities, with the spectral slope $\rho_\bullet^j$ calculated. When scaled by the mean enstrophy flux of the system, $v^{jj}(T_R)$ is found to be independent of the vertical level and type of atmospheric flow studied. The value of the eddy viscosity at truncation $v^{jj}(T_R)$, is found to decrease linearly with $T_R$, as more scales were being resolved. This is illustrated for the drain in Figure 13d. The spectral slope $\rho_\bullet^j$ increases with resolution, as shown for the drain viscosity in Figure 13e. This dependence is due to the fact that the wavenumber extent that the subgrid interactions can penetrate into the resolved scales is set by the large energy containing range. Note, Figure 13c illustrates that the number of non-zero isotropised coefficients is essentially independent of the truncation level. As $T_R$ increases a larger exponent is then required to restrict the significant values of eddy viscosity to this same wavenumber extent. Deterministic and stochastic LES, with subgrid terms, defined by either the raw anisotropic coefficients, isotropised versions, or the scaling laws, all agree closely with the DNS, on the basis of kinetic energy spectra, as shown in Figure 13f.

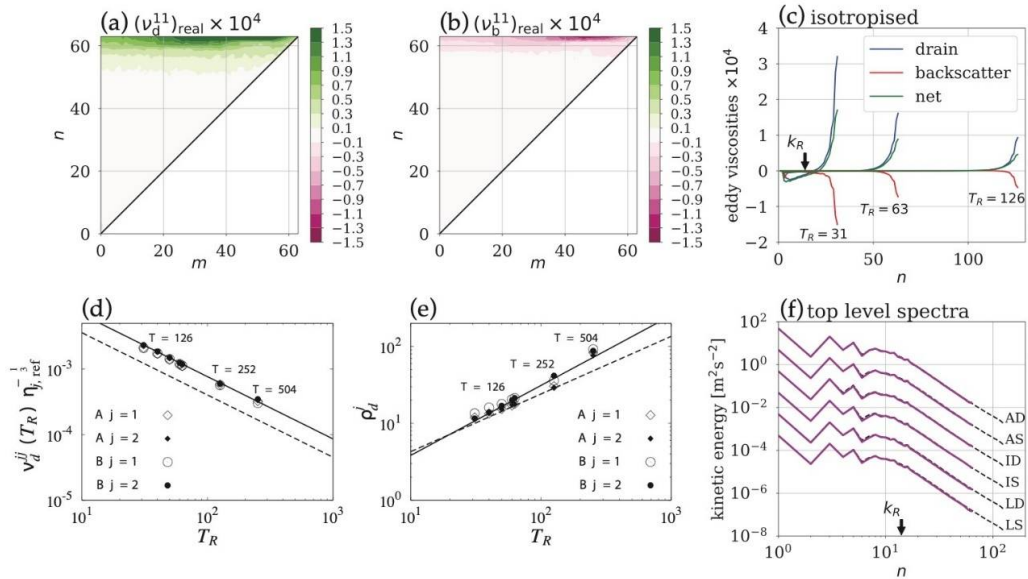

**Figure 13.** QG baroclinic atmospheric flow: (a) anisotropic drain eddy viscosity at $T_R = 63$; (b) anisotropic backscatter eddy viscosity at $T_R = 63$; (c) isotropised drain, backscatter and net eddy viscosities for multiple truncation levels; (d) value of the drain eddy viscosity at $n = T_R$ for various truncations, levels and atmospheres, with solid line the scaling law for the drain eddy viscosity, and dashed line for the net; (e) as in (d) but for the spectral slope; and (f) comparison between DNS (dashed black) and LES (solid red) using the parameterisation variants of anisotropic stochastic (AS), anisotropic stochastic (AD), isotropic stochastic (IS), isotropic stochastic (AD), scaling law stochastic (LS) and scaling law stochastic (LD). Adapted from Kitsios et al. (2012) [157].

*10.4. Oceanic Baroclinic Quasi-Geostrophic Turbulent Flows*

In the ocean, the energy containing scale $k_E$, and the energy injection scale $k_R$, are both distinct from each other. This means that all three QG regimes are present, namely the energy containing scales, the inverse energy cascade and the forward enstrophy cascade. Numerical simulations of the oceanic circulations are also more computationally challenging because the Rossby radius of deformation, of central wavenumber $k_R$, is at least an order of magnitude smaller than for the atmosphere. Thus, a higher resolution grid is required to explicitly resolve the injection of energy due to baroclinic instability. More computational resources are consequently required to generate enough samples to adequately calculate the stochastic subgrid parameterisation coefficients.

To alleviate the sampling issues, Zidikheri and Frederiksen (2010a) [158] applied the stochastic modelling approach to a simplified two-level baroclinic model of the oceanic circulation, relaxed towards a solid body rotation zonal current, and without differential rotation ($B=0$). The reference DNS had a resolution of $T=252$, corresponding to a grid size of approximately $0.5°$. They found that the kinetic energy spectrum of the DNS was reproduced, across all resolved scales, by a LES, truncated at $T_R=63$, and incorporating the calculated subgrid coefficients.

Zidikheri and Frederiksen (2010b) [159], in a second study, applied the same technique to a more realistic QG baroclinic oceanic flow, with rotation ($B=2$), and reference DNS with resolution of $T=252$. The forcings and other simulation settings were designed so that the flow broadly represented the currents in the Antarctic Circumpolar Current (ACC). The statistically steady state of the ACC has a zonal mean current centred at $60°S$. The truncation resolution for the LES was $T_R=63$, which is within the backward energy cascade regime, where baroclinic instability is not explicitly resolved. In this regime, the subgrid coefficients were strongly anisotropic and dependent upon both $m$ and $n$. The off-diagonal elements of the dissipation operators were also not negligible, indicating significant subgrid interactions between the vertical levels. For certain wavenumber pairs drain dissipations coefficients were also negative, indicating an injection of energy from the subgrid scales. This is required because baroclinic instability is not completely resolved. The LES spectra incorporating the data driven subgrid scale parametrizations were found to be in good agreement with the DNS spectra.

Kitsios et al. (2013) [160] returned to problem of LES of the ACC, but this time starting from DNS with the higher resolution of $T=504$ and significantly increased sampling. In this study, LES were performed also with large number of different resolutions that span all three regimes. For each regime, excellent agreement was obtained between the kinetic energy spectra for the DNS and the LES with calculated anisotropic matrix coefficients. Additionally, many different oceanic simulations were completed, with different ranges of the energy containing scales ($k_E$), and with baroclinic energy injection dominant at different central scales ($k_R$).

Figures 14a and 14b show the drain and backscatter eddy viscosities, for the upper diagonal element, at a LES truncation of $T_R=126$. Isotropised versions of the eddy viscosities at several resolutions are depicted in Figure 14c. As in the atmospheric study of Kitsios et al. (2012) [157], the isotropised eddy viscosities are again characterised by the power law expression in Equation (125). When converted to appropriate non-dimensional units, the magnitude and spectral slope of the eddy viscosities again follow a set of scaling laws. The scaling laws for the drain magnitude and slope are shown in Figures 14d and 14e, respectively. These scaling laws govern the eddy-eddy subgrid interactions of ocean circulations with different energy containing ranges, different scales of baroclinic instability injection, and also for different vertical levels. Scaling laws were also derived for the eddy-mean-field interactions in QG baroclinic oceanic flows in Kitsios et al. (2014) [161]. These eddy-mean-field interactions are further analysed in Section 11.

The degree of appropriate simplification of the subgrid coefficients depends on the regime in which the LES are truncated. For truncations made within the forward enstrophy containing range the off-diagonal elements are negligible, and the net and drain eddy viscosities are positive for all scales. In this range, LES, with deterministic or stochastic subgrid terms, represented by the raw anisotropic coefficients, the isotropised values or the scaling law approximation, all agree exceedingly well with DNS, in their spectral behaviour – see Figure 14f. Within the inverse energy cascade, the dissipations have non-negligible off-diagonal elements, and negative values for certain wavenumber pairs. Additional scaling laws were derived to represent the off-diagonal elements, which are detailed by Kitsios et al. (2013) [160]. Within the the forward enstrophy cascade, the LES with subgrid coefficients given by scaling laws, again give agreement with DNS, and only slightly underestimate the variability at the peak of the kinetic energy spectrum. Within the energy containing range, the subgrid coefficients are strongly anisotropic with more negative values. In this regime LES kinetic energy spectra only with the DNS results when the full anisotropic matrix eddy viscosities are used.

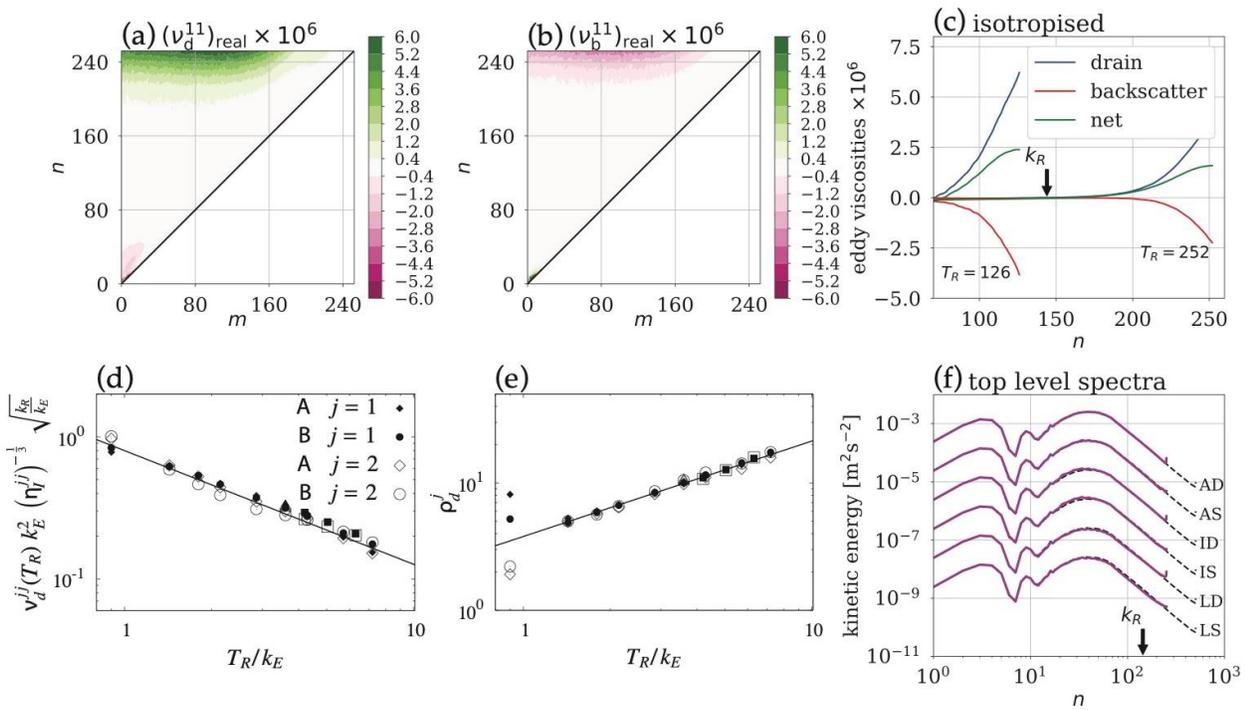

**Figure 14.** QG baroclinic oceanic flow simulations: (a) anisotropic drain eddy viscosity at $T_R = 63$; (b) anisotropic backscatter eddy viscosity at $T_R = 63$; (c) isotropised drain, backscatter and net eddy viscosities for multiple truncation levels; (d) value of the drain eddy viscosity at $n = T_R$ for various truncations, levels and oceans; (e) as in (d) but for the spectral slope; and (f) comparison between DNS (dashed black) and LES (solid red) using the parameterisation variants of anisotropic stochastic (AS), anisotropic stochastic

(AD), isotropic stochastic (IS), isotropic stochastic (AD), scaling law stochastic (LS) and scaling law stochastic (LD). Adapted from Kitsios et al. (2013) [160].

In the above studies, the stochastic subgrid modelling method [127] has produced successful LES in the QG oceanic and atmospheric simulations in all three regimes of the energy containing scales, the inverse energy cascade, and the forward enstrophy cascade. For regimes in which the kinetic energy is self-similar, the eddy viscosities were also found to be self-similar and characterised by a set of scaling laws. These scaling laws makes the parametrizations more generally applicable to LES as it removes the need to generate the coefficients from high resolution benchmark simulations, for these classes of flows. Additionally, Kitsios et al. (2016) [162] demonstrated that when non-dimensionalized appropriately, subgrid coefficients calculated from truncations made within the forward enstrophy cascade of the atmosphere and ocean are governed by the same set of scaling laws. Not only has this line of research resulted in successful LES, but it has also elucidated physical insights into the nature of QG turbulence.

*10.5. Atmospheric Primitive Equation Turbulent Flows*

The stochastic subgrid modelling approach of Frederiksen and Kepert (2006) [127] has also been applied to more complex simulations, including primitive equation atmospheric general circulation models. In the study of Frederiksen et al. (2015) [138], a nine-level primitive equation spectral model was used to determine subgrid terms for LES. The vertical structure, and wavenumber dependence, the subgrid dissipation and backscatter coefficients were calculated from the statistics of DNS at $T = 63$ DNS truncated back to $T_R = 31$. The $T = 63$ simulation consisted of a 10-year integration over the annual cycle and used relaxation physics to ensure that the base flow had typical atmospheric zonal jets, kinetic energy spectra and temperature structure.

The associated average isotropised drain, backscatter and net dissipations for the vorticity field as a function of total wavenumber ($n$) and vertical level are shown in Figures 15a, 15b and 15c respectively. The drain and backscatter dissipations are relatively level independent for a given $n$, whereas the net dissipation has a somewhat stronger dependence on the vertical coordinate. All the dissipations exhibit a cusp like structure approaching their maximum values at the truncation scale. The scale and level dependent net dissipations for divergence and temperature are depicted in Figures 15d and 15e. The temperature coefficients have similar structure to those for the vorticity. The divergence dissipation, however, is largest in the lower troposphere and tightly contained at the shortest scales. The scale dependent net dissipation of the surface pressure is shown in Figure 15f. It has negative values at low wavenumbers and rises to a positive cusp at high wavenumbers. This profile is like that obtained in the barotropic model study of Frederiksen and Davies (1997) [126] and shown in Figure 10a. This same qualitative structure is also found for all truncations made within the forward enstrophy cascade of the baroclinic atmospheric and oceanic studies discussed in the previous section.

Frederiksen et al. (2003) [81] implemented a barotropic, level independent, net eddy viscosity for each level of each field of a related global atmospheric climate model, based on the subgrid model study of Frederiksen and Davies (1997) [126]. They found that this resulted in significant improvements in the circulation and the transient kinetic energy including spectra. In hindsight, perhaps this is not surprising given the essentially equivalent barotropic vertical of the coefficients illustrated in Figure 14.

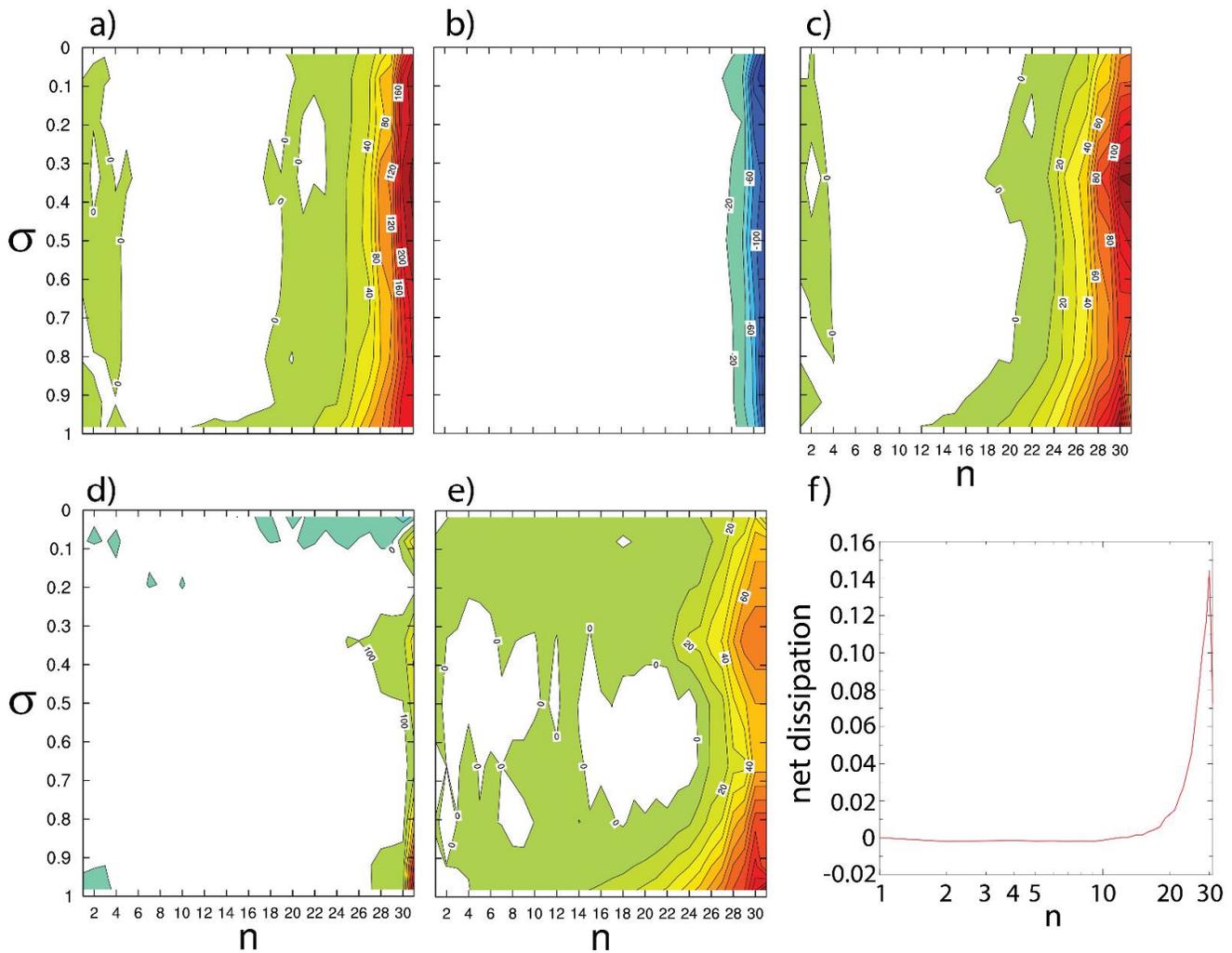

**Figure 15.** Primitive equation model based subgrid-scale parameterizations as a function of wavenumber $n$ and vertical $\sigma-$ level: (a) drain dissipation for vorticity; (b) backscatter dissipation for vorticity; (c) net dissipation for vorticity; (d) net dissipation for divergence; (e) net dissipation for temperature; and (f) net dissipation for surface pressure. Adapted from Frederiksen et al. (2015) [138].

*10.6. Three-dimensional Navier-Stokes Turbulent Flows in a Boundary Layer Channel*

In all the above studies the stochastic modelling approach [127] has been applied to atmospheric and oceanic turbulent flows, described by barotropic 2D, or baroclinic quasigeostrophic equations, or atmospheric primitive equations. The method has been shown to successfully produce low resolution LES. However, the approach has also been successfully applied in fully three-dimensional turbulent channel flow.

Kitsios et al. (2017) [163] considered the three-dimensional incompressible isothermal Navier-Stokes equations in a channel described by the three-component velocity vector field. The streamwise ($x$) and spanwise ($z$) boundaries were periodic, and the wall normal ($y$) boundaries had a no-slip condition. The prognostic variables were discretised using a Fourier discretisation in $x$ and $z$ of wavenumbers $k_x$ and $k_z$. In the wall normal direction, a collocated Chebyshev discretisation in $y$ was adopted, with polynomial index $k_y$. The DNS were performed in spectral space with a horizontal rectangular wavenumber and vertical Chebyshev polynomial set. The maximum DNS truncation wavenumber $T$ applies to both $k_x$ and $k_z$. The maximum truncation Chebyshev polynomial index is $N$. Simulations were undertaken at three friction-velocity-based Reynolds numbers defined by $\mathrm{Re}_\tau \equiv u_\tau L_h/\nu$ and taking values of 180, 550 and 950. Here, $\nu$ is the kinematic viscosity, $L_h$ is the channel half-height, and the friction velocity $u_\tau = \tau_{wall}/\rho$ with $\tau_{wall}$ the magnitude of the average wall shear stress. The domain sizes in order of increasing Reynolds number have $(L_x, L_y, L_z)/L_h$ given by $(2\pi, 2, \pi)$, $(2\pi, 2, \pi)$ and $(\pi, 2, \pi/2)$ with respective resolutions of $(T, N)$ specified by $(64, 97)$, $(128, 257)$ and $(128, 385)$. In all cases the turbulence is explicitly resolved down to the Kolmogorov scale.

LES truncations have been performed by reducing the resolution in the horizontal direction (i.e. specifying $T_R < T$) and maintaining all of the scales in the vertical direction. A pseudo-wavenumber vector was defined to be $\mathbf{k} = (k_x, k_y, k_z)$, and the state vector $\mathbf{q} = (u, v, w)$ containing the three velocity components. As such, the subgrid dissipation matrices represent the interactions between the velocity components for a given scale triplet. However, the dissipations are diagonally dominant for most $\mathbf{k}$. The diagonal elements are also symmetric about $k_z = 0$.

For the $\mathrm{Re}_\tau = 950$ case truncated back to $T_R = 63$, Figure 16 illustrates the upper diagonal element of the net dissipation for decreasing vertical scales from Figure 16a to Figure 16d. As observed in the previous applications, the coefficients are cusp-like, with the dissipation magnitude increasing as they approach the truncation boundaries. In Figures 16a to 16c the coefficients are all positive, indicating that energy is transferred from the resolved to unresolved scales. For smallest illustrated vertical in Figure 16d, the smallest resolved horizontal scales have negative dissipations, indicating energy is being transferred from the unresolved scales to these wavenumber pairs. This is consistent with the phenomenological view of a transfer of energy from small to large scales in the attached eddy hypothesis of wall bounded turbulence (Perry and Chong, 1982 [164]). The lower Reynolds number channel flows are also found to have similar properties, when non-dimenionalized in viscous units.

The DNS is compared to the LES, on the basis of the time averaged kinetic energy spectra, for the $\mathrm{Re}_\tau = 950$ case at a specific wall height. As a baseline, Figure 17a illustrates a simulation truncated at $T_R = 63$, with no subgrid model prescribed. There is an evident accumulation of energy at the smallest resolved scales. The unresolved interactions would ordinarily have transferred this energy to the smallest scales in the DNS, which killed off these eddies via viscous dissipation. Both the deterministic and stochastic subgrid models appropriately parameterise this phenomenon, as illustrated by the agreement between the LES and DNS in Figure 17b and Figure 17c respectively.

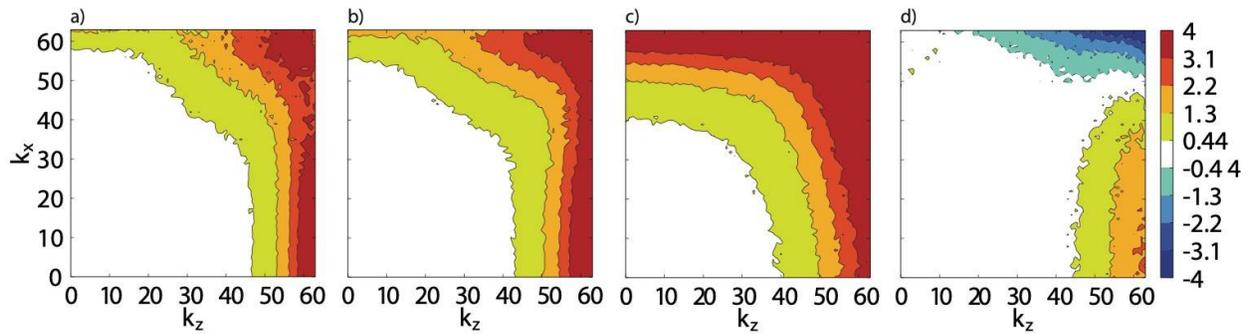

**Figure 16.** Upper diagonal component of the net subgrid dissipation matrices from Kitsios et al. (2017) [163] illustrated in the horizontal wavenumber plane ($k_x \geq 0, k_z \geq 0$) using rectangular truncation $T_R = 63$ for wall normal scales: (a) $\lambda_y/L_h = 0.1$; (b) $\lambda_y u_\tau/\nu = 10$; (c) $\lambda_y u_\tau/\nu = 2$; and (d) $\lambda_y u_\tau/\nu = 4$, where $\lambda_y = 1 - \cos(2\pi/k_y)$ with $k_y$ the Chebyshev polynomial index. All coefficients are symmetric about the line $k_x = 0$. Adapted from Kitsios et al. (2017) [163].

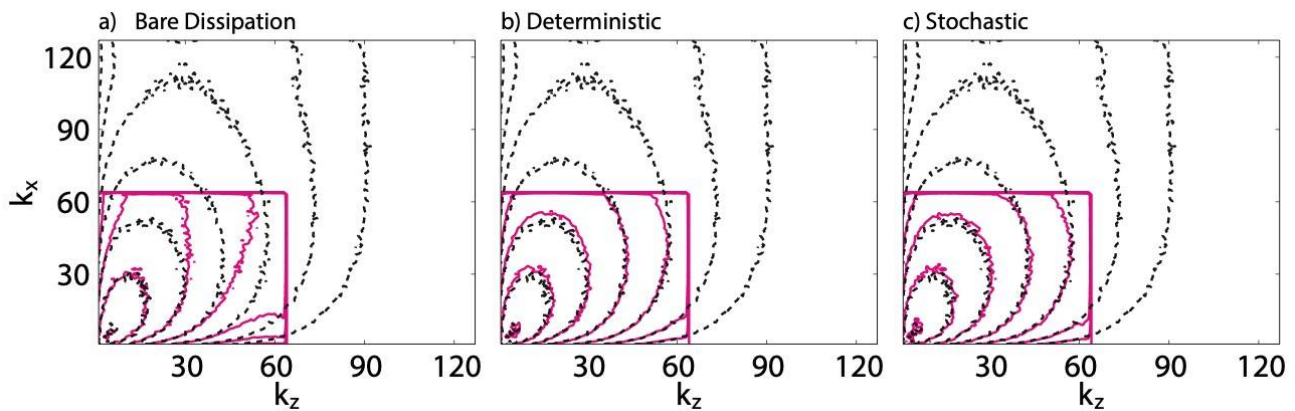

**Figure 17.** Comparison of the DNS (black dotted line) and LES (magenta solid line) on the basis of time-averaged kinetic energy spectra adapted from Kitsios et al. (2017) [163]. The two-dimensional spectrum is compared at $y^+ = 133$ for an LES with: (a) no subgrid parameterization; (b) net dissipation; and (c) drain dissipation with stochastic backscatter. Contour levels radiating out from the origin are $10^{-2.5}$, $10^{-3}$, $10^{-3.5}$, $10^{-4}$, $10^{-4.5}$, $10^{-5}$, $10^{-5.5}$. The vertical axis in (a) is applicable to (b) and (c). Adapted from Kitsios et al. (2017) [163].

## 11. Subgrid Modelling for Inhomogeneous Turbulence

*11.1 QDIA Based Subgrid Modelling*

A key advantage of the QDIA is that it no longer requires sampling to resolve the flow statistics and higher order terms rather relies on the accuracy of the prognostic equations meaning that assumptions of statistical stationarity are no longer required making time dependent subgrid scale parameterizations a possibility at least within the class of problems tractable for the QDIA formalism. Therefore, an important question for more general inhomogeneous flows is how to systematically parameterize each of the respective subgrid scale drain and injection terms of the QDIA.

In formulating subgrid scale terms for the QDIA Frederiksen (1999) [66] recognised that for a system comprising of multi-scale eddies, mean-fields and topography, the subgrid tendency comprises contributions from five possible scale interactions corresponding to specific contributions from each term in the QDIA. Specifically:

1. Eddy-eddy interactions are those between resolved and subgrid eddies and were the primary focus of Section 10.
2. Eddy-mean-field interactions are those between the resolved eddies and subgrid mean-field, and also those between the subgrid eddies and resolved mean-field.
3. Eddy-topographic interactions are those between the resolved eddies and subgrid topography, and as well those between the subgrid eddies are resolved topography.
4. The mean-field-mean-field interactions are those between the resolved and subgrid components of the mean-field.
5. Mean-field-topographic interactions are those between the resolved mean-field and subgrid topography, and as well those between the subgrid mean-field are resolved topography.

As in Appendix F the subgrid scale terms for the barotropic QDIA closure can be derived when the resolution is reduced from $T$ to $T_R < T$ where $T_R$ is the resolution of the resolved scales. We suppose in Equations (72), (73), (80) and (86) to (90) that the resolution is reduced from $\mathcal{T} = \{\mathbf{p},\mathbf{q} | p \leq T, q \leq T\}$ to $\mathcal{R} = \{\mathbf{p},\mathbf{q} | p \leq T_R, q \leq T_R\}$ with $\mathcal{S} = \mathcal{T} - \mathcal{R}$ the set of subgrid scale terms. Thus, in Equation (72) we need to make the following renormalizations:

$$\overline{f}_0(\mathbf{k},t) \to \overline{f}_r(\mathbf{k},t) = \overline{f}_0(\mathbf{k},t) + f_h(\mathbf{k},t) + j(\mathbf{k},t) \tag{126}$$

where the residual Jacobian of the subgrid scales is

$$j(\mathbf{k},t) = \sum_{(\mathbf{p},\mathbf{q}) \in \mathcal{S}} \delta(\mathbf{k},\mathbf{p},\mathbf{q})[K(\mathbf{k},\mathbf{p},\mathbf{q}) <\zeta_{-\mathbf{p}}(t)><\zeta_{-\mathbf{q}}(t)> \\ + A(\mathbf{k},\mathbf{p},\mathbf{q}) <\zeta_{-\mathbf{p}}(t)> h_{-\mathbf{q}}] \tag{127}$$

and the subgrid eddy-topographic force is

$$f_h(\mathbf{k},t) := f_\chi^{\mathcal{S}}(\mathbf{k},t) = h_\mathbf{k} \int_{t_o}^{t} ds\, \chi_\mathbf{k}^{\mathcal{S}}(t,s). \tag{128}$$

Here, the set and index $\mathcal{S}$ means that only the subgrid scales are contributing. As well we need the generalized drain matrix acting on the mean field:

$$\overline{D}_r(\mathbf{k},t) = D_0(\mathbf{k}) + \overline{D}_d(\mathbf{k},t), \tag{129}$$

with

$$\overline{N}_d(\mathbf{k},t) = \overline{D}_d(\mathbf{k},t) <\zeta_\mathbf{k}(t)> = \int_{t_o}^{t} ds\, \eta_\mathbf{k}^{\mathcal{S}}(t,s) <\zeta_\mathbf{k}(s)>. \tag{130}$$

As well, we need

$$F_r(\mathbf{k},t) = F_0(\mathbf{k},t) + F_b(\mathbf{k},t) \tag{131}$$

where

$$F_b(\mathbf{k},t) = F_S(\mathbf{k},t) + F_P(\mathbf{k},t). \tag{132}$$

In these expressions

$$F_S(\mathbf{k},t) = <\tilde{f}_S^S(\mathbf{k},t)\tilde{\zeta}(-\mathbf{k},t)> = \int_{t_o}^{t} ds <\tilde{f}_S^S(\mathbf{k},t)\tilde{f}_S^S(-\mathbf{k},s)> R_{-\mathbf{k}}(t,s)$$

$$= \int_{t_o}^{t} ds\, S_{\mathbf{k}}^S(t,s) R_{-\mathbf{k}}(t,s), \qquad (133)$$

and

$$F_P(\mathbf{k},t) = <\tilde{f}_P^S(\mathbf{k},t)\tilde{\zeta}(-\mathbf{k},t)> = \int_{t_o}^{t} ds <\tilde{f}_P^S(\mathbf{k},t)\tilde{f}_P^S(-\mathbf{k},s)> R_{-\mathbf{k}}(t,s)$$

$$= \int_{t_o}^{t} ds\, P_{\mathbf{k}}^S(t,s) R_{-\mathbf{k}}(t,s). \qquad (134)$$

We note that $\tilde{f}_S^S(\mathbf{k},t)$ and $\tilde{f}_P^S(\mathbf{k},t)$ are given in Equations (A46) and (A47) with $q \to \zeta$ and the sum $S$ over the subgrid scales instead of $T$.

We also need to consider the renormalized generalized drain matrix acting on the fluctuations given by:

$$D_r(\mathbf{k},t) = D_0(\mathbf{k}) + D_d(\mathbf{k},t) \qquad (135)$$

with

$$D_d(\mathbf{k},t) = D_\eta(\mathbf{k},t) + D_\pi(\mathbf{k},t). \qquad (136)$$

Here,

$$N_\eta(\mathbf{k},t) = D_\eta(\mathbf{k},t) C_{\mathbf{k}}(t,t) = \int_{t_o}^{t} ds\, \eta_{\mathbf{k}}^S(t,s) C_{-\mathbf{k}}(t,s) \qquad (137)$$

and

$$N_\pi(\mathbf{k},t) = D_\pi(\mathbf{k},t) C_{\mathbf{k}}(t,t) = \int_{t_o}^{t} ds\, \pi_{\mathbf{k}}^S(t,s) C_{-\mathbf{k}}(t,s). \qquad (138)$$

Now, when the resolution in Equation for the fluctuations is reduced from $T$ to $T_R < T$ so that $T$ is replaced by $R$ in the sums, we need to make the following renormalizations. The bare random forcing is replaced by,

$$\tilde{f}_0(\mathbf{k},t) \to \tilde{f}_r(\mathbf{k},t) = \tilde{f}_0(\mathbf{k},t) + f_S^S(\mathbf{k},t) + f_P^S(\mathbf{k},t) \qquad (139)$$

and the bare dissipation is replaced by

$$D_0(\mathbf{k}) \to D_0(\mathbf{k}) + \int_{t_o}^{t} ds\{\eta_{\mathbf{k}}^S(t,s) + \pi_{\mathbf{k}}^S(t,s)\}\tilde{\zeta}_{\mathbf{k}}(s). \qquad (140)$$

In this way all terms of the QDIA are consistently renormalised on reducing resolution and it is not simply the closure equations that are based on renormalization.

From the work of O'Kane and Frederiksen (2008c) [71], we show the general properties of the QDIA subgrid scale closure integral terms in $(\mathbf{k}_x, \mathbf{k}_y)$ space in Figure 18. Note that where waves are present the integral terms are complex and are in no sense balanced. Here, superscript $S$ denotes truncation from $C_{48} = T_{48}$ to $C_{24} = T_{24}$. All terms exhibit high-order function forms with the largest values near the truncation scale. The real parts of the subgrid nonlinear noise, damping, and their inhomogeneous analogues, namely the subgrid eddy-mean field, and eddy topographic noise and dissipation are relatively homogeneous across scales. In contrast, the subgrid mean-field tendencies and residual Jacobian terms are highly anisotropic.

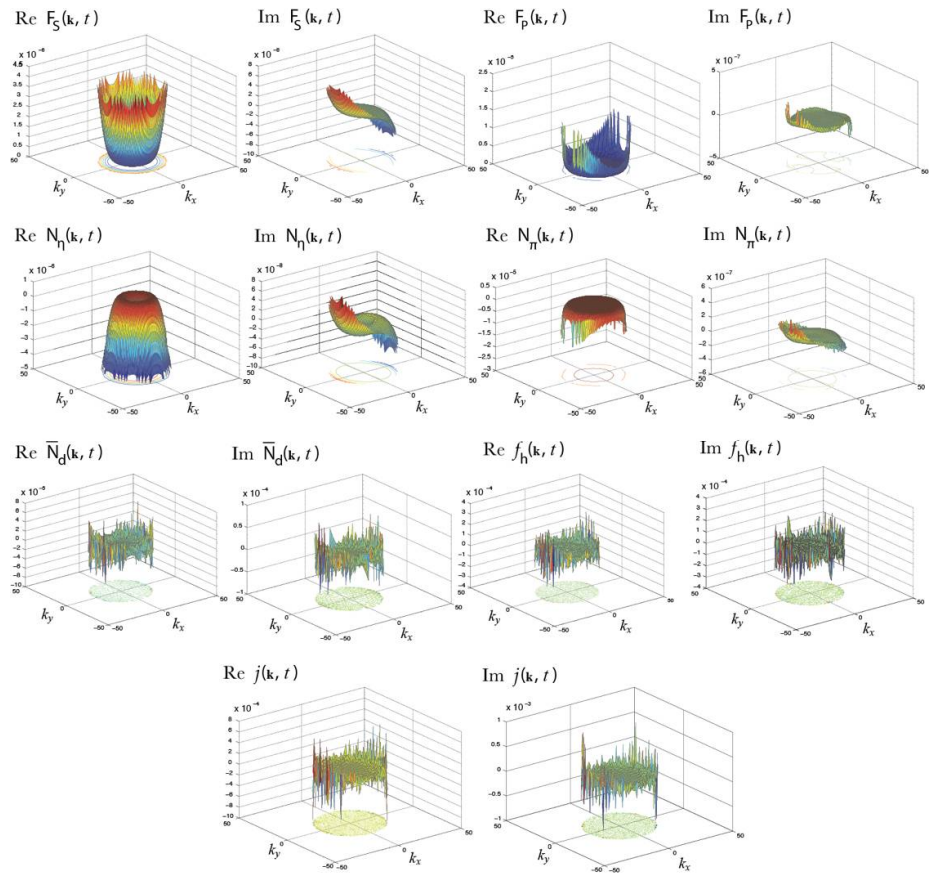

**Figure 18.** Subgrid-scale QDIA closure integral terms plotted in $(\mathbf{k}_x, \mathbf{k}_y)$-space for the general case. Adapted from O'Kane and Frederiksen (2008c) [71].

In Figure 19 we show the quadratic mean-field and mean-field-topographic terms of the residual Jacobian, converted to streamfunction, with contours of the mean zonal wind and mean streamfunction, respectively, overlaid in physical space. The spatial patterns correspond to the inverse Fourier transform of the residual Jacobian $(\mathbf{k}_x, \mathbf{k}_y)$ spectra shown in Figure 18. Several important points to note are:

1. For far from equilibrium flows as shown here, the contribution of the residual Jacobian is comparable to any of the other subgrid terms, whereas this term will vanish at canonical equilibrium.

2. The quadratic subgrid mean field term is representative of a high pass filter of the large-scale Rossby waves with largest amplitudes occurring where large gradients in the zonal winds are manifest. That is, where the midlatitude jets occur, here shown for the boreal winter.

3. The residual Jacobian mean-field topographic interactions have the largest amplitudes proximal to significant topographic features, namely, the Himalayan and Rocky Mountains and the Andes. Here we see a significant correlation between these local in space wavetrains and the mean streamfunction.

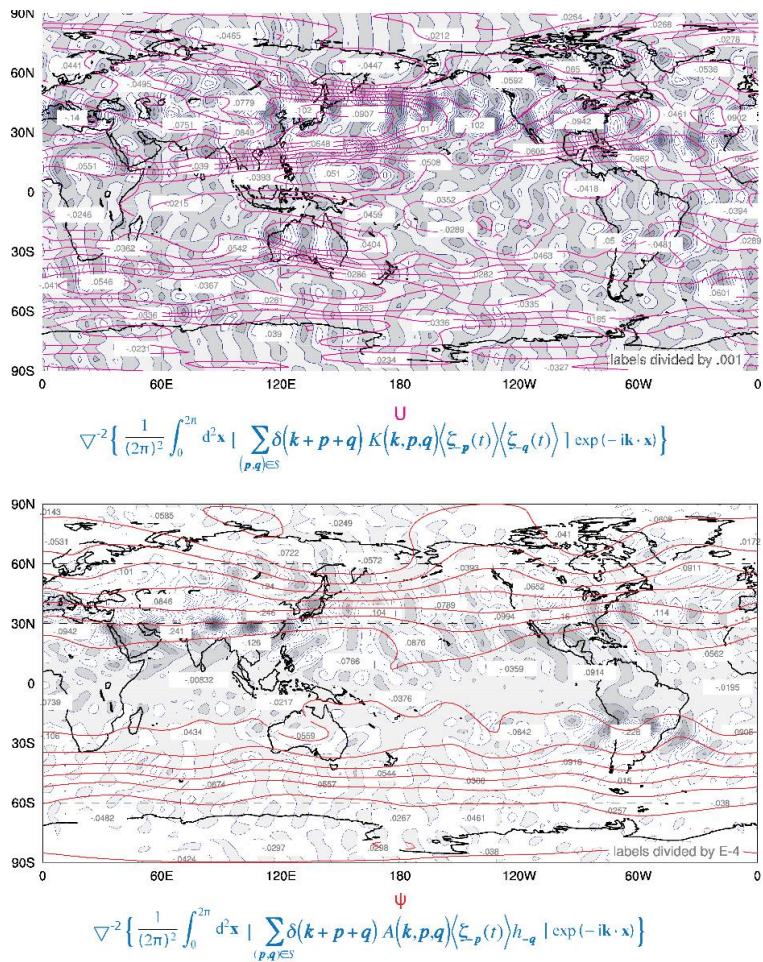

**Figure 19.** Subgrid-scale residual Jacobian terms $\nabla^{-2}J_S(\psi,\zeta)$ and $\nabla^{-2}J_S(\psi,h)$, in terms of spectral components, with the zonal wind $U$ and streamfunction $\psi$ overlaid. Adapted from O'Kane and Frederiksen (2008c) [71].

O'Kane and Frederiksen (2008c) [71] demonstrated that, for canonical equilibrium, the eddy–topographic force is proportional to the mean vorticity. It was found to be a high pass filtered form of the topography determined by the scale of the eddy-damping. The choice to relax the mean flow toward a local canonical equilibrium state where the vorticity is anti-correlated with the large structures of the topography is consistent with the minimum enstrophy hypothesis of Bretherton and Haidvogel (1976) [100] and the maximum entropy solutions of Carnevale and Frederiksen (1987) [102]. In this case however, the scale at which the eddy topographic forcing structures occurs is determined by the truncation scale and not by the dissipation operator. Energy injection in the form of stochastic backscatter is required to avoid severe underestimation of small-scale kinetic energy due to the excessive drain of energy from the retained to the subgrid scales. Specifically, at canonical equilibrium, energy injection from the subgrid to the retained scales must balance the eddy drain viscosity $v_d(\mathbf{k}) = D_d(\mathbf{k})k^{-2}$ defined in Appendix F. Of course, subgrid scale parameterizations (SSPs) based on equilibrium or statistically steady states will fail where far from equilibrium flows are manifest and where the dynamics are rapidly evolving. In this regard, the QDIA subgrid scale parameterizations allow, for the first time, scale dependent and time evolving parameterizations to be applied for general inhomogeneous flows.

*11.2 Quasi-Geostrophic Baroclinic Turbulence in the Atmospheric Simulatons with Topography*

The data-driven representation of the subgrid tendency encompassing all of the fundamental classes of subgrid interactions is given by

$$\mathbf{q}_t^S(t) = \tilde{\mathbf{q}}_t^S(t) + \overline{\mathbf{q}}_t^S(t) = \tilde{\mathbf{q}}_t^S(t) + \overline{\mathbf{f}}(t) \tag{141}$$

where $\tilde{\mathbf{q}}_t^S(t)$ is the fluctuating component defined in Equation (123) and the formulation of its parameterization is given in Section 10. The mean subgrid tendency $\overline{\mathbf{q}}_t^S(t) = \overline{\mathbf{f}}$ is represented by the form in Equation (124); namely $\overline{\mathbf{q}}_t^S(t) = \overline{\mathbf{f}} = -\overline{\mathbf{D}}\overline{\mathbf{q}} + \chi\mathbf{h} + \mathbf{J}$. The eddy-mean-field interactions are parameterised by $-\overline{\mathbf{D}}\overline{\mathbf{q}}$, where $\overline{\mathbf{D}}$ is the mean-field dissipation operator. The eddy-topographic interactions are parameterised by $\chi\mathbf{h}$, where $\mathbf{h}$ is the topography, and $\chi$ is the eddy-topographic operator. The mean-field Jacobian $\mathbf{J}$, represents the sum of the mean-field-mean-field and mean-field-topographic interactions. Again, the discrete indices, representing levels or separate fields, in Equation (5) for the field variables, result in matrix forms for the operators, but they are separable for each wavevectors. All the coefficients are directly calculated from the simulation data with no arbitrary parameters. The means by which one calculates the mean subgrid tendency is outlined in Appendix H following the methodology of Kitsios and Frederiksen (2019) [137].

In the study of Kitsios and Frederiksen (2019) [137], the above approach was used to analyse global baroclinic atmospheric flow simulations, over topography, characteristic of two climate states, one typical of January, and another typical of July. The time averaged zonal velocity at the top vertical level of the model and the realistic topography used in the simulations are respectively illustrated in Figures 20a and 20b. The high-resolution reference DNS were run with a resolution of $T = 63$. Successful LES were produced at $T_R = 31$ and $T_R = 15$, respectively using only 11.9% and 1.3% of the DNS computational effort. Both LES were able to closely reproduce the primary DNS statistics at the resolved scales. The the kinetic energy spectra, zonal jet structure, and nonzonal streamfunction were captured to high accuracy. Anisotropic matrix representations for all the subgrid coefficients were used in the LES. For the $T_R = 31$ LES the highest wavenumber is within the forward enstrophy cascade with the baroclinic energy injections explicitly resolved. However, for the lower resolution $T_R = 15$ LES, baroclinic instability was not completely resolved, but nevertheless, the parameterizations adequately captured the required injection of energy to give high accuracy results.

The components of the mean subgrid tendency, at the bottom vertical level, are illustrated in Figure 20c to Figure 20e, for the more extreme truncation of $T_R = 15$. Whilst the subgrid coefficients are calculated in spectral space, we have also transformed them into physical space as meridional accelerations, to provide additional intuition and insight. Figure 20c contains bottom level mean-field Jacobian, which shows most activity in regions with the steep topographic changes near Antarctica. The bottom level eddy–mean-field term is illustrated in Figure 20d, which shows a wavetrain of large-scale structures across the Northern Hemisphere. In spectral space, the isotropised eddy-mean-field dissipations calculated from the January and July states have very similar structures, and effectively collapse to the same form at all scales, when nondimensionalised by the enstrophy flux. The eddy–topographic term is illustrated Figure 20e, and again has wavetrain structures exhibiting the direct influence of the topography. We note the coherent structures that are clustered around the Earth's large topographic features, including the Himalayas, the Rocky Mountains and the Andes. These wavetrains are even more evident for the results with the $T_R = 31$ LES in Figure 20f of Kitsios and Frederiksen (2019) [137]. The properties of the eddy-eddy dissipations are consistent with those presented in Figure 13 Kitsios et. al. (2012) [157]. Their simulations had no topographic inhomogeneities, but the zonally averaged component of the climate state was similar.

As shown in Figure 20f Kitsios and Frederiksen (2019) [137] undertook several LES with each of the subgrid interactions progressively included in the parameterization. We note that the eddy-eddy interactions indeed represent the dominant contribution to the subgrid tendency. However, best results are achieved when all components contributing to the mean subgrid tendency are also included.

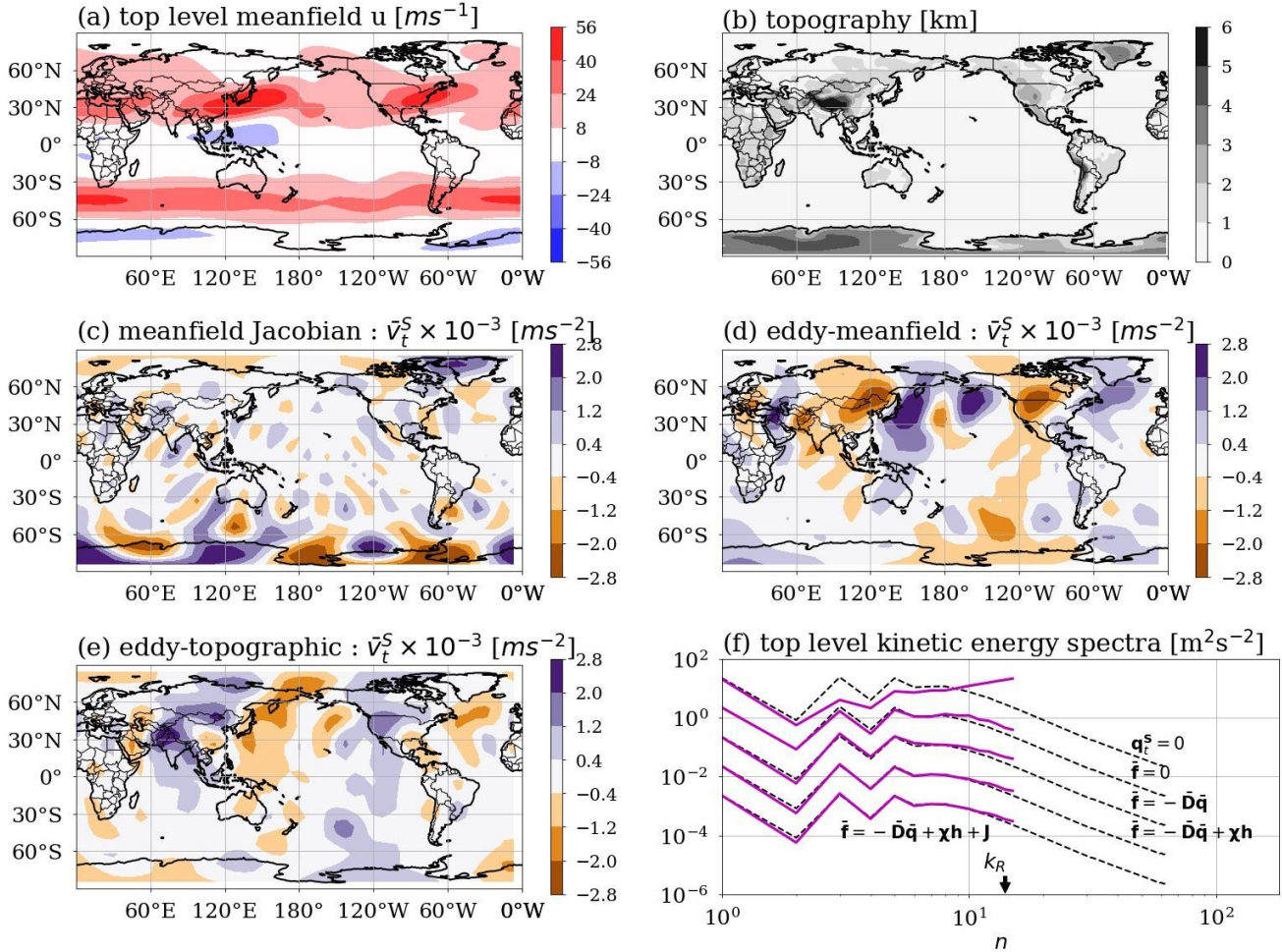

**Figure 20.** Comparison of the DNS of an atmospheric flow of $T = 63$ resolution to an LES of $T_R = 15$. Base flow properties: (a) top level mean zonal velocity; and (b) topography. At the bottom level the meridional acceleration of the: (c) mean-field Jacobian; (d) eddy-mean-field; and (e) eddy-topographic terms. (f) Top level kinetic energy spectra of DNS (dashed black) and stochastic LES (magenta). Each set of spectra has LES with the labelled terms included in the mean subgrid tendency term. Adapted from Kitsios and Frederiksen (2019) [137].

*11.3 Quasi-Geostrophic Baroclinic Turbulence in Oceanic Simulations with Bathymetry*

Kitsios et al. (2023) [155] calculated scale dependent parameterizations representing each of the fundamental classes of subgrid interaction in an oceanic flow for the first time.

It was, specifically, a baroclinic quasi-geostrophic simulation of an idealized Antarctic Circumpolar Current. The flow configuration had representative mean currents and ocean floor bathymetry. The mean zonal velocity at the top level is illustrated in Figure 21a, and the magnitude of the topographic gradient shown in Figure 21b. The eddy-eddy coefficients were calculated using the data-driven stochastic modelling approach of Frederiksen and Kepert (2006) [127], with the parameterizations governing the remaining interactions calculated using the regression approach of Kitsios and Frederiksen (2019) [137]. The reference DNS was run at a resolution of $T = 504$. In these simulations the Rossby wavenumber of deformation was $k_R = 284$, which is representative of that observed in the Southern Ocean.

Two LES were undertaken. One in which baroclinic instability was completely resolved with $T_R = 378$, and another in which it was not with $T_R = 252$. Stochastic and deterministic parameterisations were assessed for the eddy-eddy interactions, and deterministic versions for the remaining interaction classes. The eddy-eddy coefficients had similar spectral properties to those calculated in Kitsios et al. (2013) [160] for a similar (zonal) flow, but without ocean floor bathymetry. In physical space, the mean-field Jacobian, eddy-mean-field and eddy-topographic meridional accelerations at the bottom level are respectively illustrated in Figures 21c, 21d and 21e. They are all the strongest where the zonal velocities are large. The amplitudes of these terms are also enhanced over regions of strong topographic gradients in the oceans south of South America and surrounding New Zealand.

The kinetic energy spectra and mean-fields of all LES using these calculated subgrid coefficients were in good agreement with of those for the reference DNS – see Figure 21f. The eddy-eddy interactions were again found to be dominant at these truncation levels, with all interaction classes required for best results. For the $T_R = 378$ LES, where baroclinic instability was completely resolved the stochastic representations of the eddy-eddy interactions out-performed the deterministic variants for all resolved scales. For the lower resolution $T_R = 252$ LES, for which baroclinic instability was not completely resolved, the stochastic variants were in better agreement with the DNS than the deterministic cases at the large scales. However, the stochastic eddy-eddy parameterisations introduced some distortions in the kinetic energy spectra at the smallest resolved scales.

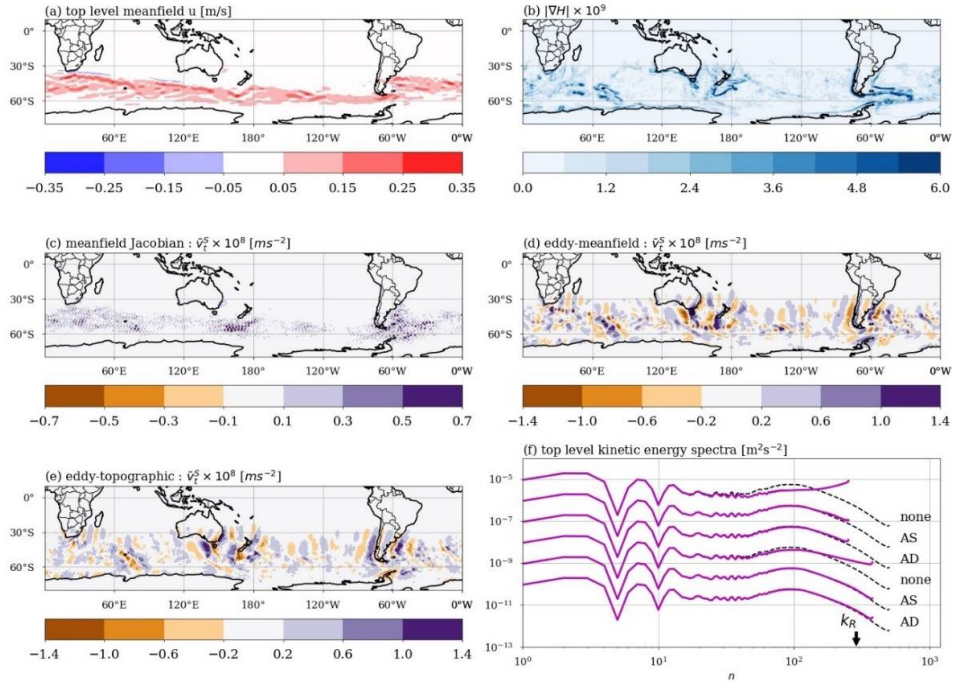

**Figure 21.** Comparison of the DNS of $T = 504$ resolution and LES of an oceanic flow. Base flow properties: (a) top level mean zonal velocity; and (b) absolute value of the topographic gradient. For a $T_R = 252$ truncation, the bottom level meridional acceleration of the: (c) mean-field Jacobian; (d) eddy-mean-field; and (e) eddy-topographic terms. (f) Top level kinetic energy spectra of DNS (dashed black) and stochastic LES (magenta). The top three spectra are for LES with $T_R = 252$ and bottom three with $T_R = 378$. Spectra are labelled with LES eddy-eddy parameterizations of no subgrid model (none), anisotropic stochastic (AS), or anisotropic deterministic (AD). The mean subgrid tendency component contains all of the terms. Adapted from Kitsios et al. (2023) [155].

## 12. Discussion: Perspectives on Closures from Renormalized Perturbation theory

The issues faced in developing closures for classical statistical field theory including turbulent systems have close parallels to those in formulating and solving the statistical dynamical equations of quantum field theory. Here, we make some comparisons between classical and quantum systems to put into perspective the achievements in turbulence closure theory.

*12.1 Quantum Electro-Dynamics: Triumph of Renormalized Perturbation Theory*

Perhaps one of the greates triumps of theoretical physics was the amazingly successful application of renormalized perturbation theory to quantum electrodynamics (QED). This was achieved using the functional equations of Schwinger (1948) [7], Dyson (1949) [8] and Tomonaga (1946) [165] and the diagrammatic methods of Feynman (1949) [21].

In QED the strength of the interaction is measured by the fine structure constant $\alpha_{QED} \approx 1/137$. This is quite weak but not as weak as for the Weak Interaction for which the strength is given by $\alpha_{WEAK} \approx 10^{-6}$ or for the Gravitational Interaction with $\alpha_{GRAVITY} \approx 10^{-39}$ but much weaker than the Strong Interaction of hadrons for which $\alpha_{STRONG} \approx 1$. The strength of the nonlinear interaction in turbulence can of course be quite variable and very large depending on the Reynolds number and the scales of interest.

The small fine structure constant makes QED very suitable for employing perturbation theory to develop statistical dynamical equations. However, the initial attempts encountered divergences of some perturbation terms in the high momentum or ultraviolet limit. This was eventually overcome, by the pioneers of QED, through mass and charge renormalization and resummation of the primitive perturbation series (Bjorken and Drell, 1964 [166], Chapter 8; 1965, [167] Chapter 19).

QED is the most succesful statistical dynamical field theory with astounding agreement with experiments. For example, the anomalous magnetic moment of the electron $a_e$, calculated to ten figures, has a difference of just one in the last figure, or one in $10^9$ with experiments (Aoyama et al. 2019 [168].

*12.2 Closure Equations for Strong Interaction Hadron Scattering*

In some ways, the steady state turbulent closures of Edwards (1963) [10], Herring (1965) [11] and McComb (1974) [6] have commonality with QED and the early approach to strong interaction hadron scattering. In these quantum systems the scattering is between asymptotic 'in' to 'out' states and the theory is an equilibrium or statistically steady state theory. Because of the strength of the interaction $\alpha_{STRONG} \approx 1$ the perturbative approach that was so successful for QED could not be applied. Strong interaction scattering was instead approached through closure equations for the scattering amplitude. Mandelstam (1958) [169] made a breakthrough with his double dispersion relation for hadron scattering. It was shown to be consistent with perturbation theory for the case of pion-pion scattering (Cutkosky 1960 [170]; Frederiksen and Wookcock 1973a, 1973 b [171,172]; Frederiksen 1974a, 1974b [173,174]) The Mandelstam closure is quadratic in the partial wave spectral scattering amplitudes and a question of interest was whether it had unique solutions that could be obtained through iteration. This was shown by reformulating the equations for pion-pion scattering and using functional analysis contraction mapping theorems (Frederiksen 1975 [175]; Atkinson et al. 1976 [176]). As a measure of the difficulties with strong interaction statistical dynamical theories, we note that it has taken until 2023 to develop a numerical iteration method to obtain the nonperturbative scattering amplitudes by Tourkine and Zhiboedov (2023) [177]

*12.3 Time-Dependent Nonequilibrium Quantum Field Theory*

Much research in quantum field theory has been concerned with equilibrium or statistical steady states or effectively climate states with scattering between 'in' states starting at minus infinity to 'out' states at plus infinity a particular focus. For these equilibrium problems the fluctuation dissipation holds and the systems are defined by Hermitian operators. Time dependent non-equilibrium quantum field theories have come under increased attention in the last 20 or so years as reviewed by Berges (2004) [178] and Calzetta and Hu (2008) [179]. This has been motivated particularly by problems in cosmology and inflation (Kofman et al. 1997 [180]; Micha and Tkachev 2003 [181]; Kofman 2008 [182]) where there is a need to understand the evolution of time-dependent quantum fluctuations. Moreover, it has been necessary to generalize quantum field theories to apply to spatially inhomogeneous fields. The evolution of quantum fluctuations in inhomogeneous fields has also found applications in researching Bose-Einstein condensation (Gasenzer 2009 [183]; Berges and Sexty 2012 [184]) and quark-gluon plasma (Arnold 2007) [185].

The theory of time-dependent nonequilibrium quantum fluctuations evolving in inhomogeneous fields was formulated, in the 1960s, in the iconic papers of Schwinger (1961) [186] and Keldysh (1965) [187]. The Schwinger-Keldysh closed time path (CTP) formalism is a path integral approach. It in fact reduces to the path integral generalization of the MSR formalism for classical fields as noted by Cooper et al. (2001) [188] and Blagoev et al. (2001) [189]. Further, as shown by Frederiksen (2017) [9] the classical path integral formalism (Jensen 1981) [26] can be used to develop the second order Schwinger Dyson equations for time-dependent nonequilibrium quantum systems. This requires suitable non-Gaussian noise terms proportional to Planck's constant squared, that account for quantum effects, to be included in the classical approach. This was documented for scalar field theories described by the Klein-Gordon equation with $g\phi^3$ or $\lambda\phi^4$ interaction Lagrangian. For the former the noise term is pure skewness and the latter pure kurtosis. Quantum systems thus result in extra quantum self-energy terms.

The major advances made in time-dependent nonequilibrium statistical dynamics with strong interactions by the pioneers in turbulence closure for classical fluid dynamics were emphasized by Mihaila et al. (2001) [190]. They make the following points:

1. "So, we, as quantum field theorists, having entered the domain of nonequilibrium phenomena, are now beset with all the problems faced by our plasma and condensed matter brethren more than 40 years ago."
2. "Most naive truncation schemes which, for example, just truncate the hierarchy of coupled correlators at a particular order, do not preserve various physical properties required of the system."
3. "A corollary of this is that in most perturbation schemes, secularity arises quickly with each term in the perturbation series, growing with higher powers of the time."
4. "In his seminal paper of 1961, Robert Kraichnan (1961) discussed in detail the key issues and obtained a partial solution to the problem by demanding that the approximations one should use should correspond to some physically realizable dynamical system. This would guarantee positivity and secularity would be avoided."

*12.4 The Problem of Dimensionality and the QDIA Closure*

The statistical dynamical theory of inhomogeneous systems is complex and computationally demanding, particularly in the common formulations, and this is so even in the bare vertex approximation. The Scwinger-Dyson [7,8] equations and the Schwinger-Keldysh [186,187] equations for quantum fields, like Kraichnan's (1972) [191] inhomogeneous DIA (IDIA) and the MSR (1973) [22] functional formalism, and path integral generalizations [25,26] for classical fields, express the mean field equation in terms of the full single-time covariance. This has $N^2$ components if the field has $N$. It in turn is coupled to closure equations for the full two-point functions - the response function and the two-time cumulant. Moreover, with the two-point function equations determined to second order in perturbation theory, as for the IDIA, the mean field equation is third order in perturbation theory in all these formalisms. Frederiksen (2012b) [73] developed the Self-Energy (SE) inhomogeneous closure which is a bare vertex approximation in which both the propagators and the mean field are second order in perturbation theory. This also has the benefit that all the nonlinear noise and damping terms are expressed as self-energies and the renormalization of the bare terms in the equations is transparent. The SE closure again consists of statistical dynamical equations for the mean field coupled to the full covariance and for the full propagators.

The QDIA closure (Frederiksen 1999, 2012a) [66,72] has the same structure as the SE closure but expresses the full response function and the two-time two-point cumulant in terms of the diagonal components in spectral space and the mean field (and possibly topography). It thus yields a system of just $N$ closure equations for the mean field and the diagonal components of the response function and the two-point cumulants. From these the full response function and full two-point cumulants can be constructed if they are needed. Frederiksen and O'Kane (2005) [67] for example calculate the associated

Jacobian terms that drive the mean-field equation. The reduction from $N^2$ equations to just $N$ makes the QDIA computationally tractable. The numerical efficiency is further increased with the Markovian MIC models and even more so with the EDMIC that has analytical forms for the relaxation functions.

**13. Conclusions and Future Prospects**

The advances made by Kraichnan, Herring, Edwards and McComb, the pioneers of turbulence closures, have been one of the great achievements in strong interaction statistical dynamical field theory (Mihaila et al. 2001) [190]. Their Eulerian closures correctly express the statics of the evolution of the energy-containing large scales but have slight inertial range deficiencies at finite Reynolds numbers. Much of the subsequent effort has aimed at correcting differences between direct numerical simulations and the closures at the smaller scales. However, even within the context of isotropic turbulence this has turned out to be an elusive goal. This is despite the heroic steps in formulating quasi-Lagrangian closures. The basic issue is that the second order closures require the correct renormalized vertex function to be accurate (Martin et al. 1973) [22]. Devising tractable equations, including for the vertices, seems an impossible objective for general strongly interacting field theories including classical turbulence. This contrasts with the remarkable triumph achieved with quantum electrodynamics where the interaction strength is quite weak. It appears that statistical dynamical closures for strongly nonlinear fields must in general rely on some empiricism. It is fortunate that for both two- and three-dimensional isotropic turbulence one parameter theoretical closure models are quite successful in reproducing the evolving statistics of fluid flows including in inertial ranges.

Another important objective of theoreticians should be to extend statistical dynamical closures from isotropic turbulence to the real world of inhomogeneous turbulence. "Ultimately, if closures are to be useful, then they must be capable of application to real-life situations." (McComb 2014) [118]. Indeed, this seems a far more tractable goal than solving or circumventing the vertex renormalization problem. However, relatively little work has been done using closures for inhomogeneous problems based on renormalized perturbation theory. The obstacle has been the shear size of the computational task with the standard approaches. For these, the mean field equations with $N$ components are coupled to equations for the generalized propagators, the two-point cumulants and response functions, with $N^2$ components. The quasi-diagonal direct interaction approximation (QDIA) closure offers a way around this problem of dimensionality when is $N$ large with a computationally tractable, realizable, approach to inhomogeneous turbulence. In this article we have reviewed some of its practical applications. Markovian reductions of the QDIA such as the Markovian Inhomogeneous Closures (MICs) and, particularly the eddy damped Markovian inhomogeneous closure (EDMIC), with analytical relaxation functions, are even more numerically efficient. Even in the case of homogeneous anisotropic turbulence the realizability of the eddy damped Markovian anisotropic closure (EDMAC), in the presence of transient waves, greatly expands the utility of the eddy damped quasi normal Markovian closure (EDQNM).

**Author Contributions:** Conceptualization, JSF; methodology, JSF, VK and TJO; formal analysis, JSF, VK and TJO; numeric implementation, JSF, VK and TJO; investigation, JSF, VK and TJO; writing—original draft preparation, JSF; writing—review and editing, JSF, VK and TJO; supervision, JSF; project administration, JSF; funding acquisition, VK and TJO. All authors have read and agreed to the published version of the manuscript.

**Funding:** VK and TJO were funded by CSIRO Environment.

**Data Availability Statement:** Not applicable since no new data were generated in this study.

**Conflicts of Interest:** The authors declare no conflict of interest.

## Appendix A: Spectral Equations for Turbulent Flows

The spectral equations for the Navier Stokes equations and the quasigeostrophic equations for geophysical flows are summarized in this Appendix.

*A.1. Spectral Equations for Navier Stokes Turbulent Flow*

We convert the Navier Stokes Equation (8) in terms of physical fields into their representation in spectral space by expanding the fields in terms of Fourier series, as in Equations (15) and (16). For the case of three-dimensional inhomogeneous turbulence, in a periodic box, $0 \leq x = x^1 \leq 2\pi$, $0 \leq y = x^2 \leq 2\pi$, $0 \leq z = x^3 \leq 2\pi$, the Navies Stokes equations take the spectral form [1]:

$$\frac{\partial}{\partial t} u_\mathbf{k}^a(t) + D_0^{a\beta}(\mathbf{k}) u_\mathbf{k}^\beta(t) = \sum_\mathbf{p} \sum_\mathbf{q} \delta(\mathbf{k},\mathbf{p},\mathbf{q}) M^{abc}(\mathbf{k}) u_{-\mathbf{p}}^b(t) u_{-\mathbf{q}}^c(t) + f_0^a(\mathbf{k},t) \quad \text{(A1)}$$

The continuous form of the equations, on the infinite domain, has the same structure, with the sums replaced by integrals [53,54]. Here, the spectral components of the velocity fields, in the $x$, $y$, $z$ directions respectively, are denoted by $u_\mathbf{k}^a(t)$, $a = 1,2,3$. They depend on the wavevector $\mathbf{k} = (k^1, k^2, k^3) = (k_x, k_y, k_z)$ and on time $t$. Forced dissipative flows are considered where $f_0^a(\mathbf{k},t)$ is the forcing term and $D_0^{a\beta}(\mathbf{k})$ is a general form of dissipation which can also include the action of waves. In Equation (A1) the interaction coefficient:

$$M^{abc}(\mathbf{k}) = i/2\{\Delta^{ab}(\mathbf{k})k^c + \Delta^{ac}(\mathbf{k})k^b\} \quad \text{(A2)}$$

and

$$\Delta^{ab}(\mathbf{k}) = \delta^{ab} - k^a k^b / |\mathbf{k}|^2. \quad \text{(A3)}$$

We note that Equations (A1) to (A3) are in standard form of Equation (5) by making the identifications $u_\mathbf{k}^a \to q_\mathbf{k}^a$
and

$$A^{abc}(\mathbf{k},\mathbf{p},\mathbf{q}) = i\Delta^{ab}(\mathbf{k})k^c. \quad \text{(A4)}$$

This means that $K^{abc}(\mathbf{k},\mathbf{p},\mathbf{q}) = M^{abc}(\mathbf{k})$ and Equation (A1) takes the standard form with $h_\mathbf{k}^a = 0$ in Equation (5).

For two-dimensional turbulence the spectral equations are again as above but with a two-dimensional wavevector $\mathbf{k} = (k^1, k^2) = (k_x, k_y)$.

*A.2. Spectral Equations for Quasigeostrophic Turbulent Flow in Planar Geometry*

The approach to devising the spectral space forms of the quasigeostrophic Equation (9) again consists of expanding the fields in Fourier series. As noted in Section 1.3.2 one can consider the case of continuous variations in the vertical (Frederiksen 2012a [72]), or multiple levels, flows in a channel, in a bounded domain and in an infinite domain [53,163]. To be specific, we focus here on two-level quasigeostrophic flow on the doubly periodic $f$-plane $0 \leq x \leq 2\pi, 0 \leq y \leq 2\pi$. Then, the interaction coefficients are given by:

$$A^{abc}(\mathbf{k},\mathbf{p},\mathbf{q}) = \begin{cases} A(\mathbf{k},\mathbf{p},\mathbf{q})V^{ab}(p) & \text{if } c = a \\ 0 & \text{otherwise,} \end{cases} \quad \text{(A5)}$$

with

$$V^{11}(p) = V^{22}(p) = \frac{p^2 + F_L}{p^2 + 2F_L}; \quad V^{12}(p) = V^{21}(p) = \frac{F_L}{p^2 + 2F_L}, \quad \text{(A6)}$$

and

$$A(\mathbf{k},\mathbf{p},\mathbf{q}) = -(p_x q_y - p_y q_x)/p^2. \quad \text{(A7)}$$

From the expression for $A^{abc}(\mathbf{k},\mathbf{p},\mathbf{q})$ in Equation (A5), $K^{abc}(\mathbf{k},\mathbf{p},\mathbf{q})$ is then defined by Equation (6) and the spectral equations are again in the standard form of Equation (5). It is of course straightforward to generalized to flows on a $\beta$-plane by including the Rossby wave dispersion relation in the generalized dissipation $D_0^{\alpha\beta}(\mathbf{k})$ in Equation (5). See for example Equation (20) for the barotropic case.

*A.2. Spectral Equations for Quasigeostrophic Turbulent Flow in Spherical Geometry*

The quasigeostrophic equations, in spherical geometry, again take the form in Equation (9) with the Jacobian:

$$J(\psi,q) = \frac{\partial \psi}{\partial \lambda}\frac{\partial q}{\partial \mu} - \frac{\partial \psi}{\partial \mu}\frac{\partial q}{\partial \lambda} \tag{A8}$$

expressed in terms of $\mu = \sin\phi$, the sin(latitude), and $\lambda$, the longitude. We again focus on the two-level QG equations. The spectral equations are for the spherical harmonic coefficients with the physical fields taking the form:

$$\begin{aligned} q^a(\lambda,\mu,t) &= \sum_m \sum_n q^a_{(m,n)}(t) P_n^m(\mu) \exp(im\lambda) \\ &= \sum_{\mathbf{k}} q^a_{\mathbf{k}}(t) P_{\mathbf{k}}(\mu) \exp(im_k \lambda). \end{aligned} \tag{A9}$$

Here, the Legendre functions

$$P_{\mathbf{k}}(\mu) = P_{n_k}^{m_k}(\mu), \tag{A10}$$

and the wavevector is expressed in terms of the zonal, $m$, and total, $n$, wavenumbers

$$\mathbf{k} = (m_k, n_k) = (m,n) \; ; \; -\mathbf{k} = (-m_k, n_k) \; ; \; k = n. \tag{A11}$$

The conversion of Equation (9) into spectral form is achieved through multiplying by the spherical harmonic $P_n^m(\mu)\exp(-im\lambda)$ and integrating over the $(\lambda,\mu)$ domain. Then using the orthogonality properties of the spherical harmonics we arrive at the following interaction coefficients:

$$\begin{aligned} A^{000}(\mathbf{k},\mathbf{p},\mathbf{q}) &= A^{101}(\mathbf{k},\mathbf{p},\mathbf{q}) \\ &= -i[n_p(n_p+1)]^{-1}\int_{-1}^{1} d\mu \, P_{\mathbf{k}} \left( m_p P_{\mathbf{p}} \frac{dP_{\mathbf{q}}}{d\mu} - m_q P_{\mathbf{q}} \frac{dP_{\mathbf{p}}}{d\mu} \right) \end{aligned} \tag{A12}$$

and

$$\begin{aligned} A^{011}(\mathbf{k},\mathbf{p},\mathbf{q}) &= A^{110}(\mathbf{k},\mathbf{p},\mathbf{q}) \\ &= -i[n_p(n_p+1)+\Gamma]^{-1}\int_{-1}^{1} d\mu \, P_{\mathbf{k}} \left( m_p P_{\mathbf{p}} \frac{dP_{\mathbf{q}}}{d\mu} - m_q P_{\mathbf{q}} \frac{dP_{\mathbf{p}}}{d\mu} \right). \end{aligned} \tag{A13}$$

We note that the remaining interaction coefficients vanish:

$$A^{010}(\mathbf{k},\mathbf{p},\mathbf{q}) = A^{001}(\mathbf{k},\mathbf{p},\mathbf{q}) = A^{100}(\mathbf{k},\mathbf{p},\mathbf{q}) = A^{111}(\mathbf{k},\mathbf{p},\mathbf{q}) = 0. \tag{A14}$$

Again, the definition of $K^{abc}(\mathbf{k},\mathbf{p},\mathbf{q})$ is given in Equation (6).

The spectral equation is then as in Equation (5) with $D_0^{\alpha\beta}(\mathbf{k}) = \nu_0^{\alpha\beta}(k)k(k+1)$ and $q_{-\mathbf{k}}^a \equiv q_{(-m,n)}^a = q_{(m,n)}^{a*} \equiv q_{\mathbf{k}}^{a*}$. On the sphere the spectral interactions are more complicated than on the plane and $\delta$ is replaced by $\delta_{Sphere}$ where

$$\delta_{Sphere}(\mathbf{k},\mathbf{p},\mathbf{q}) = \begin{cases} 1 & \text{if } m_k + m_p + m_q = 0, \\ & \text{and } n_k + n_p + n_q \text{ is odd}, \\ & \text{and } |n_p - n_q| < n_k < n_p + n_q, \\ 0 & \text{otherwise.} \end{cases} \tag{A15}$$

These selection rules are further discussed by Frederiksen and Sawford (1980) [] (Equation (2.3)). Again, $D_0^{\alpha\beta}(\mathbf{k})$ in Equation (5) can include the Rossby wave frequency due to differential rotation.

**Appendix B: Interaction Coefficients for Barotropic Flows with Large Scale Wind**

As noted by Frederiksen and O'Kane (2005) [67], the equation for the large-scale zonal flow can be combined with the Fourier spectral equations for the small scales into the single form given in Equation (18). The required interaction coefficients are:

$$A(\mathbf{k},\mathbf{p},\mathbf{q}) = -\gamma(p_x\hat{q}_y - \hat{p}_y q_x)/p^2, \tag{A16}$$

and

$$K(\mathbf{k},\mathbf{p},\mathbf{q}) = \tfrac{1}{2}[A(\mathbf{k},\mathbf{p},\mathbf{q}) + A(\mathbf{k},\mathbf{q},\mathbf{p})] = \tfrac{1}{2}\gamma[p_x\hat{q}_y - \hat{p}_y q_x](p^2 - q^2)/p^2 q^2, \tag{A17}$$

with

$$\delta(\mathbf{k},\mathbf{p},\mathbf{q}) = \begin{cases} 1 & \text{if } \mathbf{k}+\mathbf{p}+\mathbf{q}=\mathbf{0} \\ 0 & \text{otherwise.} \end{cases} \tag{A18}$$

The zero wavevector represents the large-scale zonal flow. It is included in Equation (18) by defining $\gamma$, $\hat{q}_y$ and $\hat{p}_y$ as follows:

$$\gamma = \begin{cases} -\tfrac{1}{2}k_0 & \text{if } \mathbf{k}=\mathbf{0}, \\ k_0 & \text{if } \mathbf{q}=\mathbf{0} \text{ or } \mathbf{p}=\mathbf{0}, \\ 1 & \text{otherwise,} \end{cases} \tag{A19}$$

with

$$\hat{p}_y = \begin{cases} 1 & \text{if } \mathbf{k}=\mathbf{0}, \text{ or } \mathbf{p}=\mathbf{0}, \text{ or } \mathbf{q}=\mathbf{0} \\ p_y & \text{otherwise,} \end{cases} \tag{A20}$$

and

$$\hat{q}_y = \begin{cases} 1 & \text{if } \mathbf{k}=\mathbf{0}, \text{ or } \mathbf{p}=\mathbf{0}, \text{ or } \mathbf{q}=\mathbf{0} \\ q_y & \text{otherwise.} \end{cases} \tag{A21}$$

These generalized interaction coefficients satisfy all the standard symmetry relationships:

$$A(-\mathbf{k},-\mathbf{p},-\mathbf{q}) = A(\mathbf{k},\mathbf{p},\mathbf{q}), \tag{A22}$$

and

$$K(-\mathbf{k},-\mathbf{p},-\mathbf{q}) = K(\mathbf{k},\mathbf{p},\mathbf{q}), \tag{A23}$$

as well the sum rule which is used to establish conservation laws

$$K(\mathbf{k},\mathbf{p},\mathbf{q}) + K(\mathbf{p},\mathbf{q},\mathbf{k}) + K(\mathbf{q},\mathbf{k},\mathbf{p}) = 0. \tag{A24}$$

These properties hold for all values of $\mathbf{k}$, $\mathbf{p}$ and $\mathbf{q}$.

**Appendix C: QDIA Structure and Cumulant and Response Function Relationships**

The key to the computational efficiency of the QDIA closure is the fact that the expressions for the off-diagonal elements of the two-point and three-point cumulants and response functions are represented by combinations of purely diagonal elements in addition to the mean field and topography. These relationships are documented in Frederiksen (1999) [66] and O'Kane and Frederiksen (2004) [43] and summarized here. The off-diagonal elements of the two-point cumulant and response function are

$$C_{\mathbf{k},-\mathbf{l}}^{QDIA}(t,t')$$
$$= \int_{t_0}^{t} ds\, R_{\mathbf{k}}(t,s)C_{\mathbf{l}}(s,t')[A(\mathbf{k},-\mathbf{l},\mathbf{l}-\mathbf{k})h_{(\mathbf{k}-\mathbf{l})} + 2K(\mathbf{k},-\mathbf{l},\mathbf{l}-\mathbf{k})<\zeta_{(\mathbf{k}-\mathbf{l})}(s)>]$$
$$+ \int_{t_0}^{t'} ds\, R_{-\mathbf{l}}(t',s)C_{\mathbf{k}}(t,s)[A(-\mathbf{l},\mathbf{k},\mathbf{l}-\mathbf{k})h_{(\mathbf{k}-\mathbf{l})} + 2K(-\mathbf{l},\mathbf{k},\mathbf{l}-\mathbf{k})<\zeta_{(\mathbf{k}-\mathbf{l})}(s)>]$$
$$+ R_{\mathbf{k}}(t,t_0)R_{-\mathbf{l}}(t',t_0)\tilde{K}_{\mathbf{k},-\mathbf{l}}^{(2)}(t_0,t_0) \tag{A25}$$

where $\tilde{K}_{\mathbf{k},-\mathbf{l}}^{(2)}(t_0,t_0)$ is the contribution from the initial off-diagonal second order cumulant at time $t_0$. Also,

$$R_{\mathbf{k},\mathbf{l}}^{QDIA}(t,t')$$
$$= \int_{t'}^{t} ds\, R_{\mathbf{k}}(t,s)R_{\mathbf{l}}(s,t')[A(\mathbf{k},-\mathbf{l},\mathbf{l}-\mathbf{k})h_{(\mathbf{k}-\mathbf{l})} + 2K(\mathbf{k},-\mathbf{l},\mathbf{l}-\mathbf{k})<\zeta_{(\mathbf{k}-\mathbf{l})}(s)>]. \tag{A26}$$

The three-point cumulant is expressed in terms of the diagonal two-point cumulants and response functions:

$$\left\langle \tilde{\zeta}_{-\mathbf{l}}(t)\tilde{\zeta}_{(\mathbf{l}-\mathbf{k})}(t)\tilde{\zeta}_{\mathbf{k}}(t')\right\rangle^{QDIA}$$
$$= 2\int_{t_0}^{t'} ds\, K(\mathbf{k},-\mathbf{l},\mathbf{l}-\mathbf{k})\, C_{-\mathbf{l}}(t,s)C_{(\mathbf{l}-\mathbf{k})}(t,s)R_{\mathbf{k}}(t',s)$$
$$+ 2\int_{t_0}^{t} ds\, K(-\mathbf{l},\mathbf{l}-\mathbf{k},\mathbf{k})\, R_{-\mathbf{l}}(t,s)C_{(\mathbf{l}-\mathbf{k})}(t,s)C_{\mathbf{k}}(t',s) \tag{A27}$$
$$+ 2\int_{t_0}^{t} ds\, K(\mathbf{k},-\mathbf{l},\mathbf{l}-\mathbf{k})\, C_{-\mathbf{l}}(t,s)R_{(\mathbf{l}-\mathbf{k})}(t,s)C_{\mathbf{k}}(t',s)$$
$$+ R_{-\mathbf{l}}(t,t_0)R_{(\mathbf{l}-\mathbf{k})}(t,t_0)R_{\mathbf{k}}(t',t_0)\tilde{K}_{-\mathbf{l},(\mathbf{l}-\mathbf{k}),\mathbf{k}}^{(3)}(t_0,t_0,t_0)$$

where $\tilde{K}_{-\mathbf{l},(\mathbf{l}-\mathbf{k}),\mathbf{k}}^{(3)}(t_0,t_0,t_0)$ represents initial non-Gaussian initial conditions (Frederiksen and Davies 2000 [31]; O'Kane and Frederiksen 2004 [43]). The statistical average of the product of the response function and the vorticity perturbation field is:

$$\left\langle \tilde{R}_{(\mathbf{l}-\mathbf{k})}(t,t')\tilde{\zeta}_{-\mathbf{l}}(t)\right\rangle^{QDIA} = 2\int_{t'}^{t} ds\, K(\mathbf{l}-\mathbf{k},-\mathbf{l},\mathbf{k})\, C_{-\mathbf{l}}(t,s)R_{(\mathbf{l}-\mathbf{k})}(t,s)R_{\mathbf{k}}(s,t'). \tag{A28}$$

These relationships are derived by Frederiksen (1999, 2003) [48,66] and form the basis of the cumulant update method of integrating the QDIA closure on the $f-$plane (O'Kane and Frederiksen 2004) [43] and $\beta-$plane (Frederiksen and O'Kane 2005) [67].

**Appendix D: Multi-Field QDIA Closure for Inhomogeneous Turbulent Flows**

In this Appendix we briefly summarise the more general expressions for the QDIA closure equations for multiple fields and possibly also at multiple levels. The multiple field QDIA equations are derived by Frederiksen (2012a) [72]. We again start by representing the field $q_{\mathbf{k}}^a$ in terms of its ensemble mean $<q_{\mathbf{k}}^a>$ and deviation $\tilde{q}_{\mathbf{k}}^a$ and similarly for the bare forcing $f_0^a(\mathbf{k})$:

$$q_{\mathbf{k}}^a = <q_{\mathbf{k}}^a> + \tilde{q}_{\mathbf{k}}^a \tag{A29}$$

And

$$f_0^a(\mathbf{k}) = <f_0^a(\mathbf{k})> + \tilde{f}_0^a(\mathbf{k}) = \overline{f}_0^a(\mathbf{k}) + \tilde{f}_0^a(\mathbf{k}). \tag{A30}$$

The prognostic closure equation for the mean field equation then takes the form:

$$\frac{\partial}{\partial t}<q_{\mathbf{k}}^{a}(t)>+D_{0}^{\alpha\beta}(\mathbf{k})<q_{\mathbf{k}}^{\beta}(t)>$$
$$=\sum_{\mathbf{p}}\sum_{\mathbf{q}}\delta(\mathbf{k},\mathbf{p},\mathbf{q})K^{abc}(\mathbf{k},\mathbf{p},\mathbf{q})<q_{-\mathbf{p}}^{b}(t)><q_{-\mathbf{q}}^{c}(t)>$$
$$+\sum_{\mathbf{p}}\sum_{\mathbf{q}}\delta(\mathbf{k},\mathbf{p},\mathbf{q})A^{abc}(\mathbf{k},\mathbf{p},\mathbf{q})<q_{-\mathbf{p}}^{b}(t)>h_{-\mathbf{q}}^{c} \quad \text{(A31)}$$
$$-\int_{t_{o}}^{t}ds\,\eta_{\mathbf{k}}^{\alpha\beta}(t,s)<q_{\mathbf{k}}^{\beta}(s)>+\overline{f}_{0}^{a}(\mathbf{k},t)+f_{\chi}^{a}(\mathbf{k},t).$$

Here, $\eta_{\mathbf{k}}^{\alpha\beta}$ is the nonlinear damping that results from the interaction of the eddies with the mean field and $f_{\chi}^{a}(\mathbf{k},t)$ is the eddy-topographic force due to the interaction of the eddies with the topography. These terms are defined by:

$$\eta_{\mathbf{k}}^{a\alpha}(t,s)$$
$$=-4\sum_{\mathbf{p}}\sum_{\mathbf{q}}\delta(\mathbf{k},\mathbf{p},\mathbf{q})K^{abc}(\mathbf{k},\mathbf{p},\mathbf{q})K^{\beta\gamma\alpha}(-\mathbf{p},-\mathbf{q},-\mathbf{k})R_{-\mathbf{p}}^{b\beta}(t,s)C_{-\mathbf{q}}^{c\gamma}(t,s), \quad \text{(A32)}$$

and

$$f_{\chi}^{a}(\mathbf{k},t)=h_{\mathbf{k}}^{\alpha}\int_{t_{o}}^{t}ds\,\chi_{\mathbf{k}}^{a\alpha}(t,s), \quad \text{(A33)}$$

with

$$\chi_{\mathbf{k}}^{a\alpha}(t,s)$$
$$=2\sum_{\mathbf{p}}\sum_{\mathbf{q}}\delta(\mathbf{k},\mathbf{p},\mathbf{q})K^{abc}(\mathbf{k},\mathbf{p},\mathbf{q})A^{\beta\gamma\alpha}(-\mathbf{p},-\mathbf{q},-\mathbf{k})R_{-\mathbf{p}}^{b\beta}(t,s)C_{-\mathbf{q}}^{c\gamma}(t,s). \quad \text{(A34)}$$

Throughout, we use the following short-hand notation for the diagonal components of the two-point cumulant and response function:

$$C_{\mathbf{k}}^{ab}(t,s)=C_{\mathbf{k},-\mathbf{k}}^{ab}(t,s)\;;\;R_{\mathbf{k}}^{ab}(t,s)=R_{\mathbf{k},\mathbf{k}}^{ab}(t,s), \quad \text{(A35)}$$

where

$$C_{\mathbf{k},-\mathbf{l}}^{ab}(t,s)=<\tilde{q}_{\mathbf{k}}^{a}(t)\tilde{q}_{-\mathbf{l}}^{b}(s)>, \quad \text{(A36)}$$

and

$$R_{\mathbf{k},\mathbf{l}}^{ab}(t,s)=\left\langle\frac{\delta\tilde{q}_{\mathbf{k}}^{a}(t)}{\delta\tilde{f}_{0}^{b}(\mathbf{l})(s)}\right\rangle. \quad \text{(A37)}$$

The mean field equation involves the diagonal two-time two-point cumulant and response function. Thus, we need closure evolution equations for these terms. The cumulant equation is

$$\frac{\partial}{\partial t}C_{\mathbf{k}}^{ab}(t,t')+D_{0}^{\alpha\beta}(\mathbf{k})C_{\mathbf{k}}^{\beta b}(t,t')$$
$$=\int_{t_{o}}^{t'}ds\,\{S_{\mathbf{k}}^{\alpha\beta}(t,s)+P_{\mathbf{k}}^{\alpha\beta}(t,s)+F_{0}^{\alpha\beta}(\mathbf{k},t,s)\}R_{-\mathbf{k}}^{b\beta}(t',s) \quad \text{(A38)}$$
$$-\int_{t_{o}}^{t}ds\,\{\eta_{\mathbf{k}}^{\alpha\beta}(t,s)+\pi_{\mathbf{k}}^{\alpha\beta}(t,s)\}C_{-\mathbf{k}}^{b\beta}(t',s).$$

Here, the bare noise covariance is:

$$F_{0}^{a\alpha}(\mathbf{k},t,s)=<\hat{f}_{0}^{a}(\mathbf{k},t)\hat{f}_{0}^{\alpha}(-\mathbf{k},s)>, \quad \text{(A39)}$$

and the nonlinear noise due to eddy-eddy interactions is:

$$S_{\mathbf{k}}^{a\alpha}(t,s) = 2\sum_{\mathbf{p}}\sum_{\mathbf{q}}\delta(\mathbf{k},\mathbf{p},\mathbf{q})K^{abc}(\mathbf{k},\mathbf{p},\mathbf{q})K^{\alpha\beta\gamma}(-\mathbf{k},-\mathbf{p},-\mathbf{q})C_{-\mathbf{p}}^{b\beta}(t,s)C_{-\mathbf{q}}^{c\gamma}(t,s), \quad (A40)$$

while that from eddy-mean-field and eddy-topographic interactions is:

$$P_{\mathbf{k}}^{a\alpha}(t,s) = \sum_{\mathbf{p}}\sum_{\mathbf{q}}\delta(\mathbf{k},\mathbf{p},\mathbf{q})C_{-\mathbf{p}}^{b\beta}(t,s)[2K^{abc}(\mathbf{k},\mathbf{p},\mathbf{q})<q_{-\mathbf{q}}^{c}(t)> + A^{abc}(\mathbf{k},\mathbf{p},\mathbf{q})h_{-\mathbf{q}}^{c}] \quad (A41)$$
$$\cdot [2K^{\alpha\beta\gamma}(-\mathbf{k},-\mathbf{p},-\mathbf{q})<q_{\mathbf{q}}^{\gamma}(s)> + A^{\alpha\beta\gamma}(-\mathbf{k},-\mathbf{p},-\mathbf{q})h_{\mathbf{q}}^{\gamma}].$$

Note also that the nonlinear damping from eddy-mean-field and eddy-topographic interactions is

$$\pi_{\mathbf{k}}^{a\alpha}(t,s) = -\sum_{\mathbf{p}}\sum_{\mathbf{q}}\delta(\mathbf{k},\mathbf{p},\mathbf{q})R_{-\mathbf{p}}^{b\beta}(t,s)[2K^{abc}(\mathbf{k},\mathbf{p},\mathbf{q})<q_{-\mathbf{q}}^{c}(t)> + A^{abc}(\mathbf{k},\mathbf{p},\mathbf{q})h_{-\mathbf{q}}^{c}] \quad (A42)$$
$$\cdot [2K^{\beta\alpha\gamma}(-\mathbf{p},-\mathbf{k},-\mathbf{q})<q_{\mathbf{q}}^{\gamma}(s)> + A^{\beta\alpha\gamma}(-\mathbf{p},-\mathbf{k},-\mathbf{q})h_{\mathbf{q}}^{\gamma}],$$

with eddy-eddy term, $\eta_{\mathbf{k}}^{\alpha\beta}$, given in Equation (A32). Both the nonlinear noise terms $S_{\mathbf{k}}^{a\alpha}(t,s)$ and $P_{\mathbf{k}}^{a\alpha}(t,s)$ are positive semi-definite. The integro-differential equation for the response function is:

$$\frac{\partial}{\partial t}R_{\mathbf{k}}^{ab}(t,t') + D_{0}^{a\beta}(\mathbf{k})R_{\mathbf{k}}^{\beta b}(t,t') = -\int_{t'}^{t}ds\{\eta_{\mathbf{k}}^{a\beta}(t,s) + \pi_{\mathbf{k}}^{a\beta}(t,s)\}R_{\mathbf{k}}^{\beta b}(s,t') \quad (A43)$$

and the initial condition is $R_{\mathbf{k}}^{ab}(t,t) = \delta^{ab}$ with $\delta^{ab}$ the Kronecker delta function. The single-time cumulant is obtained from

$$\frac{\partial}{\partial t}C_{\mathbf{k}}^{ab}(t,t) = \lim_{t' \to t}\{\frac{\partial}{\partial t}C_{\mathbf{k}}^{ab}(t,t') + \frac{\partial}{\partial t'}C_{-\mathbf{k}}^{ba}(t',t)\} \quad (A44)$$

and this then forms the complete set of the QDIA closure equations.

**Appendix E: Langevin Equation for Multi-Field QDIA Closure**

It is most important that closure equations are realizable to avoid the possibility of unphysical behaviour. Realizability can be guaranteed by an underpinning Langevin equation from which the closure can be exactly reconstructed. For the QDIA closure this generalized Langevin equation is:

$$\frac{\partial}{\partial t}\tilde{q}_{\mathbf{k}}^{a}(t) + D_{0}^{a\beta}(\mathbf{k})\tilde{q}_{\mathbf{k}}^{\beta}(t)$$
$$= -\int_{t_o}^{t}ds\{\eta_{\mathbf{k}}^{a\beta}(t,s) + \pi_{\mathbf{k}}^{a\beta}(t,s)\}\tilde{q}_{\mathbf{k}}^{\beta}(s) + \tilde{f}_{0}^{a}(\mathbf{k},t) + f_{S}^{a}(\mathbf{k},t) + f_{P}^{a}(\mathbf{k},t) \quad (A45)$$

where:

$$f_{S}^{a}(\mathbf{k},t) = \sqrt{2}\sum_{(\mathbf{p},\mathbf{q})\in\mathcal{T}}\delta(\mathbf{k},\mathbf{p},\mathbf{q})K^{abc}(\mathbf{k},\mathbf{p},\mathbf{q})W_{-\mathbf{p}}^{(1)b}(t)W_{-\mathbf{q}}^{(2)c}(t), \quad (A46)$$

and

$$f_{P}^{a}(\mathbf{k},t) = \sum_{(\mathbf{p},\mathbf{q})\in\mathcal{T}}\delta(\mathbf{k},\mathbf{p},\mathbf{q})[2K^{abc}(\mathbf{k},\mathbf{p},\mathbf{q})<q_{-\mathbf{q}}^{c}(t)> + A^{abc}(\mathbf{k},\mathbf{p},\mathbf{q})h_{-\mathbf{q}}^{c}]$$
$$\times W_{-\mathbf{p}}^{(3)b}(t)w_{-\mathbf{q}}^{c}(t). \quad (A47)$$

Here $\mathcal{T}$ could be any domain of interest but for convenience we suppose that $\mathcal{T} = \{\mathbf{p},\mathbf{q} | p \leq T, q \leq T\}$ as in Equation (19). As well, $W_{\mathbf{k}}^{(j)a}(t)$, where $j$ = 1, 2 or 3, and $w_{\mathbf{k}}^{a}(t)$ are statistically independent random variables such that:

$$< W_{\mathbf{k}}^{(j)a}(t) W_{-\mathbf{l}}^{(j')b}(t') > = \delta^{jj'} \delta_{\mathbf{kl}} C_{\mathbf{k}}^{ab}(t,t'), \tag{A48}$$

with

$$< w_{\mathbf{k}}^{a}(t) w_{-\mathbf{l}}^{b}(t') > = \delta_{\mathbf{kl}} \tag{A49}$$

and

$$< \tilde{q}_{\mathbf{k}}^{a}(t) \tilde{q}_{-\mathbf{k}}^{b}(t') > = C_{\mathbf{k}}^{ab}(t,t'). \tag{A50}$$

In Equation (A48) and Equation (A49), $\delta$ is the Kronecker delta function.

The Langevin Equation (A45) guarantees realizability for the elements of the covariance matrices that are diagonal in spectral space in the quasi-diagonal closure equations.

## Appendix F: Subgrid Parameterizations from Multi-Field QDIA Closure

In this Appendix we are now in a position to derive expressions for subgrid scale terms when the resolution is reduced from $T$ to $T_R < T$ where $T_R$ is the resolution of the resolved scales. In planar geometry we define $\mathcal{T} = \{\mathbf{p},\mathbf{q} | p \leq T, q \leq T\}$, $\mathcal{R} = \{\mathbf{p},\mathbf{q} | p \leq T_R, q \leq T_R\}$ and $\mathcal{S} = \mathcal{T} - \mathcal{R}$. Here, $\mathcal{S}$ is the set of subgrid scale terms. The corresponding sets for spherical geometry are defined in Equations (2.6), (3.2) and (3.3) of Frederiksen and Kepert (2006) [127]. The subgrid scale parametrisations modify or renormalise the bare forcing and dissipation terms in the dynamical and closure equations. We define the renormalised mean force determined from Equation (A31) as

$$\overline{f}_r^a(\mathbf{k},t) = \overline{f}_0^a(\mathbf{k},t) + \overline{f}_h^a(\mathbf{k},t) + j^a(\mathbf{k},t) \tag{A51}$$

where the residual Jacobian of the subgrid scales is

$$j^a(\mathbf{k},t) = \sum_{(\mathbf{p},\mathbf{q}) \in \mathcal{S}} \sum \delta(\mathbf{k},\mathbf{p},\mathbf{q})[K^{abc}(\mathbf{k},\mathbf{p},\mathbf{q}) < q_{-\mathbf{p}}^b(t) >< q_{-\mathbf{q}}^c(t) > \\ + A^{abc}(\mathbf{k},\mathbf{p},\mathbf{q}) < q_{-\mathbf{p}}^b(t) > h_{-\mathbf{q}}^c] \tag{A52}$$

and the subgrid eddy-topographic force is

$$f_h^a(\mathbf{k},t) := f_\chi^{\mathcal{S}a}(\mathbf{k},t) = h_{\mathbf{k}}^\alpha \int_{t_o}^{t} ds \, \chi_{\mathbf{k}}^{\mathcal{S}a\alpha}(t,s). \tag{A53}$$

Here, the set and index $\mathcal{S}$ indicates that the subgrid set of wavevectors is summed over.

The renormalized random forcing covariance is defined by:

$$F_r^{\alpha\beta}(\mathbf{k},t) = F_0^{\alpha\beta}(\mathbf{k},t) + F_b^{\alpha\beta}(\mathbf{k},t) \tag{A54}$$

where

$$F_b^{\alpha\beta}(\mathbf{k},t) = F_S^{\alpha\beta}(\mathbf{k},t) + F_P^{\alpha\beta}(\mathbf{k},t). \tag{A55}$$

Here,

$$F_S^{\alpha\beta}(\mathbf{k},t) = \int_{t_o}^{t} ds \, S_{\mathbf{k}}^{\mathcal{S}\alpha\gamma}(t,s) R_{-\mathbf{k}}^{\beta\gamma}(t,s) \tag{A56}$$

and

$$F_P^{\alpha\beta}(\mathbf{k},t) = \int_{t_o}^{t} ds \, P_{\mathbf{k}}^{\mathcal{S}\alpha\gamma}(t,s) R_{-\mathbf{k}}^{\beta\gamma}(t,s), \tag{A57}$$

with

$$F_0^{\alpha\beta}(\mathbf{k},t) = \int_{t_o}^{t} ds \, F_0^{\alpha\gamma}(\mathbf{k},t,s) R_{-\mathbf{k}}^{\beta\gamma}(t,s). \tag{A58}$$

The renormalized generalized drain matrix acting on the mean field is defined by:
$$\bar{D}_r^{\alpha\beta}(\mathbf{k},t) = D_0^{\alpha\beta}(\mathbf{k}) + \bar{D}_d^{\alpha\beta}(\mathbf{k},t), \tag{A59}$$

where

$$\bar{N}_d^{\alpha\beta}(\mathbf{k},t) = \bar{D}_d^{\alpha\beta}(\mathbf{k},t) <q_\mathbf{k}^\beta(t)> = \int_{t_o}^t ds\, \eta_\mathbf{k}^{\mathcal{S}\alpha\beta}(t,s) <q_\mathbf{k}^\beta(s)>. \tag{A60}$$

Similarly, the renormalized generalized drain matrix acting on the fluctuations is:
$$D_r^{\alpha\beta}(\mathbf{k},t) = D_0^{\alpha\beta}(\mathbf{k}) + D_d^{\alpha\beta}(\mathbf{k},t) \tag{A61}$$

with
$$D_d^{\alpha\beta}(\mathbf{k},t) = D_\eta^{\alpha\beta}(\mathbf{k},t) + D_\pi^{\alpha\beta}(\mathbf{k},t). \tag{A62}$$

Here,
$$N_\eta^{\alpha\beta}(\mathbf{k},t) = D_\eta^{\alpha\gamma}(\mathbf{k},t) C_\mathbf{k}^{\gamma\beta}(t,t) = \int_{t_o}^t ds\, \eta_\mathbf{k}^{\mathcal{S}\alpha\gamma}(t,s) C_{-\mathbf{k}}^{\beta\gamma}(t,s) \tag{A63}$$

and
$$N_\pi^{\alpha\beta}(\mathbf{k},t) = D_\pi^{\alpha\gamma}(\mathbf{k},t) C_\mathbf{k}^{\gamma\beta}(t,t) = \int_{t_o}^t ds\, \pi_\mathbf{k}^{\mathcal{S}\alpha\gamma}(t,s) C_{-\mathbf{k}}^{\beta\gamma}(t,s). \tag{A64}$$

The generalized dissipation matrix due to backscatter is defined by:
$$N_b^{\alpha\beta}(\mathbf{k},t) = D_b^{\alpha\gamma}(\mathbf{k},t) C_\mathbf{k}^{\gamma\beta}(t,t) = F_b^{\alpha\beta}(\mathbf{k},t) \tag{A65}$$

and the generalized net and renormalized net dissipation matrices are given by:
$$D_n^{\alpha\beta}(\mathbf{k},t) = D_0^{\alpha\beta}(\mathbf{k}) + D_b^{\alpha\beta}(\mathbf{k},t), \tag{A66}$$
$$D_{rn}^{\alpha\beta}(\mathbf{k},t) = D_0^{\alpha\beta}(\mathbf{k}) + D_n^{\alpha\beta}(\mathbf{k},t). \tag{A67}$$

The corresponding generalized viscosity matrices are
$$\bar{\nu}_\bullet^{ab}(\mathbf{k}) = \bar{D}_\bullet^{ab}(\mathbf{k}) k^{-2}, \tag{A68}$$
$$\nu_\bullet^{ab}(\mathbf{k}) = D_\bullet^{ab}(\mathbf{k}) k^{-2} \tag{A69}$$

and $K = k$ on the plane and $k^2 = n(n+1)$ on the sphere. The symbol $\bullet$ denotes any of $0, b, d, n, r, rn, \eta, \pi$ as appropriate.

## Appendix G: Stochastic Modelling of Fluctuating Subgrid Tendency

Here the method of Frederiksen and Kepert (2006) is used to parameterize $\tilde{\mathbf{q}}_t^\mathcal{S}$ via the stochastic equation

$$\tilde{\mathbf{q}}_t^\mathcal{S}(t) = -\mathbf{D}_d \tilde{\mathbf{q}}(t) + \tilde{\mathbf{f}}(t) \tag{A70}$$

where $\mathbf{D}_d$ is the drain dissipation matrix, $\tilde{\mathbf{q}}(t)$ is the fluctuating component of $\mathbf{q}(t)$, and $\tilde{\mathbf{f}}$ is a random forcing vector. The state vector, $\tilde{\mathbf{q}}(t)$, contains the prognostic variables for a given scale. For example, in a simulation with one prognostic variable and two vertical levels, for each horizontal wavenumber pair, $\mathbf{D}_d$ is a $2 \times 2$ matrix, and $\tilde{\mathbf{f}}$ is a two-element column vector. One can also use scale-based decompositions in the vertical direction. For example, in Kitsios et al. (2017) [163] a turbulent channel flow simulation was decomposed using Chebyshev polynomials in the vertical direction, Fourier in the horizontal, with matrix formulations capturing the relationships between the three prognostic velocity components. In this case $\mathbf{D}_d$ was a $3 \times 3$ matrix for each pseudo wavenumber triplet.

Regardless of the specific structure of $\tilde{\mathbf{q}}$, one determines $\mathbf{D}_d$ by post-multiplying Equation (A70) by $\tilde{\mathbf{q}}^\dagger(t_0)$, integrating over the turbulent decorrelation period $\tau$, ensemble averaging , and then rearranging to return

$$\mathbf{D_d} = -\left[\int_{t_0}^{t} d\sigma \left\langle \tilde{\mathbf{q}}_t^S(\sigma)\tilde{\mathbf{q}}^\dagger(t_0)\right\rangle\right]\left[\int_{t_0}^{t} d\sigma \left\langle \tilde{\mathbf{q}}(\sigma)\tilde{\mathbf{q}}^\dagger(t_0)\right\rangle\right]^{-1} \quad (A71)$$

where $\dagger$ denotes the Hermitian conjugate. The angle brackets denote averaging over the realizations taken at different times. The integrals are evaluated over the time window from $t_0$ to $t_0 + \tau$, where $t_0$ is shifted forward by the time step size $\Delta t$ as one progresses through the realizations or samples in time. This can either be done offline if the samples are output to disk, or online by accumulating the appropriate statistics. The turbulence decorrelation time, $\tau$, must be chosen sufficiently large to capture the memory effects of the turbulence – see Figure 7 of Kitsios et al. (2012) [157].

The representation for the stochastic force $\tilde{\mathbf{f}}$ is given by calculating the noise covariance matrix $\mathbf{F_b} = \mathbf{\bar{F}_b} + \mathbf{\bar{F}_b^\dagger}$ where $\mathbf{\bar{F}_b} = <\tilde{\mathbf{f}}(t)\tilde{\mathbf{q}}^\dagger(t)>$. Post-multiplying Equation (A70) by $\tilde{\mathbf{q}}^\dagger(t)$, adding the conjugate transpose of Equation (A70), pre-multiplied by $\tilde{\mathbf{q}}(t)$, and averaging produces the result:

$$\left\langle \tilde{\mathbf{q}}_t^S(t)\tilde{\mathbf{q}}^\dagger(t)\right\rangle + \left\langle \tilde{\mathbf{q}}(t)\tilde{\mathbf{q}}_t^{S\dagger}(t)\right\rangle = -\mathbf{D_d}\left\langle \tilde{\mathbf{q}}(t)\tilde{\mathbf{q}}^\dagger(t)\right\rangle - \left\langle \tilde{\mathbf{q}}(t)\tilde{\mathbf{q}}^\dagger(t)\right\rangle\mathbf{D_d^\dagger} + \mathbf{F_b}. \quad (A72)$$

One can rearrange this equation to solve for $\mathbf{F_b}$. From the successful stochastic LES in our previous publications, we have shown that it is sufficient to assume that $\tilde{\mathbf{f}}$ can be approximated as white noise $<\tilde{\mathbf{f}}(t)\tilde{\mathbf{f}}^\dagger(t')> = \mathbf{F_b}\delta(t-t')$. The backscatter dissipation is given by

$$\mathbf{D_b} = -\mathbf{F_b} <\tilde{\mathbf{q}}(t)\tilde{\mathbf{q}}^\dagger(t)>^{-1} \quad (A73)$$

An eigenvalue decomposition of $\mathbf{F_b}$ is adopted to implement the stochastic representation for $\tilde{\mathbf{f}}$ into the LES (Zidikheri and Frederiksen, 2009) [156]. The finite difference approximation of the stochastic backscatter variance is

$$<\tilde{\mathbf{f}}(t)\tilde{\mathbf{f}}^\dagger(t')> = \mathbf{F_b}\delta(t-t') \approx \mathbf{F_b}/(2\Delta t) \quad (A74)$$

where $(2\Delta t)^{-1}$ is the discrete approximation of the delta function. In general, $\mathbf{F_b}$ is not diagonal. This means that the stochastic perturbation of a prognostic variable (or level) should be correlated with the forcing applied to the other variables (or levels) represented by the vector $\mathbf{q}$. Since $\mathbf{F_b}$ is Hermitian, it can be represented by the eigenvalue decomposition

$$\mathbf{F_b} = \mathbf{PGP}^\dagger \quad (A75)$$

where $\mathbf{G}$ is a diagonal matrix with elements defined by the real eigenvalues $\lambda_i$ i=1,...,N of $\mathbf{F_b}$. The unitary matrix $\mathbf{P}$ comprises of the associated eigenvectors. One then defines the vector $\mathbf{g}$ with elements $\sqrt{\lambda_i w_i}$, where $w_i$ is a random number drawn from a Gaussian distribution of zero mean and unit variance. The stochastic perturbation is then given by:

$$\tilde{\mathbf{f}} = \mathbf{Pg}/(2\Delta t) \quad (A76)$$

which satisfies Equation (A74).

The drain and stochastic backscatter can also be combined into a simplified deterministic parameterisation:

$$\tilde{\mathbf{q}}_t^S(t) = -\mathbf{D_{net}}\tilde{\mathbf{q}}(t) \quad (A77)$$

where the net dissipation defined as

$$\mathbf{D_{net}} \equiv \mathbf{D_n} = \mathbf{D_d} + \mathbf{D_b}. \quad (A78)$$

**Appendix H: Data Driven Modelling of Mean Subgrid Tendency**

The mean subgrid tendency can be broken down into its eddy-mean-field, eddy-topographic, mean-field-mean-field, and mean-field-topographic components. The data-driven decomposition method of Kitsios and Frederiksen (2019) [137] was designed to be consistent with the QDIA subgrid structural forms in Appendix F. This decomposition captures the relationship that the mean subgrid tendency $\overline{\mathbf{q}}_t^S = \overline{\mathbf{f}}$ has with the topography and mean field $\overline{\mathbf{q}}$. For each particular scale

$$\overline{\mathbf{f}} = -\overline{\mathbf{D}}\overline{\mathbf{q}} + \chi\mathbf{h} + \mathbf{J}, \quad (A79)$$

where $\overline{\mathbf{D}}$ is a eddy-mean-field dissipation matrix, $\chi$ is the eddy-topographic matrix, $\mathbf{h}$ is the topography, and $\mathbf{J}$ is the mean field Jacobian representing the subgrid interactions involving the mean field and topography. For example, let us consider a simulation with a horizontal scale-based decomposition and two vertical levels. In this case for each wavenumber pair $(m,n)$, $\overline{\mathbf{D}}$ is a 2 × 2 matrix, $\chi$ is the 2 × 2 matrix, $\mathbf{J}$ is a two-element column vector, and $\mathbf{h}$ has elements $(0, h_{mn})^T$, where $h_{m,n}$ are the spectral coefficients of the topography. This means that $\chi_{mn}^{11}$ and $\chi_{mn}^{21}$ are taken as being zero because there is no topography at the top level.

In order to calculate the terms in Equation (A79) we create an ensemble of average states. A given ensemble member, $e$, satisfies

$$\overline{\mathbf{f}}_e = -\overline{\mathbf{D}}\overline{\mathbf{q}}_e + \chi\mathbf{h} + \mathbf{J}_e + \boldsymbol{\varepsilon}_e. \quad (A80)$$

The fields $\overline{\mathbf{f}}_e$ and $\overline{\mathbf{q}}_e$ are time averaged over non-overlapping windows of length $\tau_M$. The mean field Jacobian, $\mathbf{J}_e$, is calculated from $\overline{\mathbf{q}}_e$ according to

$$J_{mn}^j = i \sum_{(p,q,r,s) \in \mathcal{S}} K_{nqs}^{mpr} \left( \overline{\psi}_{-pq}^j \overline{q}_{-rs}^j + \overline{\psi}_{-pq}^j h_{-rs} \delta^{j2} \right) \quad (A81)$$

where the subscript notation $e$ has been dropped here for clarity, and the summations are performed over the subgrid wavenumber set $\mathcal{S}$. The additional vector, $\boldsymbol{\varepsilon}_e$, in Equation (A80) is an error term. Note, $\overline{\mathbf{D}}$, $\chi$ and $\mathbf{h}$ are all time-independent and hence also independent of the ensemble member. This means that the ensemble average of Equation (A80) is equivalent to Equation (A79), where $<\boldsymbol{\varepsilon}_e>= 0$ by definition.

The matrix $\overline{\mathbf{D}}$ is solved by subtracting Equation (A80) from Equation (A79), post-multiplying by $(\overline{\mathbf{q}}_e - \overline{\mathbf{q}})^\dagger$, and ensemble averaging both sides. This produces:

$$\overline{\mathbf{D}} = -\left[ \left\langle (\overline{\mathbf{f}}_e - \mathbf{J}_e) \overline{\mathbf{q}}_e^\dagger \right\rangle - (\overline{\mathbf{f}} - \mathbf{J}) \overline{\mathbf{q}}^\dagger \right] \left( \left\langle \overline{\mathbf{q}}_e \overline{\mathbf{q}}_e^\dagger \right\rangle - \overline{\mathbf{q}} \overline{\mathbf{q}}^\dagger \right)^{-1}. \quad (A82)$$

The error $\boldsymbol{\varepsilon}_e$ is assumed to be is uncorrelated with $(\overline{\mathbf{q}}_e - \overline{\mathbf{q}})^\dagger$. Since $\overline{\mathbf{D}}$ and $\mathbf{J}$ are now both known, one we can calculate the eddy-topographic coefficients term by rearranging Equation (A79) such that

$$\left( \chi_{mn}^{12}, \chi_{mn}^{22} \right)^T = \left[ \overline{\mathbf{f}} + \overline{\mathbf{D}}\overline{\mathbf{q}} - \mathbf{J} \right] / h_{mn}. \quad (A83)$$

Effectively, the relationship between $\overline{\mathbf{f}} - \mathbf{J}$, and $\overline{\mathbf{q}}$ is established using multi-variate linear regression.